%% file: main.tex
\documentclass{article}

\usepackage[a4paper,right=.75in,left=.75in,top=.75in,bottom=.75in]{geometry}
\RequirePackage{natbib}
\usepackage[title]{appendix}
\usepackage{xcolor}
\usepackage{amssymb}
\usepackage{amsmath}
\usepackage{graphicx}
\usepackage{gensymb}
\usepackage{url}
\usepackage{bm}
\usepackage[parfill]{parskip}
\usepackage{xr} 
\usepackage{subcaption}
\usepackage{cleveref}
\usepackage{multirow}
\usepackage{authblk}
\usepackage{pdflscape}
\usepackage{algorithm}
\usepackage{algpseudocode}
\usepackage{comment}
\usepackage{enumitem}

\newcommand{\bs}{\boldsymbol}

\usepackage{lineno}

\usepackage{xr}
\title{Towards Improved Heliosphere Sky Map Estimation with Theseus}
\author[1,*]{Dave Osthus}
\author[1]{Brian P. Weaver}
\author[1,2]{Lauren J. Beesley}
\author[1]{Kelly R. Moran}
\author[1]{Madeline A. Ausdemore}
\author[4]{Eric J. Zirnstein}
\author[5,6]{Paul H. Janzen}
\author[3]{Daniel B. Reisenfeld}
\affil[1]{Statistical Sciences Group, Los Alamos National Laboratory, Los Alamos, New Mexico, USA}
\affil[2]{Information Systems and Modeling, Los Alamos National Laboratory, Los Alamos, New Mexico, USA}
\affil[3]{Space Science and Applications Group, Los Alamos National Laboratory, Los Alamos, New Mexico, USA}
\affil[4]{Department of Astrophysical Sciences, Princeton University, Princeton, New Jersey, USA}
\affil[5]{Department of Physics and Astronomy, University of Montana, Missoula, Montana, USA}
\affil[6]{The New Mexico Consortium, Los Alamos, New Mexico, USA}
\affil[*]{Corresponding author: Dave Osthus, dosthus@lanl.gov}

\date{}

\begin{document}

\maketitle

\abstract{The Interstellar Boundary Explorer (IBEX) satellite has been in orbit since 2008 and detects energy-resolved energetic neutral atoms (ENAs) originating from the heliosphere. Different regions of the heliosphere generate ENAs at different rates. It is of scientific interest to take the data collected by IBEX and estimate spatial maps of heliospheric ENA rates (referred to as sky maps) at higher resolutions than before. These sky maps will subsequently be used to discern between competing theories of heliosphere properties that are not currently possible.
The data IBEX collects present challenges to sky map estimation. The two primary challenges are noisy and irregularly spaced data collection and the IBEX instrumentation's point spread function. In essence, the data collected by IBEX are both noisy and biased for the underlying sky map of inferential interest. In this paper, we present a two-stage sky map estimation procedure called Theseus. In Stage 1, Theseus estimates a blurred sky map from the noisy and irregularly spaced data using an ensemble approach that leverages projection pursuit regression and generalized additive models. In Stage 2, Theseus deblurs the sky map by deconvolving the PSF with the blurred map using regularization. Unblurred sky map uncertainties are computed via bootstrapping. We compare Theseus to a method closely related to the one operationally used today by the IBEX Science Operation Center (ISOC) on both simulated and real data. Theseus outperforms ISOC in nearly every considered metric on simulated data, indicating that Theseus is an improvement over the current state of the art.}

\section{Introduction}
\label{sec:intro}

Launched in 2008, the Earth-orbiting Interstellar Boundary Explorer (IBEX) mission has been continuously providing energy-resolved measurements of energetic neutral atoms (ENAs) from the boundary between our solar system and what lies beyond (i.e., the boundary between the heliosphere and interstellar space) \citep{mccomas2009interstellar}.
ENAs are fast moving particles with no charge. 
They are produced when a particle with a net electric charge in the heliosphere undergoes charge exchange with a particle with a net neutral charge in interstellar space.
The particle that previously had a net electric charge but now has a net neutral charge is called an ENA. 
One property of an ENA is that it no longer interacts with magnetic fields and will travel in a ballistic trajectory (i.e., a nearly straight line at the energies being observed) from the time and location it was neutralized. 
Some of the ENAs created in the heliosphere travel in a ballistic trajectory towards Earth and can be detected by IBEX.  
This is what IBEX is counting: ENAs created in the heliosphere.
By continuously counting energy-resolved ENAs, IBEX is able to create spatial maps of ENA rates at different energies over time, effectively creating dynamic images of the heliosphere.
These maps of ENA rates called \emph{sky maps} have been successfully used to deduce the global structure and dynamics of the heliosphere, leading to groundbreaking advances in our understanding of the interstellar boundary (see Table 1 in \citet{mccomas2017SevenYears}, \citet{mccomas2020SolarCycle}, and studies since then).

The heliosphere is a large cavity shaped like an elongated egg that encapsulates our solar system. 
For context, the outer boundary of the heliosphere, called the heliopause, is over 100 astronomical units from the Sun, where 1 astronomical unit is the mean distance between Earth and the Sun.
The heliosphere travels through interstellar space like a boat traveling through water; the front of the boat traveling forward with the back of the boat following along. 
The compressed front of the heliosphere is called the \emph{nose} and the elongated back is called the \emph{tail}.
The ENAs observed by IBEX primarily come from two different populations.
The primary population is called the \emph{globally distributed flux (GDF)}.
These particles arrive from all directions in the sky, but arrive at higher rates toward the nose and tail of the heliosphere  \citep[e.g.,][]{mccomas2009global,schwadron2014separation}.
The second ENA population is called the \emph{ribbon}.
It is a relatively narrow band of enhanced ENA emissions that form a nearly complete circular band across much of the sky \citep{fuselier2009width}.
The ribbon was entirely unanticipated by any model or theory prior to its discovery by IBEX: a bona fide success for discovery science \citep{mccomas2009global}.

As the ribbon was unknown prior to IBEX's launch in 2008, theories to explain its origin are actively being developed. 
To date, over a dozen theories have been posited to explain the ribbon’s origin (see list in \citet{mccomas2014RibbonSources} and more recently, \citet{zirnstein2018weak, zirnstein2019strong, zirnstein2021RibbonGeometry}), the details of which are outside the scope of this paper.
What is relevant is that different ribbon origin theories can lead to different ribbon profiles or shapes.
For example, \citet{zirnstein2019strong} shows how two different ribbon origin theories~\textemdash~the Weak Scattering model and the Spatial Retention model~\textemdash~result in distinct ribbon shapes.
If those shapes could be discerned with observations, that could add support to one model over the other.
The current state-of-the-art sky map rendering approach used by the IBEX Science Operations Center (ISOC), however, produces maps in which the ribbon profile is not well enough resolved to discern between competing ribbon formation theories. 
Furthermore, finer-scale GDF structures are also not clearly resolved (e.g., enhancements that appear at higher energies in the tailward direction \citep{zirnstein2016HeliosheathGeom}).
Improved sky map estimates are needed.

 The principal instrument on IBEX for making the observations used to construct ENA sky maps is the IBEX-Hi ENA imager \citep{funsten2009interstellar}.  
 Prior to the launch of IBEX, there was no expectation that ENA sources as narrow as the ribbon would be present.
 The IBEX-Hi ENA imager has a roughly circular instantaneous field-of-view (FOV) having a $6.5\degree$ full-width at half maximum (FWHM) angular resolution \citep{funsten2009interstellar}. What that means is when IBEX-Hi detects an ENA, even though IBEX-Hi records the direction of the boresight of the FOV in the sky at the time of detection (i.e., the location in the sky IBEX-Hi was aimed at the time of detection), that does not mean the detected ENA originated from that same boresight location. 
 The detected ENA could have originated from any location ``near" IBEX-Hi's boresight, where ``near" is probabilistically defined by the IBEX-Hi point spread function (PSF): a peaked, roughly conical-shaped distribution that describes the likelihood of ENA detection as a function of angle off the boresight.  
 In practice, the PSF creates a blurring effect on the sky maps; a blurring effect that should be accounted for and removed in order to resolve more granular sky map features (e.g., the ribbon's fine structure).
 Currently, the ISOC sky maps do not account for the PSF or attempt to remove its blurring effect.

Another limitation of ISOC sky maps is resolution.
In principle, a sky map is a continuous spatial map where every location in the spatial domain has a corresponding ENA rate. In practice, a sky map is a spatial map discretized into non-overlapping pixels where every pixel has a corresponding ENA rate.
ISOC sky maps are made at a fixed $6\degree$ pixel resolution and are unable to be reliably created at finer resolutions, resulting in a pixelated aesthetic.

In this paper we introduce a two-stage sky map estimation procedure called \emph{Theseus} to address both the need to discriminate between competing ribbon theories and to resolve narrow GDF features and the opportunity to improve sky maps by accounting for and removing the blurring effect of the PSF. 
In Stage 1, Theseus \emph{estimates} the as-observed, or blurred, sky map from the noisy and irregularly spaced binned direct event data using an ensemble approach that leverages projection pursuit regression and generalized additive models. 
In Stage 2, Theseus \emph{deblurs} the sky map by deconvolving the PSF with the blurred map using regularization. 
Sky map uncertainties are computed via bootstrapping. 
For Theseus, the sky map resolution is user-defined, allowing for more resolved sky maps than the fixed $6\degree$ ISOC sky maps. 

The contributions made in this paper are four-fold. 
1) Never before have science operations estimated sky maps been compared to simulated sky maps, allowing us, for the first time, to \emph{quantitatively} evaluate different sky map estimation procedures. 
2) Theseus is the first procedure used to estimate sky maps that directly accounts for and removes the blurring effect of the PSF. 
3) Theseus makes maps at a user-defined resolution and executes principled interpolation; the current state of the art makes sky maps at a fixed $6\degree$ resolution and leaves ``holes" in the sky maps where data are missing. 
4) We show that Theseus sky maps are better than current ISOC sky maps when compared over a variety of evaluation metrics on a collection of simulated sky maps.

The paper proceeds as follows. In Section \ref{sec:datadescription}, we describe the binned direct event data collected by IBEX-Hi. In Section \ref{sec:methodology}, we describe the sky map making methodology, Theseus. In Section \ref{sec:simstudy}, we compare Theseus to the current state-of-the-art sky map estimation procedure used by the ISOC on simulated maps capturing different ribbon models. In Section \ref{sec:realdataexample}, we fit Theseus and the ISOC method to real data collected by IBEX-Hi. Finally in Section \ref{sec:discussion}, we provide a discussion of future work.

\section{The IBEX-Hi Binned Direct Event Data}
\label{sec:datadescription}
IBEX collects temporally-, spatially-, and energy-resolved ENA measurements from the heliosphere (i.e., IBEX images the entire sky).
IBEX rotates at $\sim 4$ revolutions per minute around a Sun-pointing spin axis.
The IBEX-Hi imager's boresight is aligned perpendicular to the spacecraft spin axis, thus over the course of a spin, IBEX-Hi images ENAs arriving from a roughly $6.5\degree \times 360\degree$  circular band of the sky. IBEX orbits the Earth with an orbital period of $\sim 9$ days, and the spin axis is repointed toward the sun roughly every 4.5 days, leading to bands of sky viewing centered $\sim 4.5\degree$ apart.\footnote{For the first few years of the IBEX mission, IBEX reorientation occurred approximately every 7 days, or once per orbit, not 4.5 days. In June 2011, the IBEX orbit was changed to a 9-day period orbit and since then, repoints are done twice per orbit.}  Two different bands of sky viewing occur during an orbit, one on the ascending arc of the orbit, and one on the descending arc; thus, each band is referred to as an \emph{arc} of data collection.  The intensity of ENAs of heliospheric origin does not change appreciably over the course of months, and thus the average ENA rate arriving from a specific direction averaged over the course of an arc is considered a true measure of the ENA rate from that location in the sky. 
It takes IBEX roughly 180 days to image the whole sky. 
As such, sky maps are made in 6-month cadences.\footnote{Heliospheric ENA rates vary on the order of months and are treated as constant for each 6-month sky map.} 
The maps corresponding roughly to the first part of the calendar year are labeled ``A" maps; the latter half ``B" maps.

The IBEX-Hi instrument is continuously detecting energy-resolved ENAs in the range of 500 eV to 6 keV. The instrument covers this range by stepping through six discrete energy settings of its electrostatic analyzer (ESA).
ESA step 2 is the lowest step used for science, and ESA step 6 is the highest.\footnote{Due to data quality issues, ESA 1~\textemdash~the lowest energy step~\textemdash~is not used.} 
Each IBEX-Hi measurement event is referred to as a \emph{direct event}, and occurs when particle detectors within the instrument are triggered in a specified sequence.  When a direct event is measured, numerous pieces of information are recorded including the \emph{time} the direct event occurred, the \emph{spatial location} in the sky the IBEX-Hi boresight was pointed, and the \emph{ESA setting} IBEX-Hi was set to at the time of the direct event measurement. The direct events amount to energy-resolved, spatial-temporal point process data \citep{gonzalez2016spatio}.

While direct events are the raw measurements made by IBEX-Hi, that is not the data product we work with. 
Instead, we work with a curated data product called the \emph{binned direct event data}. 
The binned direct event data are in essence a partitioned, aggregated, and processed version of the direct events. 
The process to convert direct events to binned direct events is standard for the IBEX mission.
Essentially, the direct events measured in an arc are partitioned into 360 $1\degree$ bins for each ESA step.
The direct events assigned to each ESA step/$1\degree$ bin amounts to temporal point process data, where the total number of direct events would ideally equal the number of ENAs measured by the IBEX-Hi instrument when set to a specified ESA step and pointed at a $1\degree$ bin of the sky.
Unfortunately, not all direct events are ENAs of heliospheric origin.
A direct event could be a heliospheric ENA or it could be a background particle (i.e., a high-energy penetrating radiation particle, a non-heliospheric ENA, or another particle local to the spacecraft environment that enters the instrument aperture mimicing a heliospheric ENA). 
A differentiating behavior of background direct events are their degree of spatial uniformity and the time scale of their dynamics. 
Much of the background of local origin varies much more rapidly in time, and varies with viewing direction.
A standard IBEX data processing step segments the temporal point process direct event data for each ESA step/$1\degree$ bin and discards the direct events in the time segments where it is determined that heliospheric ENA direct events are either overwhelmed by background particles, background rates change too quickly, or background particles are associated with unusual local solar wind conditions.
Generally speaking, the time segments that are kept in this IBEX-standard culling step are times when the background rate is relatively constant across locations, slowly varying in time, and the background rates are not appreciably larger than their corresponding ENA rates. 
The total duration of the time segments kept after the culling step is referred to as the binned direct event datum's \emph{exposure time}. 

A binned direct event datum corresponds to a location in the sky (i.e., $1\degree$ bin) and an ESA setting for a given 6-month map. A binned direct event datum includes the IBEX-Hi boresight direction, the exposure time, the number of measured direct events, and a background rate (which is another standard IBEX data product). 
The background rate is technically an estimate, but with small uncertainties. 
In this paper, the background rate is treated as known. 

\begin{figure}[!ht]
\centering
\includegraphics[width=.24\linewidth]{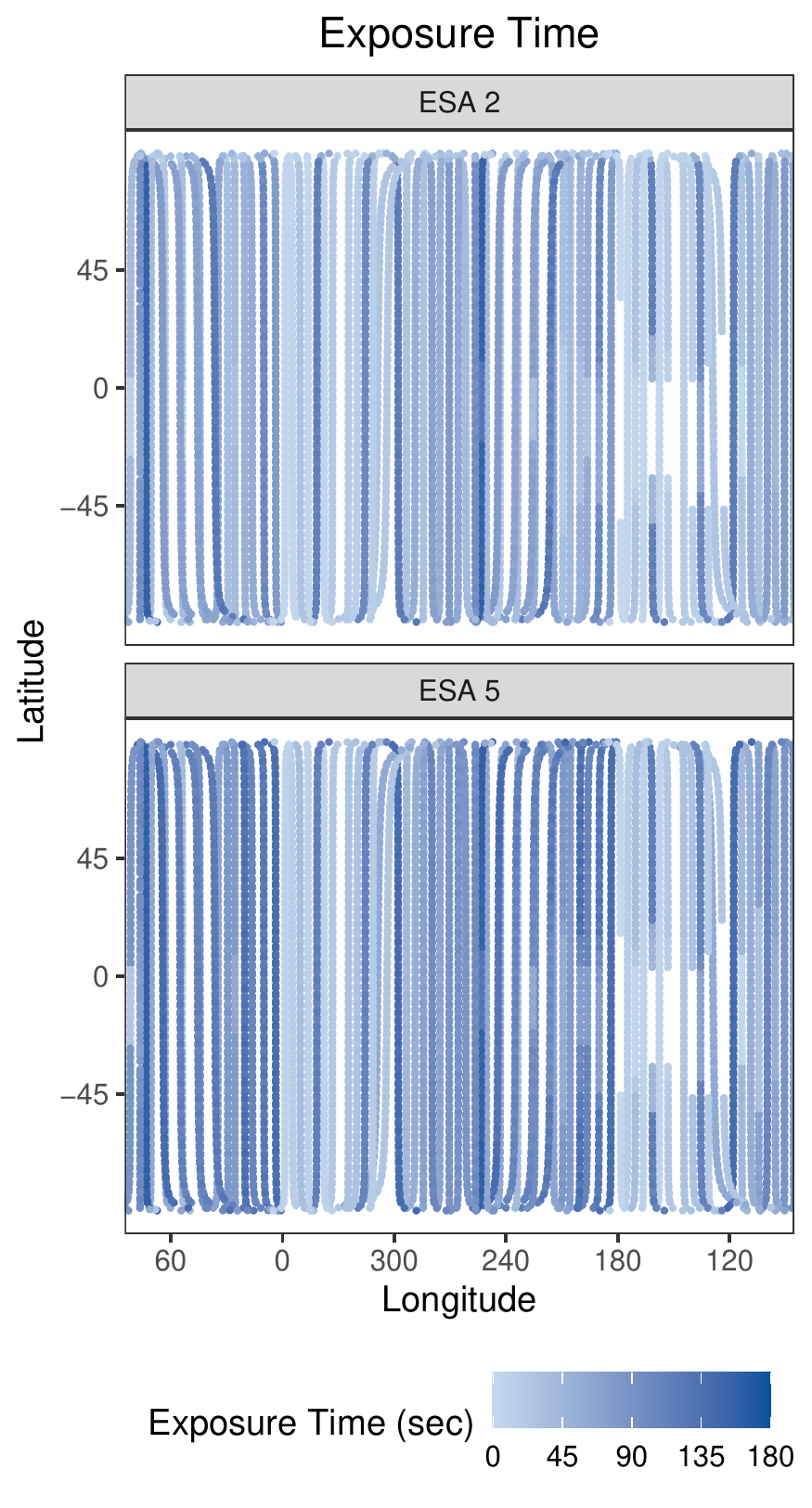}
\includegraphics[width=.24\linewidth]{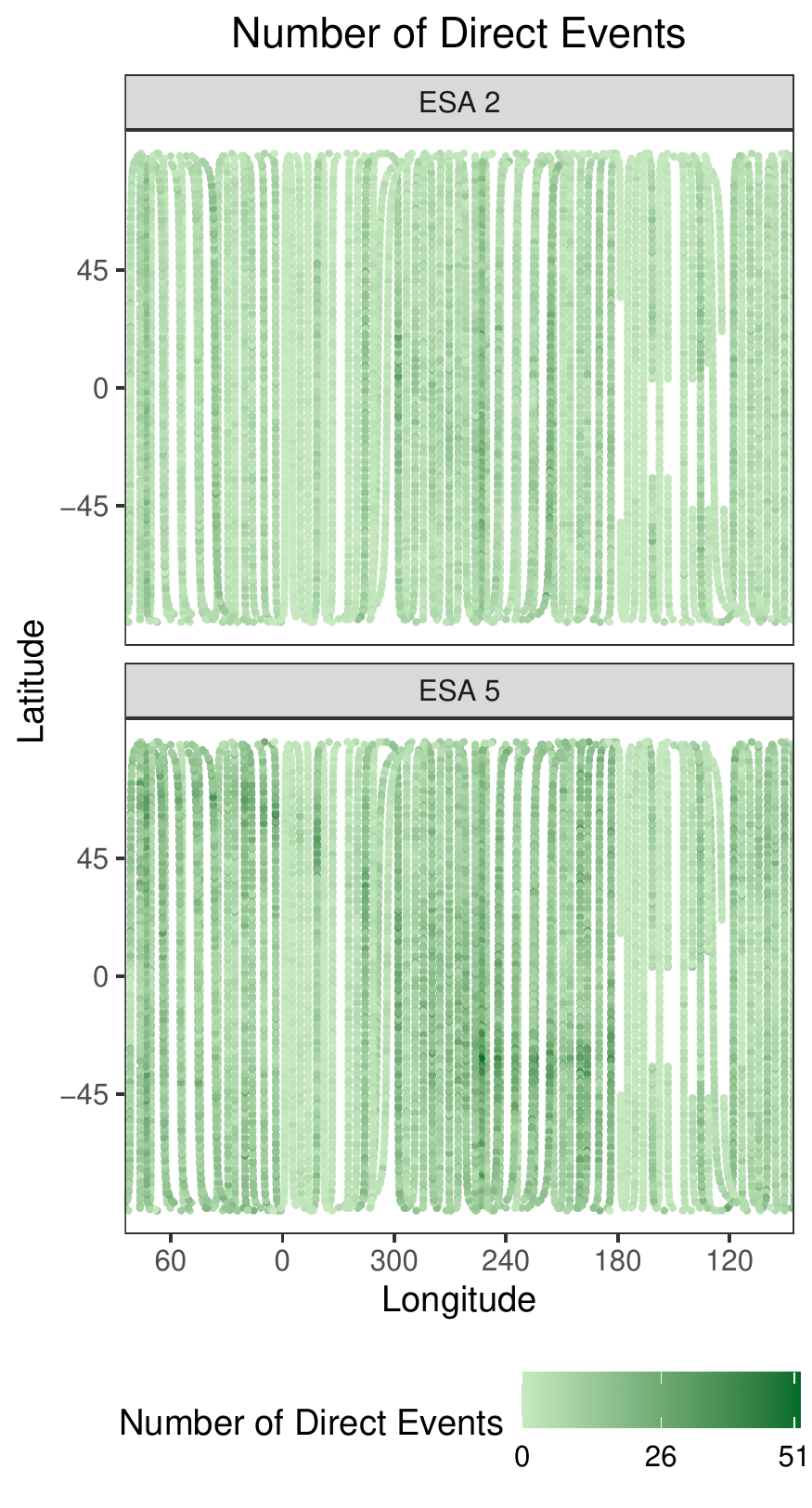}
\includegraphics[width=.24\linewidth]{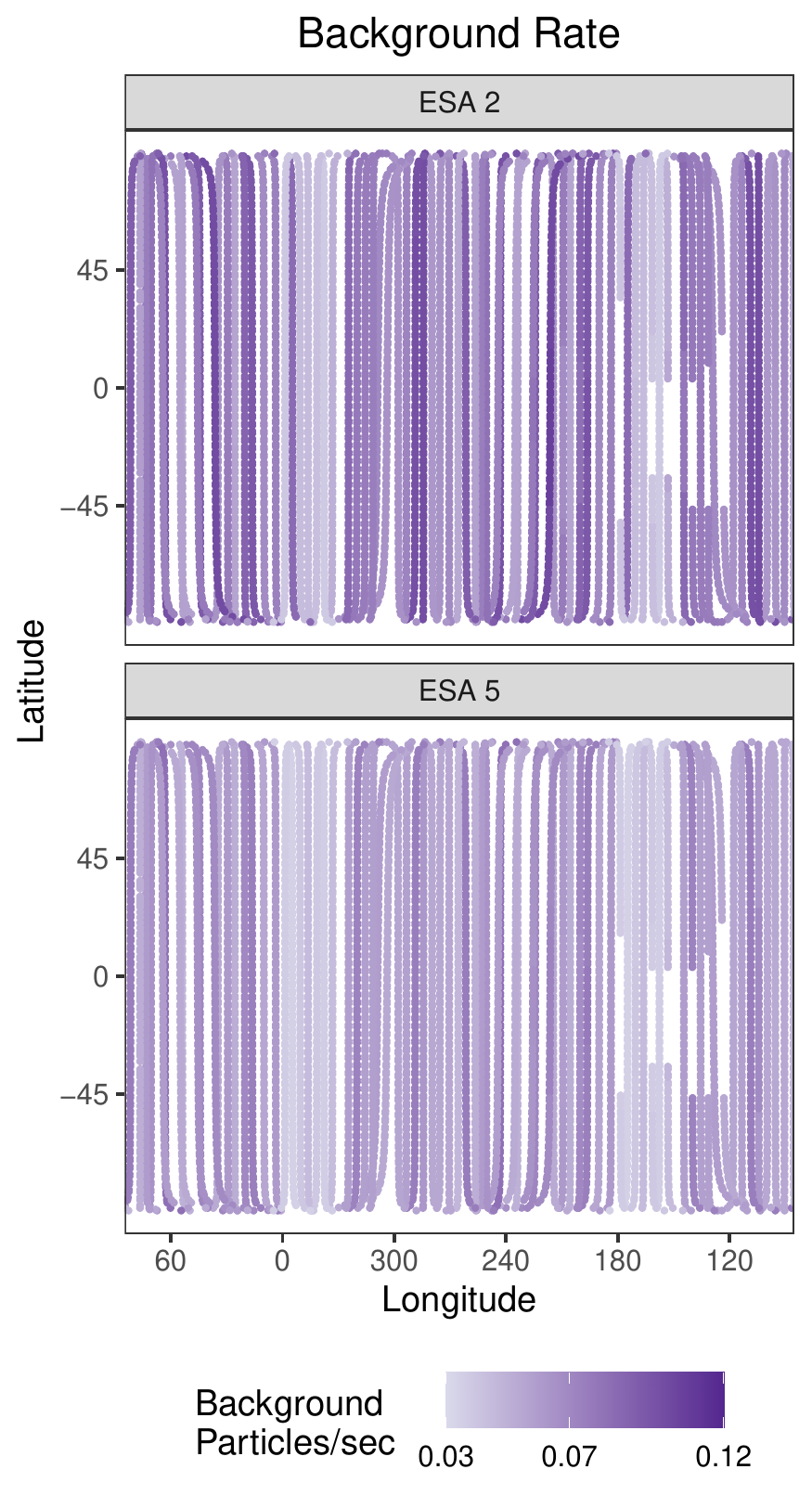}
\includegraphics[width=.24\linewidth]{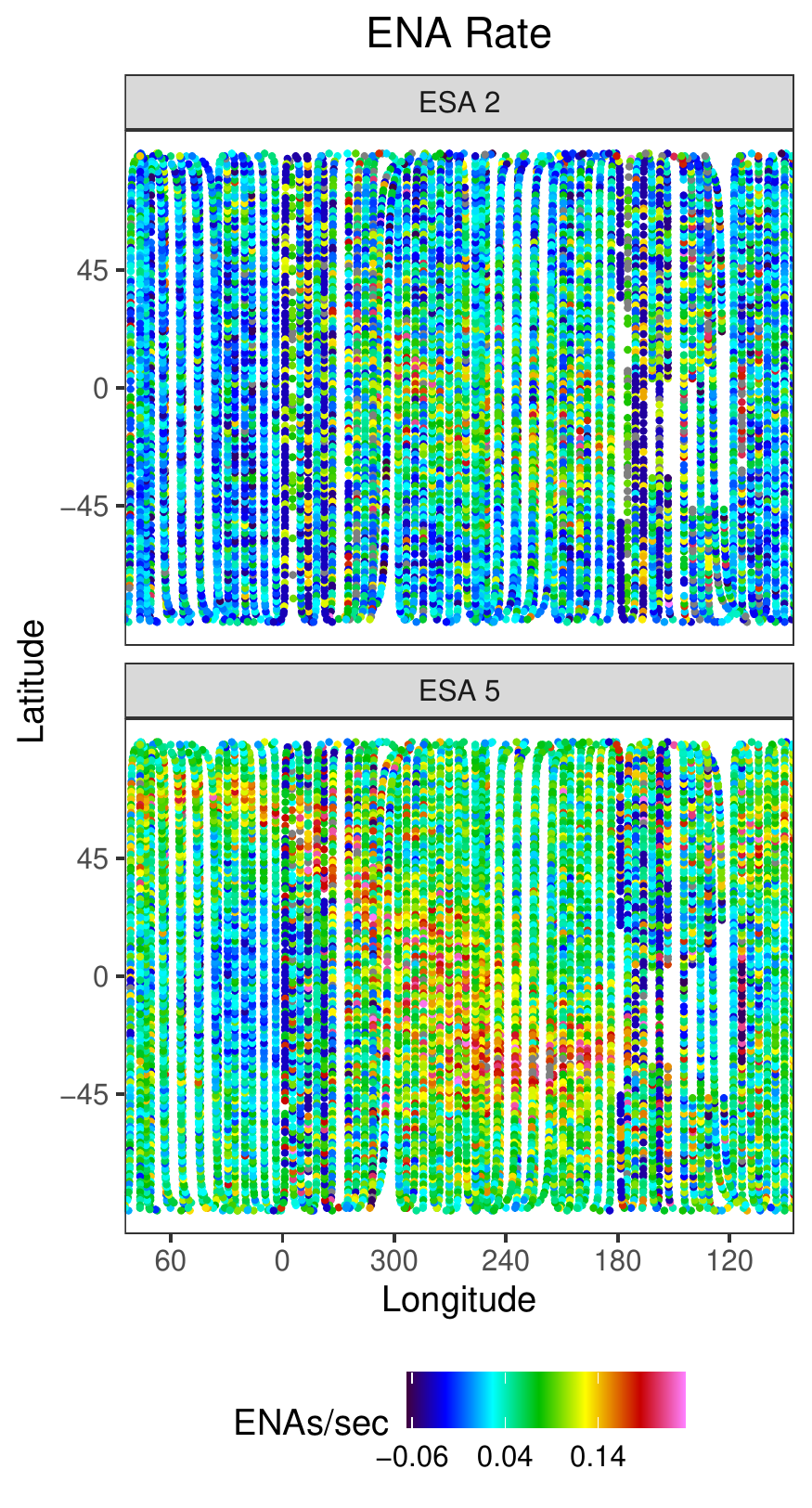}
\caption{\small Binned direct event data ecliptic longitude and latitude locations (points) for ESA steps 2 and 5 (rows) corresponding to the 2013A sky map. From left to right, the columns show the exposure time, number of direct events, background rate, and ENA rate, where ENA rate is computed as the number of direct events/exposure time - background rate. The ribbon can be seen in ESA step 5's ENA rate plot as the streak of red and yellow points starting at longitude and latitude (60, 70), extending down and right to roughly (240, -40), and extending up and right to roughly (130, 45).}
\label{fig:eda}
\end{figure}

For illustration, Figure \ref{fig:eda} plots the exposure time, the number of direct events, the background rate, and an estimated ENA rate (number of direct events/exposure time - background rate) for ESA steps 2 and 5 for the 2013A sky map.
The plots in Figure \ref{fig:eda} and throughout this paper are centered on longitude 265, with longitudes in reverse order (i.e., larger longitude values are on the left and smaller ones on the right, before centering). 
The centering on longitude 265 corresponds to a ``nose-centered" frame, where the nose of the heliosphere is placed roughly in the center of the map.
The plotting of longitudes in reverse order before centering is standard in astronomy, as the sky is being viewed from \emph{inside} the heliosphere.
There are 13,716 and 13,740 binned direct event data points for ESA steps 2 and 5, respectively.
The exposure time varies by location, with exposure times nominally between 0 and 180 seconds.\footnote{180 seconds $\times$ 6 ESA steps $\times$ 360 $1\degree$ bins $=$ 388,800 seconds $=$ 4.5 days, which is the duration of an arc.} Locations that have missing binned direct event data correspond to ESA steps/bins where the entire time series of direct events was culled. 
This creates gaps in the binned direct event data. 
Those gaps are correlated across ESA steps, as the reason for culling at one energy step is often the same at other energy steps. 
Roughly two-thirds of the possible collection time is discarded by this standard IBEX data processing step. 
In general, lower energy ESA steps tend to have more data discarded than higher energy ESA steps.
For ESA steps 2 and 5 of the 2013A sky map, the number of direct events per bin is between 0 and 51.
The number of measured direct events is correlated with exposure time; the larger the exposure time, the more measured direct events.
The background rate varies between 0.03 and 0.12 background particles per second and is typically common for all bins within an ESA and arc.
Finally, the estimated ENA rate varies by ESA step.
The ribbon can be seen in ESA step 5 as the streak of red and yellow points roughly connecting the longitude and latitude points (60,70), (240, -40), and (130, 45).
There is appreciable noise in the crude ENA rate estimates, with many negative estimated ENA rates (which are physically impossible).

\section{Theseus Methodology}
\label{sec:methodology}
In this section, we outline Theseus~\textemdash~our sky map estimation methodology. 
In Section \ref{subsec:datamodel}, we describe the data model, define terms and lay out notation.
Theseus is a two-stage estimation procedure that, for a specified ESA and 6-month map (e.g., the 2013A, ESA 2 sky map), takes a binned direct event data set $\bs{D}$ as input and outputs an estimated \emph{unblurred sky map} (i.e., a sky map where the blurring effect of the PSF has been removed) and \emph{blurred sky map} (i.e., a sky map where the blurring effect of the PSF has not been removed).
More specifically, in Stage 1, Theseus estimates a blurred sky map from a binned direct event data set (Section \ref{subsec:stage1}). 
In Stage 2, Theseus estimates an unblurred sky map from the estimated blurred sky map in Stage 1 (Section \ref{subsec:stage2}). 

Sky map uncertainties are estimated via percentile bootstrap intervals \citep[e.g.,][Chapter 13]{efron1994introduction}. 
As such, Theseus is run $N_B$ times, resulting in $N_B$ estimated unblurred and blurred sky maps corresponding to $N_B$ different bootstrap binned direct event data sets $\bs{D}_b$ for bootstrap sample $b=1,2,.\ldots,N_B$.
In this paper, $N_B = 1,000$.
$\bs{D}_b$ is constructed by first sampling the rows of $\bs{D}$ with replacement, where a row corresponds to a binned direct event datum (recall Section \ref{sec:datadescription}), and then simulating the number of direct events from its fitted Poisson data model consistent with Equation \ref{eq:datamodel}.
While each estimated unblurred and blurred sky map from Theseus is properly indexed by ESA, 6-month map, and bootstrap sample, those indices will be suppressed throughout the remainder of Section \ref{sec:methodology} for notational simplicity. In what follows, we describe how Theseus estimates an unblurred and blurred sky map for a specified ESA,  6-month map, and bootstrap sample.

Finally, for clarity, we explicitly list some of the notational conventions used throughout this paper. 
\begin{itemize}
    \item Spatial quantities indexed by $i$ will correspond to locations of binned direct event data while spatial quantities indexed by $j$ will correspond to locations of sky map pixel centers (recall from Section \ref{sec:intro} that, in practice, a sky map is defined by a discrete collection of mutually exclusive and exhaustive pixels).
    \item Non-bolded quantities will be used for scalars, while bolded quantities will be used for non-scalars (e.g., vectors and matrices).
    \item Numerous sky map-related quantities will be estimated throughout the paper. Final estimates of quantities will be denoted with ``hats" (e.g., $\hat{\theta}$), while intermediate estimates needed to arrive at the final estimated quantities will be denoted with ``dots" (e.g., $\dot{\theta}$) or ``double dots" (e.g., $\ddot{\theta}$).
\end{itemize}

\subsection{Data Model}
\label{subsec:datamodel}
Let $\bs{s}_i = (\text{lon}_i,\text{lat}_i)$ be the ecliptic longitude/latitude coordinates of location $\bs{s}_i$ for binned direct event datum $i = 1,2,\ldots,N_o$ where $N_o$ is the number of binned direct event data observations, $\text{lon}_i \in [0, 360)$, and $\text{lat}_i \in [-90, 90]$. 
For the basic data model, we have
\begin{linenomath}\begin{align}
    \label{eq:datamodel}
    y(\bs{s}_i)~|~t(\bs{s}_i),\theta(\bs{s}_i), b(\bs{s}_i) &\sim \text{Poisson}\big(t(\bs{s}_i)\big[\theta(\bs{s}_i) + b(\bs{s}_i)\big]\big)\\
    \label{eq:blurringfn}
    \theta(\bs{s}_i) &= \bs{K}(\bs{s}_i) \bs{\Theta}
\end{align}\end{linenomath}
\noindent where 
\begin{itemize}
    \item $y(\bs{s}_i) \in 0,1,2, \ldots $~is the number of direct events (ENAs $+$ background particles) counted by IBEX-Hi
    \item $t(\bs{s}_i) > 0$ is the known exposure time (seconds)
    \item $b(\bs{s}_i) \geq 0$ is the known background rate (background particles/second)
    \item $\theta(\bs{s}_i) \geq 0$ is the unknown ENA rate (ENAs/second) corresponding to the blurred sky map
    \item $\bs{K}(\bs{s}_i)$ is a known $1 \times N_p$ row vector. It is the \emph{blurring operator} encoding the properties of the IBEX-Hi PSF (those properties explained further below) corresponding to the boresight location $\bs{s}_i$. $N_p$ is the number of mutually exclusive and exhaustive pixels the sky map has been partitioned into (e.g., for a $2\degree$~sky map, $N_p = (180/2)*(360/2) = 16,200$ pixels). $\bs{K}(\bs{s}_i)$ has two properties: 
    \begin{enumerate}
        \item $\bs{K}(\bs{s}_i)_j \geq 0$ for all $j \in 1,2,\ldots,N_p$
        \item $\sum_{j=1}^{N_p} \bs{K}(\bs{s}_i)_j = 1$
    \end{enumerate}
    \noindent where $\bs{K}(\bs{s}_i)_j$ is the $j^{\text{th}}$ entry of the $\bs{K}(\bs{s}_i)$ row vector.
    \item $\bs{\Theta}$ is the \emph{unblurred sky map}: an unknown, non-negative $N_p \times 1$ column vector of ENA rates.
\end{itemize}
\noindent \emph{The unblurred sky map $\bs{\Theta}$ is the quantity of inferential interest}. Our goal is to estimate $\bs{\Theta}$, with uncertainty, given $\bs{y} = \{y(\bs{s}_1), y(\bs{s}_2), \ldots, y(\bs{s}_{N_o})\}$, $\bs{t} = \{t(\bs{s}_1), t(\bs{s}_2), \ldots, t(\bs{s}_{N_o})\}$, and $\bs{b} = \{b(\bs{s}_1), b(\bs{s}_2), \ldots, b(\bs{s}_{N_o})\}$.

One of the primary challenges with estimating $\bs{\Theta}$ is the obscuring role the blurring operator $\bs{K}(\bs{s}_i)$ plays. Equation \ref{eq:datamodel} models the number of direct events counted by IBEX-Hi after being pointed at location $\bs{s}_i$ for $t(\bs{s}_i)$ seconds. On average, $t(\bs{s}_i)b(\bs{s}_i)$ direct events are background particles and $t(\bs{s}_i)\theta(\bs{s}_i)$ are ENAs. However, not all ENAs counted by IBEX-Hi when pointed at location $\bs{s}_i$ in the sky originated from location $\bs{s}_i$ because of the IBEX-Hi PSF. $\bs{K}(\bs{s}_i)$ probabilistically describes the mapping between the true ENA origin location to the observed ENA location $\bs{s}_i$. Specifically, $\bs{K}(\bs{s}_i)_j$ is the probability an ENA originated from within pixel $j$ of the sky map given it was counted by the IBEX-Hi instrument when pointed at location $\bs{s}_i$. The PSF is a property of the IBEX-Hi instrument and must be accounted for in the statistical model. How $\bs{K}(\bs{s}_i)$ is constructed is detailed in Supplementary Materials \ref{appendix:K}.

\subsection{Theseus Stage 1: Estimating the Blurred Sky Map $\bs{\theta}$}
\label{subsec:stage1}
The first stage of Theseus is to estimate the true, blurred sky map $\bs{\theta}$ where $\bs{\theta} = \bs{K} \bs{\Theta}$ and $\bs{K}$ is a $N_p \times N_p$ matrix where the $j^{\text{th}}$ row of $\bs{K}$ equals $\bs{K}(\bs{s}_j)$ for pixel center location $\bs{s}_j$. The true, blurred sky map $\bs{\theta}$ is a spatial surface wrapped on a sphere, where the binned direct event datum at location $\bs{s}_i$ provides an unbiased but noisy estimate of $\theta(\bs{s}_i)$. Recall from Equation \ref{eq:datamodel} that 
\begin{linenomath}\begin{align}
    \text{E}(y(\bs{s}_i)~|~t(\bs{s}_i), \theta(\bs{s}_i), b(\bs{s}_i)) = t(\bs{s}_i)\big[\theta(\bs{s}_i) + b(\bs{s}_i)\big].
\end{align}\end{linenomath}
\noindent Let
\begin{linenomath}\begin{align}
    z(\bs{s}_i) &= y(\bs{s}_i)/t(\bs{s}_i) - b(\bs{s}_i)
\end{align}\end{linenomath}
\noindent be the method of moments estimate of ENAs per second at location $\bs{s}_i$, treated as an initial estimate of $\theta(\bs{s}_i)$. 

For Theseus Stage 1, let $\bs{s}_i$ be the ecliptic longitude/latitude location in \emph{spherical coordinates}. Theseus Stage 1 uses spherical coordinates rather than ecliptic coordinates for estimation to naturally account for spherical wrapping and prevent edge effects. We model $z(\bs{s}_i)$ as a noisy realization from $\theta(\bs{s}_i)$:
\begin{linenomath}\begin{align}
    z(\bs{s}_i) &= \theta(\bs{s}_i) + \epsilon(\bs{s}_i),
\end{align}\end{linenomath}
\noindent where $\epsilon(\bs{s}_i)$ is a mean 0 error term. Our goal is to estimate the spatial surface, $\bs{\theta}$. We estimate $\bs{\theta}$ in three steps:
\begin{enumerate}[wide, labelindent=0pt]
    \item[] \textbf{Step 1}: Estimate candidate blurred sky map $\dot{\bs{\theta}}_l$ for $l=1,2,\ldots,N_L$.
    \item[] \textbf{Step 2}: Estimate a single, initial blurred sky map $\dot{\bs{\theta}}$ as an ensemble of the candidate blurred sky maps $\dot{\bs{\theta}}_l$.
    \item[] \textbf{Step 3}: Estimate the residual-adjusted blurred sky map $\ddot{\bs{\theta}}$ as a residual-adjusted version of $\dot{\bs{\theta}}$.
\end{enumerate}

These three steps provide the high-level roadmap for Theseus Stage 1. There are, however, many subjective decisions that must be made in order to operationalize these steps. It is difficult to make ``optimal" decisions in a discovery science setting where minimal assumptions about the underlying structure of the sky map are available. Therefore in this manuscript we focused on developing procedures that protect against sub-optimal decisions rather than seeking optimal ones. We will show that the collective set of decisions we made ultimately led to improved sky map estimates.

\textbf{Step 1: Estimate $\dot{\bs{\theta}}_l$.}
The goal of this step is to estimate a diverse and computationally inexpensive collection of candidate blurred sky maps. We consider two different regression approaches to estimate $\bs{\theta}$: projection pursuit regression (PPR) \citep{friedman1981projection} and generalized additive models (GAMs) \citep{hastie1987generalized}. PPR and GAMs are both general purpose, non-parametric regression models, useful for estimating continuous surfaces of varying degrees of smoothness. They do so by estimating $\bs{\theta}$ as the sum of smooth functions of inputs (e.g., spherical locations $\bs{s}_i$). Numerous choices are required before fitting can commence, including the choice of smoothing functions and observation weighting. The details of the PPR and GAM fits can be found in Supplementary Materials \ref{appendix:stage1details}.  

\begin{figure}[!ht]
\centering
\includegraphics[width=1\linewidth]{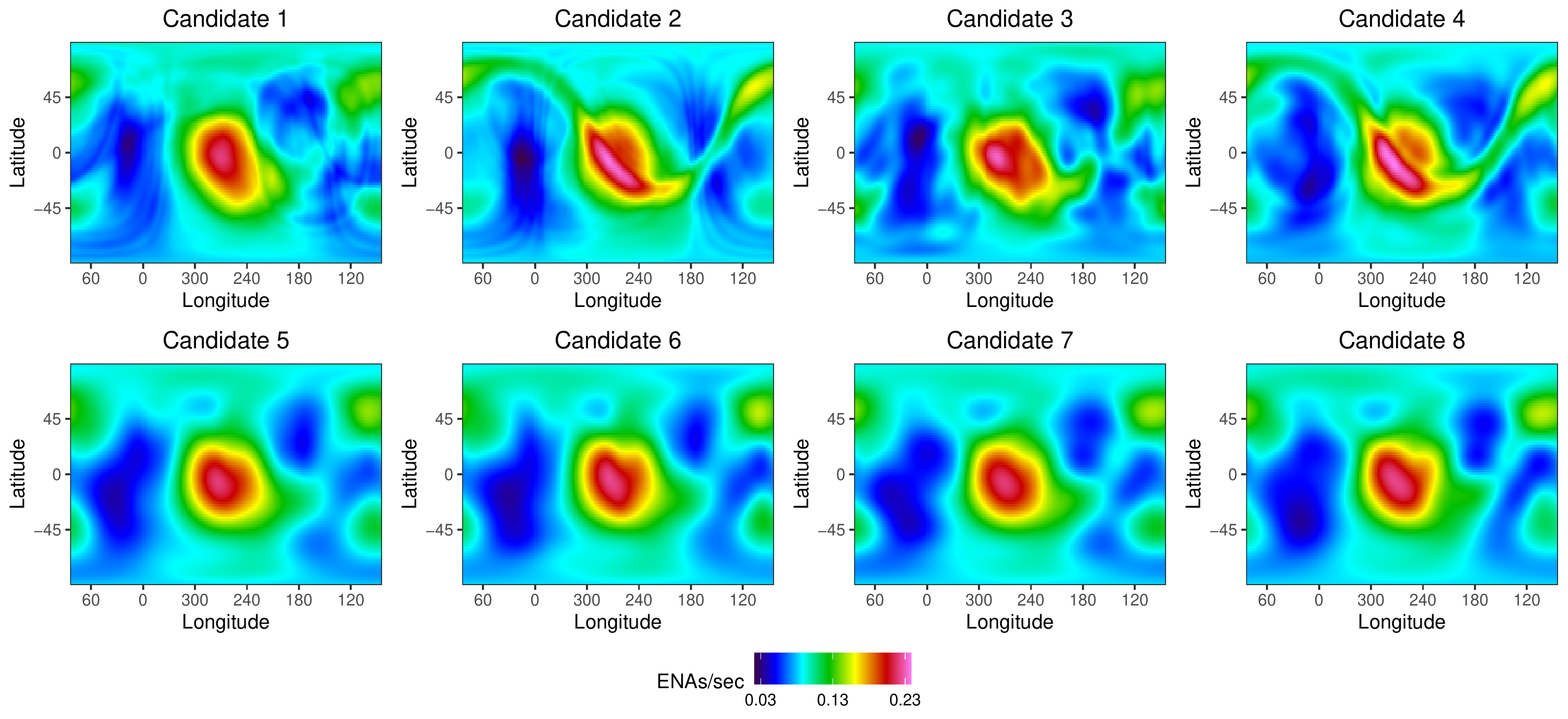}
\caption{\small $N_L=8$ candidate blurred sky map estimates. The top row correspond to the PPR-generated candidate blurred sky maps; the bottom row corresponds to the GAM-generated ones.}
\label{fig:stage1step1components}
\end{figure}

To assist in describing Theseus, we will present a working example in Sections \ref{subsec:stage1} and \ref{subsec:stage2} corresponding to a single bootstrapped data set of a simulated binned direct event data set generated from a simulated Spatial Retention sky map (details in Section \ref{sec:simstudy}).
Figure \ref{fig:stage1step1components} shows $N_L = 8$ candidate blurred sky map estimates: 4 PPR-generated and 4 GAM-generated. At a high-level, each candidate estimates a similar blurred sky map: they all estimate the highest ENA rate around ecliptic longitude/latitude (265, -15), and two other lobes of elevated ENA rates near (100, 50) and (90, -45). The specifics of the estimated blurred sky maps, however, noticeably vary from candidate to candidate. For instance, Candidates 5 through 8 (the GAM-generated candidates) tend to be smoother but do not obviously estimate a ribbon, while Candidates 2 and 4 (the PPR-generated candidates using exposure time weighting) do visually estimate a ribbon. Banding artifacts are present in some of the PPR-generated candidate sky maps (e.g., Candidates 1 and 2) as a result of the internal PPR rotations and ridge function estimates. As Figure \ref{fig:stage1step1components} makes clear, the particular choices of smoothing functions, weightings, and regression approaches are consequential and challenging to select \emph{a priori}. Rather than try to determine what the ``optimal" smoothing function/weighting/regression approach is, we combine the candidate blurred sky maps to form a single, initial blurred sky map estimate.

\textbf{Step 2: Estimate $\dot{\bs{\theta}}$.}
The goal of Step 2 is to combine the $N_L$ candidate blurred sky maps $\dot{\bs{\theta}}_l$ into a single, initial blurred sky map estimate $\dot{\bs{\theta}}$.
A straightforward way to do this would be to learn a weight $w_l \in [0,1]$ for each candidate map such that $\sum_{l=1}^{N_L} w_l = 1$ and the resulting initial blurred sky map estimate $\dot{\bs{\theta}} = \sum_{l=1}^{N_L} w_l \dot{\bs{\theta}}_l$ minimizes some distance between the binned direct event data and $\dot{\bs{\theta}}$. This approach, however, is restricted to assign a single weight to each candidate blurred sky map rather than assign different weights to different parts of candidate blurred sky maps. 
From Figure \ref{fig:stage1step1components}, it appears different candidate blurred sky maps may capture some parts of the blurred sky map better than others. 
For instance, the PPR-generated candidates might be better suited to capture the ribbon, while the GAM-generated candidates might be better suited to the GDF.

\begin{figure}[!ht]
\centering
\includegraphics[width=1\linewidth]{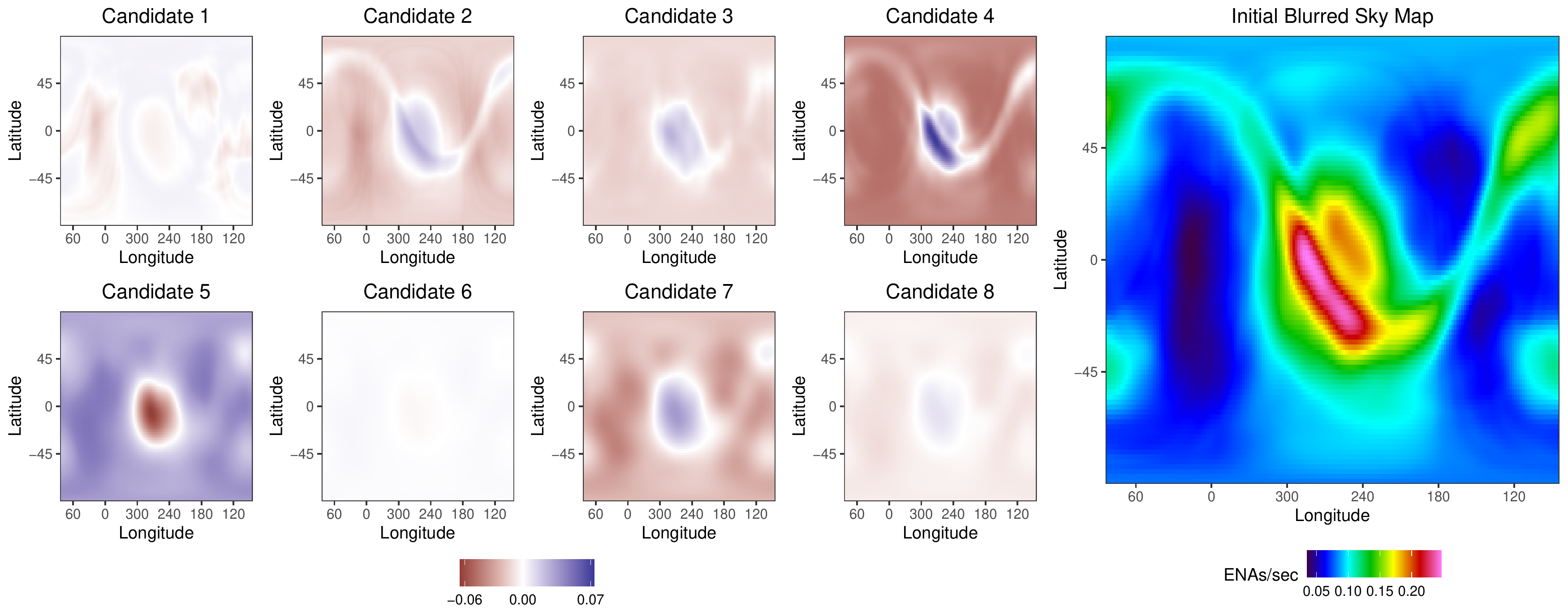}
\caption{\small The candidate blurred sky map components $f_l\big(\dot{\theta}_l(\bs{s}_i)\big)$ (left) and the initial blurred sky map estimate $\dot{\bs{\theta}}$ (right). Following Equation \ref{eq:gambetadot}, the initial blurred sky map estimate is calculated as the sum of the candidate blurred sky map components $f_l\big(\dot{\theta}_l(\bs{s}_i)\big)$ plus the estimated constant $\alpha_0 \approx 0.15$ ENAs/sec.}
\label{fig:stage1step2gamcomponents}
\end{figure}

As different weights for different parts of candidate blurred sky maps are desired, we choose to combine the candidate blurred sky maps into a single, initial blurred sky map estimate by treating the candidate blurred sky maps as covariates in a GAM regression model. Specifically, we fit the model 
\begin{linenomath}\begin{align}
    \label{eq:theseusstage1step2datamodel}
    y(\bs{s}_i)~|~t(\bs{s}_i), \dot{\theta}_1(\bs{s}_i),\ldots,\dot{\theta}_{N_L}(\bs{s}_i), b(\bs{s}_i) &\sim \text{Poisson}\bigg(t(\bs{s}_i)\bigg[\dot{\theta}(\bs{s}_i) + b(\bs{s}_i)\bigg] \bigg)\\
    \label{eq:gambetadot}
    \dot{\theta}(\bs{s}_i) &= \alpha_0 + \sum_{l=1}^{N_L} f_l\bigg(\dot{\theta}_l(\bs{s}_i)\bigg)\\
    \label{eq:theseusstage1step2basisfunctions}
    f_l\bigg(\dot{\theta}_l(\bs{s}_i)\bigg) &= \sum_{b=1}^{N_{lb}} \alpha_{lb} g_{lb}\bigg(\dot{\theta}_l(\bs{s}_i)\bigg).
\end{align}\end{linenomath}
\noindent Cubic regression spline basis functions were used for $g_{lb}()$. We use functions within the \texttt{mgcv} package \citep{wood2016smoothing} in \texttt{R} \citep{R} to carry out the estimation. Figure \ref{fig:stage1step2gamcomponents} shows the contribution each candidate blurred sky map makes to the initial blurred sky map estimate, $\dot{\bs{\theta}}$. 
We see that the initial blurred sky map estimate $\dot{\bs{\theta}}$ largely resembles the ribbon of Candidates 2 and 4 and the smooth GDF of Candidates 5 through 8. By combining the candidate blurred sky maps via the GAM-regression, the banding artifacts over the GDF from  PPR-generated Candidates 2 and 4 (Figure \ref{fig:stage1step1components}) are greatly diminished. 

\textbf{Step 3: Estimate $\ddot{\bs{\theta}}$.}
The final step is to correct any systematic lack of fit that may remain between the initial blurred sky map estimate $\dot{\bs{\theta}}$ and the observed binned direct event data $\bs{y}$. This step is included as a layer of protection. If $\dot{\bs{\theta}}$ is consistent with $\bs{y}$ (as is desired), then the residual-adjusted blurred sky map $\ddot{\bs{\theta}}$~\textemdash~which is an estimate of $\bs{\theta}$, but not Theseus's final estimate of $\bs{\theta}$~\textemdash~will be similar to $\dot{\bs{\theta}}$. If, however, $\dot{\bs{\theta}}$ is inconsistent with $\bs{y}$, that inconsistency can be remedied.

Computing $\ddot{\bs{\theta}}$ proceeds as follows. First, we compute the residuals for each observation:
\begin{linenomath}\begin{align}
    \dot{r}(\bs{s}_i) &= y(\bs{s}_i) - \dot{\lambda}(\bs{s}_i)\\
    \label{eq:lambdaestimate}
    \dot{\lambda}(\bs{s}_i) &= t(\bs{s}_i)\bigg[\dot{\theta}(\bs{s}_i) + b(\bs{s}_i)\bigg].
\end{align}\end{linenomath}
\noindent We then fit a GAM regression model to $\dot{r}(\bs{s}_i)$ as a function of $\dot{\theta}(\bs{s}_i)$ and $\bs{s}_i$, again using cubic regression spline bases functions, resulting in GAM-fitted residuals $\hat{r}(\bs{s}_i)$. 
The final step is to fit another GAM model like in Step 2 (Equations \ref{eq:theseusstage1step2datamodel} through \ref{eq:theseusstage1step2basisfunctions}), replacing $\dot{\theta}_l(\bs{s}_i)$ in Equations \ref{eq:gambetadot} and \ref{eq:theseusstage1step2basisfunctions} with $\dot{\theta}(\bs{s}_i)$ and $\hat{r}(\bs{s}_i)$. 

\begin{figure}[!ht]
\centering
\includegraphics[width=1\linewidth]{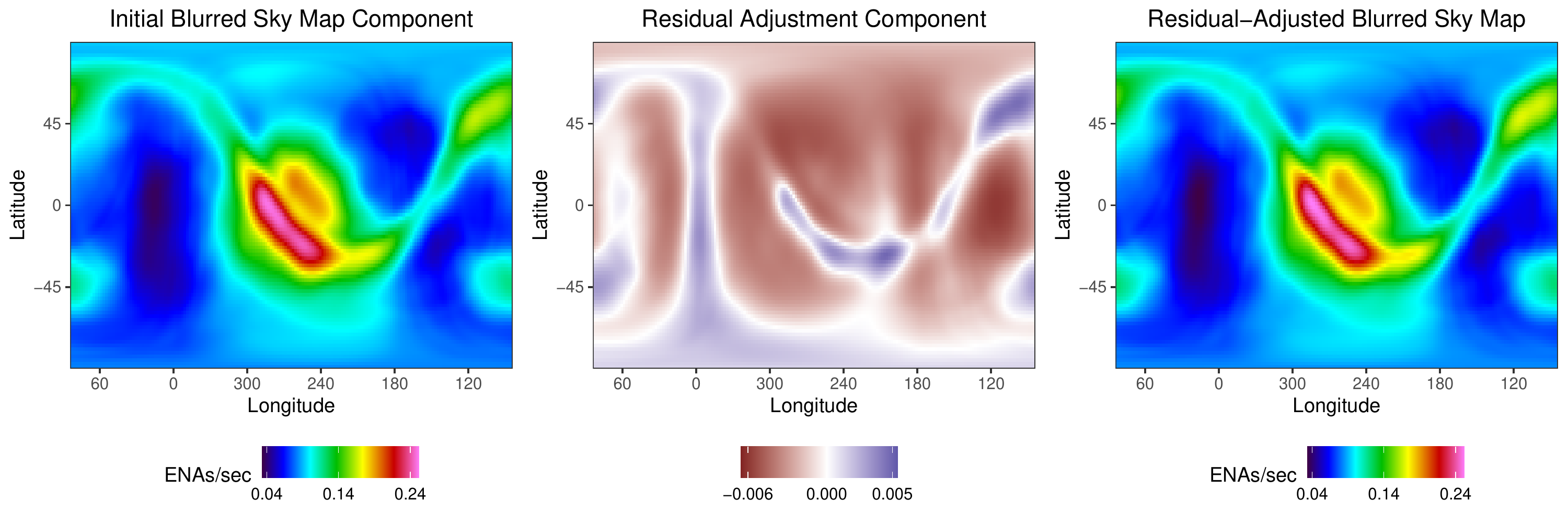}
\caption{\small The initial blurred sky map component (left), the residual-adjustment component (middle) and the residual-adjusted blurred sky map estimate $\ddot{\bs{\theta}}$ (right). $\ddot{\bs{\theta}}$ is computed by adding the left and middle panels. Relatively minor residual adjustments were made to the initial blurred sky map component, as noted by the scale of the residual-adjustment component compared to the scale of the initial blurred sky map component.}
\label{fig:stage1step3final}
\end{figure}

Figure \ref{fig:stage1step3final} shows the residual-adjusted blurred sky map estimate, $\ddot{\bs{\theta}}$, as the sum of the initial blurred sky map estimate component (left) and the fitted residual surface component (middle). 
The right of Figure \ref{fig:stage1step3final} shows the residual-adjusted blurred sky map estimate $\ddot{\bs{\theta}}$.
The residual adjustment was quite minor, as the initial blurred sky map estimate $\dot{\bs{\theta}}$ was already consistent with the binned direct event data $\bs{y}$.
In general, it is overwhelmingly common, but not guaranteed, that $\dot{\bs{\theta}} \approx \ddot{\bs{\theta}}$.

\subsection{Theseus Stage 2: Estimate an Unblurred Sky Map $\bs{\Theta}$}
\label{subsec:stage2}

Recall that the $N_p \times 1$ blurred sky map $\bs{\theta}$ is related to the $N_p \times 1$ unblurred sky map $\bs{\Theta}$ through the $N_p \times N_p$ PSF matrix $\bs{K}$, where $\bs{\theta} = \bs{K} \bs{\Theta}$ and $\bs{K}(\bs{s}_j)$~\textemdash~the $j^{\text{th}}$ row of $\bs{K}$~\textemdash~is the blurring operator applied to the $j^{\text{th}}$ pixel center $\bs{s}_j$. For this application, $\bs{K}^{-1}$ exists, suggesting that the unblurred sky map $\bs{\Theta}$ can be easily computed given the blurred sky map $\bs{\theta}$ as follows: $\bs{\Theta} = \bs{K}^{-1} \bs{\theta}$. While this relationship holds for the true unblurred and true blurred sky maps, there is a problem when $\bs{\theta}$ is replaced with an estimate $\hat{\bs{\theta}}$.\footnote{Recall the notational conventions of Section \ref{sec:methodology} and note here that the ``hat" notation of $\hat{\bs{\theta}}$ is being used as a generic estimate of $\bs{\theta}$, not as Theseus's final estimate of $\bs{\theta}$.} This problem is related to the large condition number of the PSF matrix $\bs{K}$ (over ten million), thus the inverse problem is ill-conditioned \citep{rice1966theory}. A large condition number for the matrix $\bs{K}$ indicates that small differences between an estimated blurred sky map $\hat{\bs{\theta}}$ and the true blurred sky map $\bs{\theta}$ can result in large differences between their inverses (i.e., between $\hat{\bs{\Theta}} = \bs{K}^{-1} \hat{\bs{\theta}}$ and $\bs{\Theta} = \bs{K}^{-1} \bs{\theta}$). 

\begin{figure}
\centering
\includegraphics[width=1\linewidth]{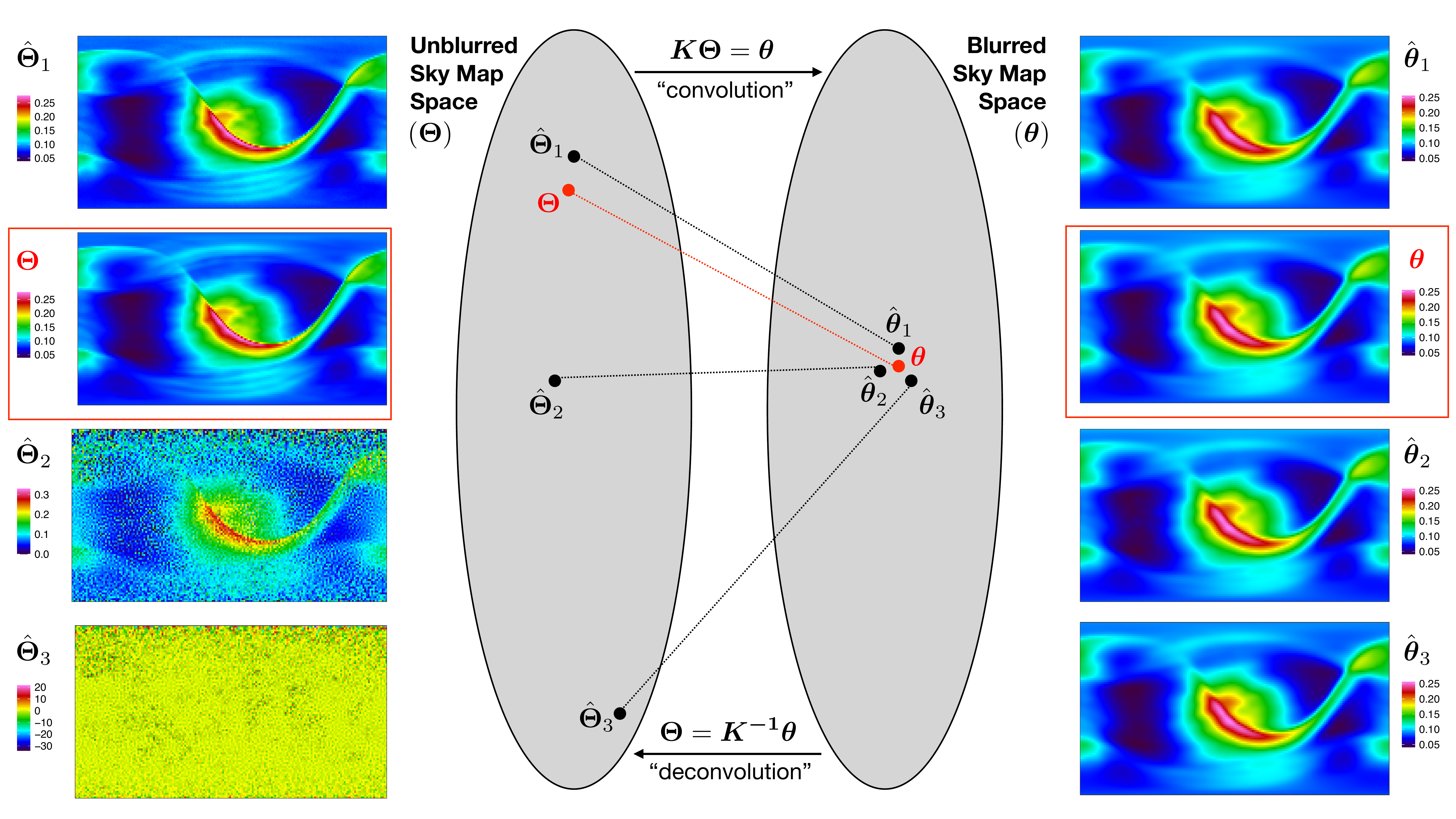}
\caption{\small A diagram illustrating the blurring/unblurring (convolution/deconvolution) process. Unblurred sky map space is on the left; blurred sky map space is on the right. The true unblurred/blurred sky map is outlined in red, while the estimated pairs of unblurred/blurred sky maps are in black with hats. For all unblurred/blurred sky map pairs (true or estimates), it is true that $\bs{\theta} = \bs{K}\bs{\Theta}$ and $\bs{K}^{-1} \bs{\theta} = \bs{\Theta}$. The point spread function matrix $\bs{K}$ has a large condition number of over ten million. A large condition number poses a challenge for deconvolution, as small differences between to $\hat{\bs{\theta}}$ and $\bs{\theta}$ can result in large differences between $\hat{\bs{\Theta}}$ and $\bs{\Theta}$.}
\label{fig:deconvolution}
\end{figure}

This phenomenon is illustrated in Figure \ref{fig:deconvolution}. Four similar but not identical blurred sky maps are shown on the right and their unblurred counterparts are shown on the left. For each pair of unblurred/blurred sky maps, it is true that the $\bs{K}$ times the unblurred sky map equals the blurred map, and $\bs{K}^{-1}$ times the blurred sky map equals the unblurred sky map. Figure \ref{fig:deconvolution} makes clear, however, that even if an estimated blurred sky map $\hat{\bs{\theta}}$ is nearly identical to the true blurred sky map $\bs{\theta}$, computing the unblurred sky map as $\hat{\bs{\Theta}} = \bs{K}^{-1} \hat{\bs{\theta}}$ is unreliable. 
This immediately raises challenges for unblurring $\ddot{\bs{\theta}}$, the Theseus Stage 1 estimate of $\bs{\theta}$.

To address the challenges with unblurring (also referred to as ``unfolding" \citep{lyons2011phystat}), we turn to regularization \citep[e.g.,][]{kuusela2015statistical, kuusela2017shape}. Regularization requires making some assumption about how the unblurred and blurred sky maps relate to each other. This is challenging in general, but particularly so in our discovery science setting where we are willing to make few assumptions about the sky maps. However, Figure \ref{fig:deconvolution} makes clear that if we are not willing to assume anything about how the unblurred and blurred sky maps are related to one another, we are forced to accept some wildly impractical unblurred sky maps as plausible: an unwise choice. Thus, we assume that the unblurred and blurred sky maps are ``close" to one another. That is, the blurred sky map is just a less crisp version of the unblurred sky map as is the case with $\hat{\bs{\Theta}}_1$ and $\bs{\Theta}$ in Figure \ref{fig:deconvolution}. By making this assumption, we could estimate $\bs{\Theta}$ by minimizing the following equation:
\begin{linenomath}\begin{align}
    \label{eq:ridge1}
    \hat{\bs{\Theta}} &= \underset{\bs{\Theta}}{\arg\min} ||\hat{\bs{\theta}} - \bs{K} \bs{\Theta}||^2_2 + \lambda||\hat{\bs{\theta}} - \bs{\Theta}||^2_2,
\end{align}\end{linenomath}
\noindent where $||\hat{\bs{\theta}} - \bs{K} \bs{\Theta}||^2_2$ encourages the convolution of the unblurred sky map $\bs{\Theta}$ with the PSF matrix $\bs{K}$ to be close to a blurred sky map estimate $\hat{\bs{\theta}}$, while $\lambda||\hat{\bs{\theta}} - \bs{\Theta}||^2_2$ encourages the difference between the unblurred sky map and a blurred sky map estimate to be small. Equation \ref{eq:ridge1} corresponds to a generalized ridge regression problem \citep[Section 3.4]{van2015lecture}.

\begin{figure}[!ht]
\centering
\includegraphics[width=1\linewidth]{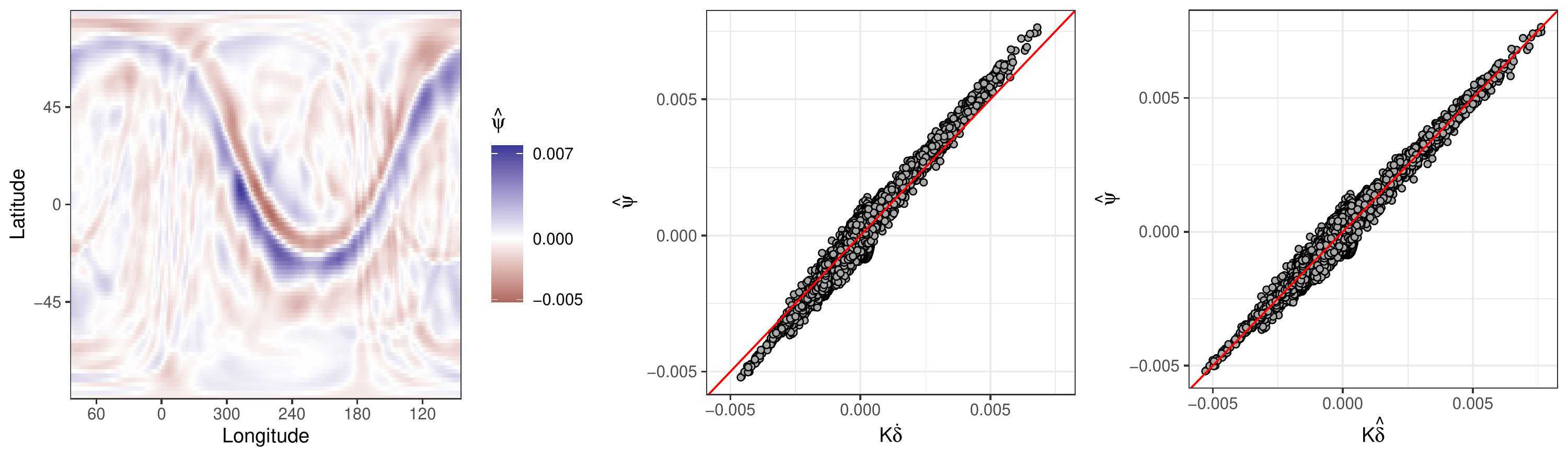}
\caption{\small (left) The difference between the once and twice blurred residual-adjusted blurred sky map estimate: $\hat{\bs{\psi}} = \ddot{\bs{\theta}} - \bs{K}\ddot{\bs{\theta}}$. The ribbon is present as the largest deviations from 0. (middle) $\hat{\bs{\psi}}$ versus $\bs{K} \dot{\bs{\delta}}$ and (right) $\hat{\bs{\psi}}$ versus  $\bs{K} \hat{\bs{\delta}}$. The red line is the $y=x$ line. $\bs{K} \dot{\bs{\delta}}$ is a biased estimate of $\hat{\bs{\psi}}$, where it underestimates large, positive values of $\hat{\bs{\psi}}$ and overestimates large, negative values of $\hat{\bs{\psi}}$. This bias is a result of the regularization imposed by ridge regression. $\hat{\bs{\delta}}$ corrects for this bias, resulting in a $\bs{K} \hat{\bs{\delta}}$ that is unbiased for $\hat{\bs{\psi}}$.}
\label{fig:stage2delta}
\end{figure}

The relationship between the blurred and unblurred sky map can be rewritten. Let $\bs{\Theta} = \bs{\theta} + \bs{\delta}$, where $\bs{\delta}$ is an $N_p \times 1$ vector called the \emph{sharpening map} that equals $\bs{\Theta} - \bs{\theta}$. It then follows that
\begin{linenomath}\begin{align}
    \label{eq:originalrelationship}
    \bs{\theta} &= \bs{K}\bs{\Theta}\\
    \label{eq:rewrittenrelationship}
    &= \bs{K} (\bs{\theta} + \bs{\delta})\\
    \label{eq:psianddelta}
    \implies \bs{\psi} \equiv \bs{\theta} - \bs{K} \bs{\theta} &= \bs{K} \bs{\delta}.
\end{align}\end{linenomath}
\noindent The vector $\bs{\psi}$ is the once blurred sky map $\bs{\theta}$ minus the twice blurred sky map $\bs{K} \bs{\theta}$ and is estimated as $\hat{\bs{\psi}} = \ddot{\bs{\theta}} - \bs{K}\ddot{\bs{\theta}}$.
The PSF matrix $\bs{K}$ plays the role of dampening and smearing out the sharp, peaked features in the blurred sky map (e.g., the ribbon) but leaves the more gradually varying parts of the blurred sky map relatively unchanged (e.g., the GDF). 
Consequently, the outline of the ribbon can be seen by plotting $\hat{\bs{\psi}}$, as is done in the left panel of Figure \ref{fig:stage2delta}.

Under this new parameterization where $\bs{\Theta} = \bs{\theta} + \bs{\delta}$, assuming $||\bs{\theta} - \bs{\Theta}||^2_2$ is small is equivalent to assuming that $||\bs{\delta}||^2_2$ is small. Thus, Equation \ref{eq:ridge1} can be rewritten as a standard ridge regression problem \citep{hoerl1970ridge}. Specifically, $\bs{\Theta}$ could be estimated as
\begin{linenomath}\begin{align}
\label{eq:ridge2}
    \dot{\bs{\Theta}} &= \ddot{\bs{\theta}} + \dot{\bs{\delta}}
\end{align}\end{linenomath}
\noindent where
\begin{linenomath}\begin{align}
    \label{eq:deltaridge}
    \dot{\bs{\delta}} &= \underset{\bs{\delta}}{\arg\min} ||\hat{\bs{\psi}} - \bs{K} \bs{\delta}||^2_2 + \lambda||\bs{\delta}||^2_2.
\end{align}\end{linenomath}

\noindent Equation \ref{eq:ridge2} says that the unblurred sky map can be estimated as the sum of the residual-adjusted blurred sky map estimate $\ddot{\bs{\theta}}$ and an initial sharpening map estimate $\dot{\bs{\delta}}$. 
The initial sharpening map estimate $\dot{\bs{\delta}}$ is the solution to the ridge regression problem.
The regularization part of Equation \ref{eq:deltaridge}, namely $\lambda||\bs{\delta}||^2_2$, encourages $\bs{\delta}$ to be small (i.e., encourages the unblurred and blurred sky maps to be ``close").
The tuning parameter $\lambda \geq 0$ controlling the degree of regularization in Equation \ref{eq:deltaridge} is learned via cross-validation. We use the \texttt{glmnet} package \citep{friedman2010regularization} in \texttt{R} to perform the ridge regression, imposing constraints such that $\ddot{\bs{\theta}} + \dot{\bs{\delta}} \geq \bs{0}$.

Estimating $\bs{\delta}$ via ridge regression, however, creates a bias between $\hat{\bs{\psi}}$ and $\bs{K} \dot{\bs{\delta}}$. 
From Equation \ref{eq:psianddelta}, we have that $\bs{\psi} = \bs{K}\bs{\delta}$. 
The middle panel of Figure \ref{fig:stage2delta} plots $\hat{\bs{\psi}}$ versus $\bs{K}\dot{\bs{\delta}}$ where a clear bias can be seen. 
$\bs{K}\dot{\bs{\delta}}$ underestimates $\hat{\bs{\psi}}$ for large, positive values and overestimates $\hat{\bs{\psi}}$ for large, negative values. 
The implication of using $\dot{\bs{\delta}}$ in Equation \ref{eq:ridge2} to estimate the unblurred sky map $\bs{\Theta}$ is that the base and peak of the ribbon may be over- and underestimated, respectively.

To correct for this bias, we estimate the sharpening map as a bias-adjusted version of $\dot{\bs{\delta}}$. Specifically, we estimate the sharpening map as
\begin{linenomath}\begin{align}
    \label{eq:deltahat}
    \hat{\bs{\delta}} &= \bs{A}\hat{\bs{\alpha}}
\end{align}\end{linenomath}
\noindent where $\hat{\bs{\alpha}}_{3 \times 1} = (\hat{\alpha}_0, \hat{\alpha}_1, \hat{\alpha}_2)'$, $\bs{A}_{N_p \times 3} = [\bs{1}, \dot{\bs{\delta}}_{-}, \dot{\bs{\delta}}_{+}]$, $\bs{1}$ is a vector of 1s, and the vectors $\dot{\bs{\delta}}_{-}$ and $\dot{\bs{\delta}}_{+}$ are hinge function representations of the vector $\dot{\bs{\delta}}$, where all entries of $\dot{\bs{\delta}}$ greater than 0 ($\dot{\bs{\delta}}_{-}$) or less than 0 ($\dot{\bs{\delta}}_{+}$) are set to 0. 
Equation \ref{eq:deltahat} just says that the relationship between $\hat{\bs{\delta}}$ and $\dot{\bs{\delta}}$ is linear, but a different linear relationship is possible between $\hat{\bs{\delta}}$ and the negative part of $\dot{\bs{\delta}}$ ($\dot{\bs{\delta}}_{-}$) than with the positive part of $\dot{\bs{\delta}}$ ($\dot{\bs{\delta}}_{+}$).
The different linear relationships, however, are required to match when an entry of $\dot{\bs{\delta}}$ equals 0, preventing a jump discontinuity at 0 in Equation \ref{eq:deltahat}.
We estimate $\bs{\alpha}$ as
\begin{linenomath}\begin{align}
    \hat{\bs{\alpha}} &= \underset{\bs{\alpha}}{\arg\min} ||\hat{\bs{\psi}} - \bs{K} \bs{A} \bs{\alpha}||^2_2.
\end{align}\end{linenomath}
\noindent The right panel of Figure \ref{fig:stage2delta} plots $\hat{\bs{\psi}}$ versus $\bs{K}\hat{\bs{\delta}}$. As can be seen, the bias that was present when using $\dot{\bs{\delta}}$ has been corrected when using $\hat{\bs{\delta}}$. 

\begin{figure}[!ht]
\centering
\includegraphics[width=1\linewidth]{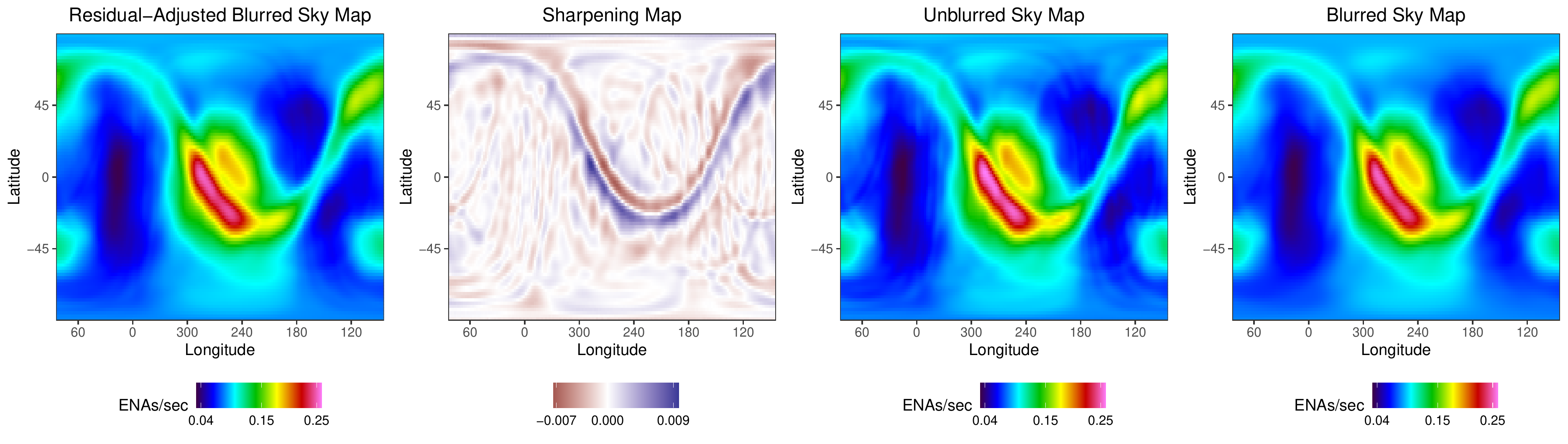}
\caption{\small From left to right, the residual-adjusted blurred sky map estimate $\ddot{\bs{\theta}}$, the sharpening map estimate $\hat{\bs{\delta}}$, the final unblurred sky map estimate $\hat{\bs{\Theta}} = \ddot{\bs{\theta}} + \hat{\bs{\delta}}$, and the final blurred sky map estimate $\hat{\bs{\theta}} = \bs{K}\hat{\bs{\Theta}}$.}
\label{fig:stage2final}
\end{figure}

Figure \ref{fig:stage2final} shows the final unblurred and blurred sky map estimates, where the \emph{final unblurred sky map estimate} is
\begin{linenomath}\begin{align}
    \label{eq:finalunblurredestimate}
    \hat{\bs{\Theta}} &= \ddot{\bs{\theta}} + \hat{\bs{\delta}}
\end{align}\end{linenomath}
\noindent and the \emph{final blurred sky map estimate} is
\begin{linenomath}\begin{align}
    \label{eq:finalunblurredestimate2}
    \hat{\bs{\theta}} &= \bs{K}\hat{\bs{\Theta}}.
\end{align}\end{linenomath}

Figure \ref{fig:stage2final} (and all the illustrative figures in Section \ref{sec:methodology}) correspond to one bootstrapped binned direct event data set $\bs{D}_b$. 
That is, for a given ESA step (e.g., ESA step 2), a given 6-month map (e.g., 2013A), and a given bootstrapped binned direct event data set $\bs{D}_b$, Theseus estimates an unblurred sky map $\hat{\bs{\Theta}}_b$ and a blurred sky map $\hat{\bs{\theta}}_b$.
The final point estimate unblurred sky map $\hat{\bs{\Theta}}$ and point estimate blurred sky map $\hat{\bs{\theta}}$ for a given ESA and 6-month map are computed as averages over all $N_B = 1,000$ bootstrap estimates:
\begin{linenomath}\begin{align}
    \label{eq:bsavgunblurredmap}
    \hat{\bs{\Theta}} &= N_B^{-1}\sum_{b=1}^{N_B} \hat{\bs{\Theta}}_b\\
    \label{eq:bsavgblurredmap}
    \hat{\bs{\theta}} &= N_B^{-1}\sum_{b=1}^{N_B} \hat{\bs{\theta}}_b.
\end{align}\end{linenomath}

Estimating $\hat{\bs{\Theta}}$ as the average over all bootstrap samples (Equation \ref{eq:bsavgunblurredmap}) rather than a single run of Theseus applied to the original binned direct event data set $\bs{D}$ can greatly reduce the banding artifacts initially caused by the PPR regression in step 1 of Theseus Stage 1 that can remain in $\hat{\bs{\Theta}}_b$ (Figure \ref{fig:stage2final}). 
This is because, though PPR very often has banding artifacts, those artifacts are not always found in the same location across bootstrap samples.
By averaging over all $N_B$ unblurred sky map estimates, the banding artifacts are often (but not always) blended out of the final unblurred sky map estimate $\hat{\bs{\Theta}}$.
Further strategies for addressing the banding artifacts are discussed in Section \ref{sec:discussion}.


\section{Simulation Study}
\label{sec:simstudy}
To investigate the inferential properties of Theseus and compare Theseus to a near facsimile of the current state-of-the-art sky map estimation procedure used by the ISOC, we perform a simulation study. A description of the implemented ISOC sky map estimation methodology is provided in Supplementary Materials \ref{appendix:isoc}.

\subsection{Simulation Study Setup}
\label{subsec:setup}
\begin{figure}[!ht]
\centering
\includegraphics[width=.24\linewidth]{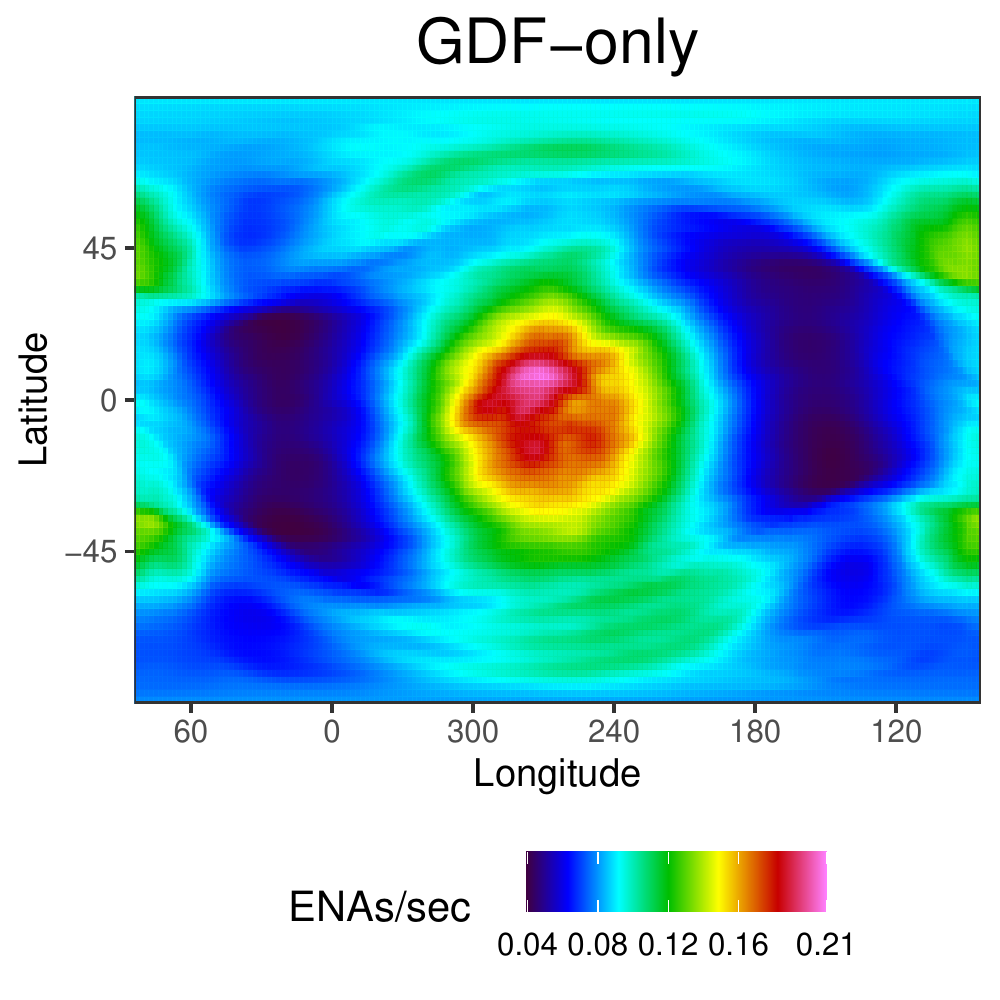}
\includegraphics[width=.24\linewidth]{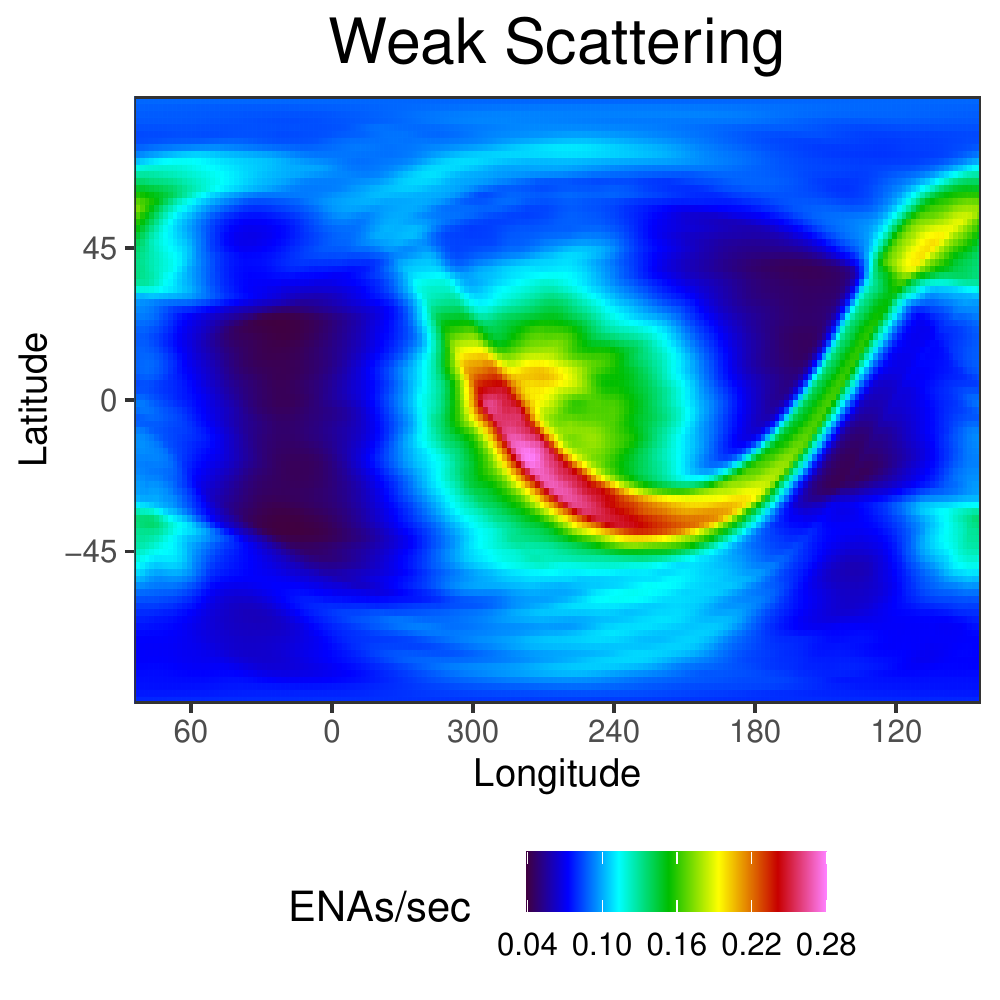}
\includegraphics[width=.24\linewidth]{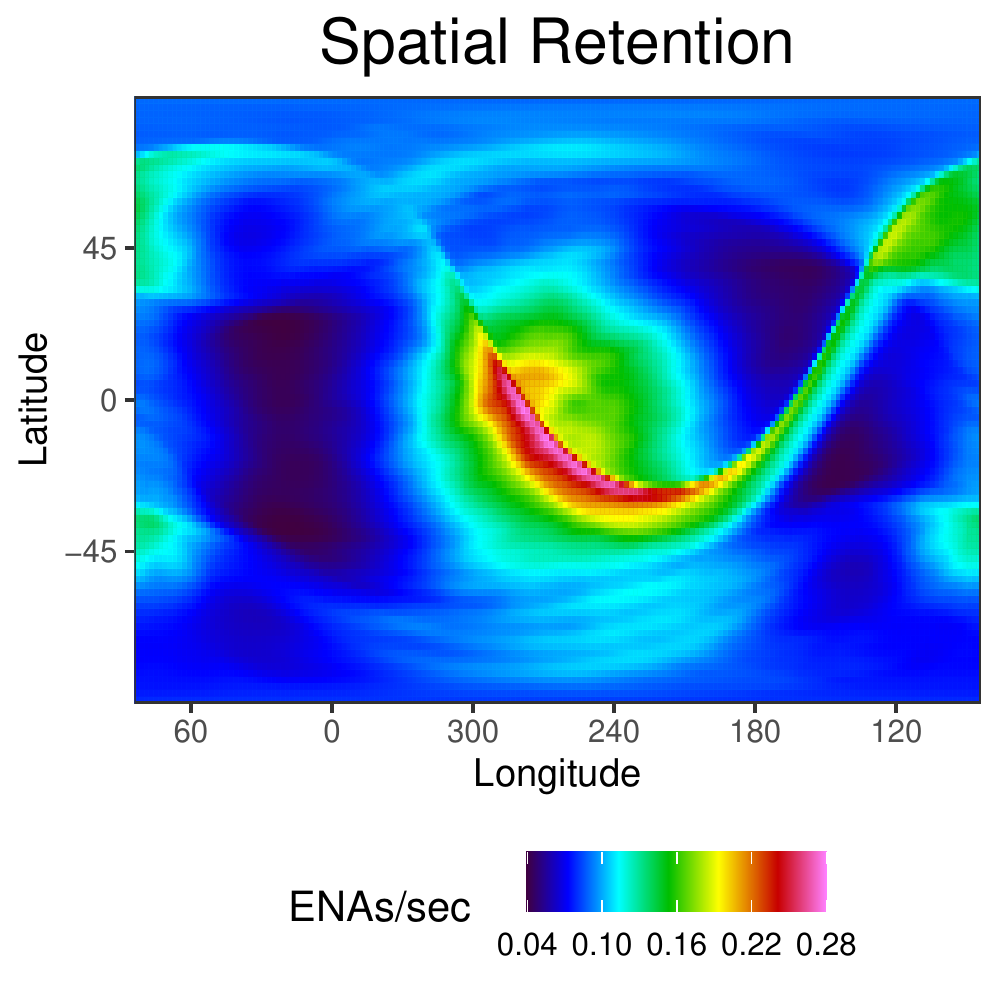}
\includegraphics[width=.24\linewidth]{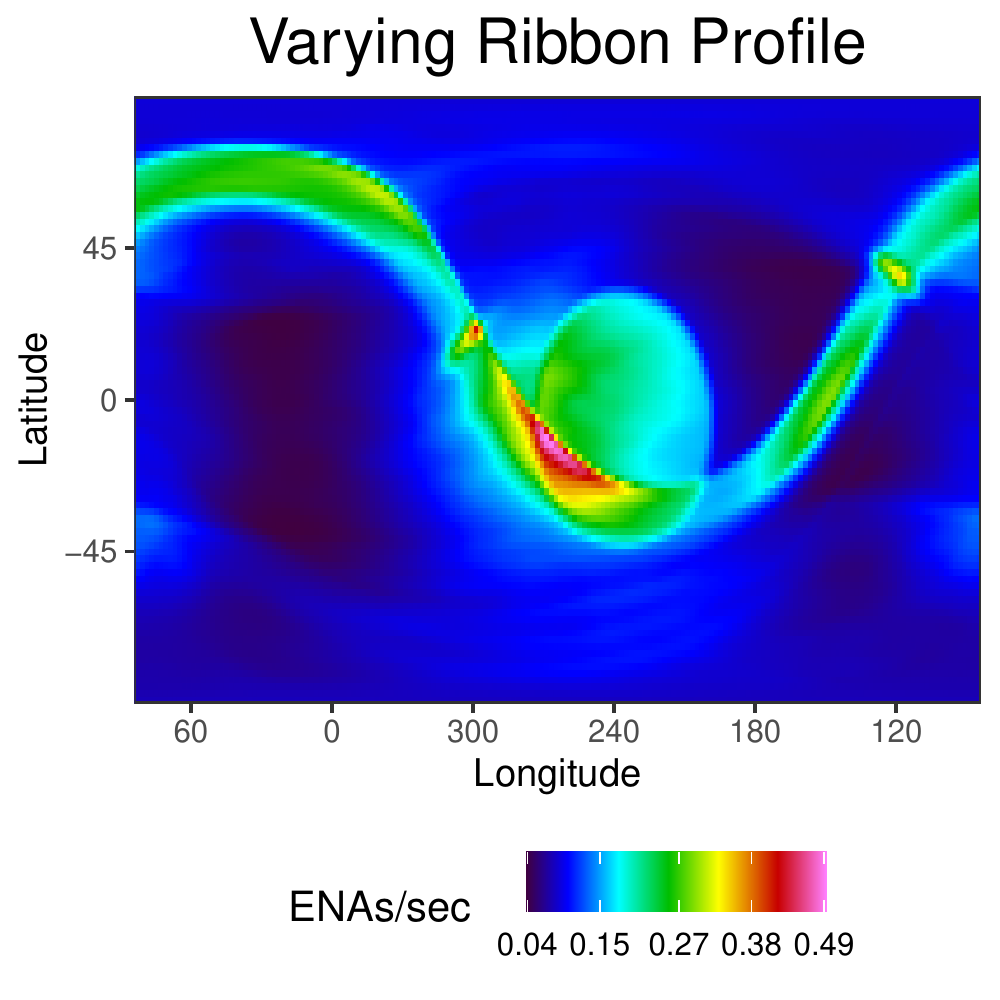}
\caption{\small From left to right, the simulated sky maps for the GDF-only, Weak Scattering, Spatial Retention, and Varying Ribbon Profile models. The GDF-only sky map has elevated ENA rates at the nose (longitude 265) and tail (longitude 85). The Weak Scattering and Spatial Retention sky maps are equal to the GDF-only sky map plus a ribbon. The Varying Ribbon Profile sky map is the only intentionally unrealistic sky map. It has more exaggerated features than are anticipated to exist in the heliosphere, but is included in the simulation study to test the limits of Theseus and ISOC.}
\label{fig:simmaps}
\end{figure}

We consider four simulated sky maps with different ribbon models resembling those from the literature:

\begin{itemize}
    \item \textbf{GDF-only}: a sky map with no ribbon, just globally distributed flux \citep{zirnstein2017heliotail}. Elevated ENA rates are present at the nose and tail.
    \item \textbf{Weak Scattering}: the GDF-only sky map plus a ribbon with heterogeneous brightening and a relatively smooth, symmetric ribbon profile \citep{zirnstein2018weak}.
    \item \textbf{Spatial Retention}: the GDF-only sky map plus a ribbon with heterogeneous brightening and an asymmetric ribbon profile \citep{schwadron2013spatial,zirnstein2019strong}.
    \item \textbf{Varying Ribbon Profile}: the GDF-only sky map plus a ribbon that transitions from the Weak Scattering ribbon to the Spatial Retention ribbon with highly-concentrated pockets of higher ENA rates added at locations on the ribbon as well as a disk over the nose that appears suddenly. \emph{This model is intentionally unrealistic}, but is useful for testing the limits of Theseus and ISOC.
\end{itemize}
\noindent The different simulated unblurred sky maps are shown in Figure \ref{fig:simmaps}. 
Each model is simulated for ENAs at $\sim$1.7 keV, corresponding to IBEX-Hi ESA step 4. 
The Weak Scattering and Spatial Retention ribbon model results shown in Figure 8 are extracted from a single cross-sectional cut across the model ribbon and then copied across the sky to mimic a ribbon with consistent cross-section shape. The change in ribbon intensity across the sky is artificially varied to mimic reality.
Binned direct event counts $\bs{y}$ were simulated from Equations \ref{eq:datamodel} and \ref{eq:blurringfn} using a conical PSF with a FWHM of 6.5 degrees. Five binned direct event data sets $\bs{D}$ were simulated for each ribbon model, corresponding to the locations, exposure times, and background rates of ESA steps 2 through 6 of the 2018B map. This amounts to 5 different simulated binned direct event data sets per ribbon model.

\subsection{Simulation Study Results}
\label{subsec:results}
Theseus and ISOC were fit to each simulated binned direct event data set corresponding to each of the 20 ESA/ribbon model combinations. 
In Section \ref{subsubsec:visual}, we visually compare the simulated sky maps to Theseus and ISOC estimated sky maps. 
In Section \ref{subsubsec:datacomp}, we assess the agreement between the simulated binned direct event data $\bs{y}$ and the Theseus and ISOC estimated sky maps. 
In Section \ref{subsubsec:pixelcomp}, we assess the accuracy, skill, sharpness, and reliability of the Theseus and ISOC estimated sky maps. 
Finally in Section \ref{subsubsec:skew}, we assess Theseus's and ISOC's ability to estimate the shape of the simulated ribbon profiles. 

\subsubsection{Visual Comparison}
\label{subsubsec:visual}

\begin{figure}[!ht]
\centering
\includegraphics[width=1\linewidth]{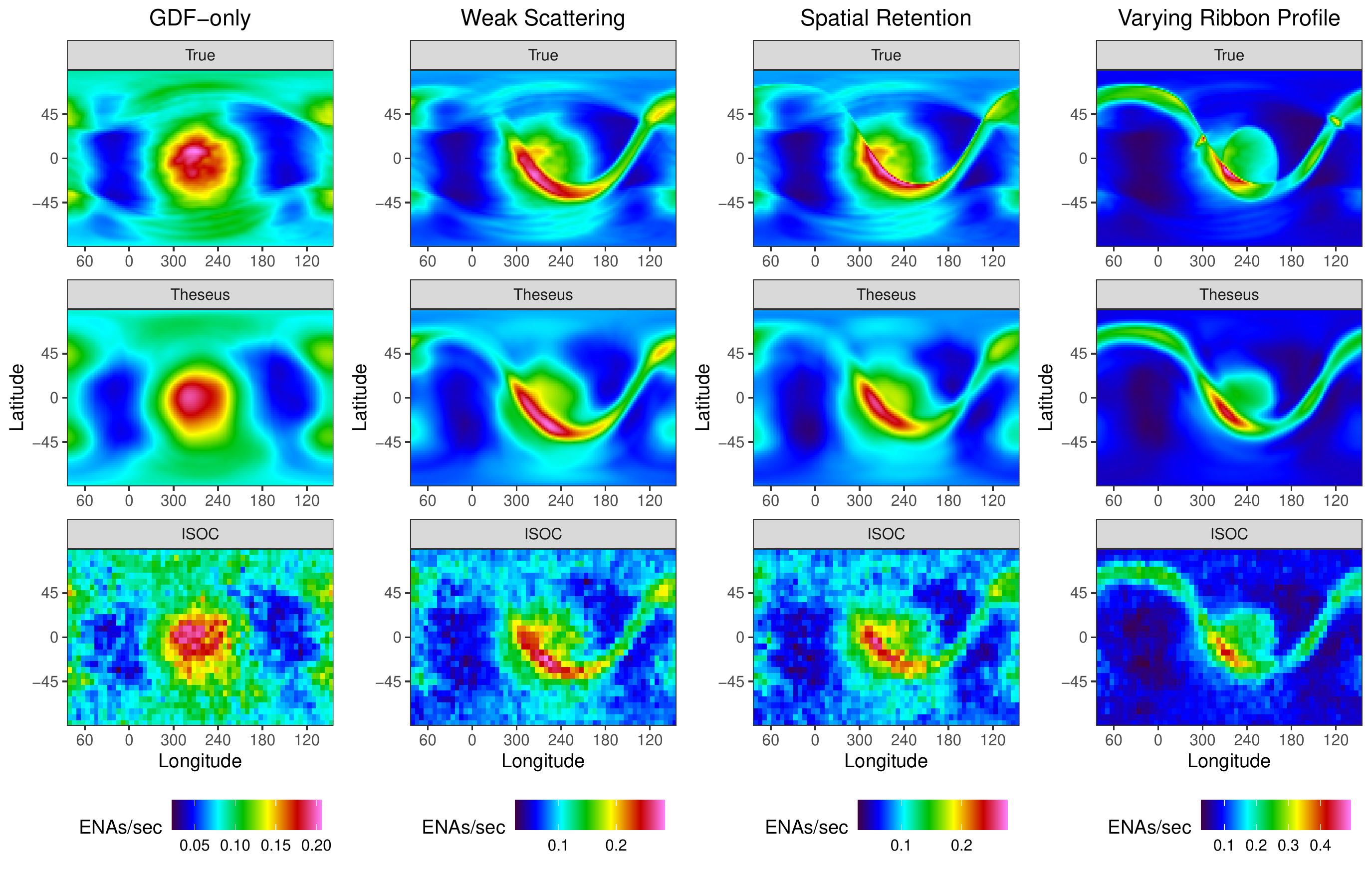}
\caption{\small The true and estimated sky maps from Theseus and ISOC (rows) for the different simulated ribbon models (columns) for ESA 6. Visually, there appears to be better agreement between Theseus and the true sky maps than ISOC and the true sky maps.}
\label{fig:unblurredestimatedmaps}
\end{figure}

Figure \ref{fig:unblurredestimatedmaps} plots unblurred sky map estimates for the different ribbon models based on ESA 6 binned direct event data. 
Visually, there appears to be better agreement between Theseus and the true sky maps than ISOC and the true sky maps.
ISOC sky maps appear grainier than Theseus because ISOC makes sky maps at a fixed $6\degree$~resolution, while the resolution is a user-defined parameter for Theseus ($2\degree$~maps are shown in Figure \ref{fig:unblurredestimatedmaps}). 
ISOC sky map estimates are done on a pixel-by-pixel basis, and thus can suffer from erratic pixel estimates in regions with little to no binned direct event data. 
Theseus, on the other hand, borrows information over space via the PPR and GAM smoothing, resulting in less erratic pixel-to-pixel sky map estimates. 
The knots present in the Varying Ribbon Profile near ecliptic longitude and latitude (120, 40) and (300, 20) are not captured by either Theseus or ISOC.

\begin{figure}[!ht]
\centering
\includegraphics[width=1\linewidth]{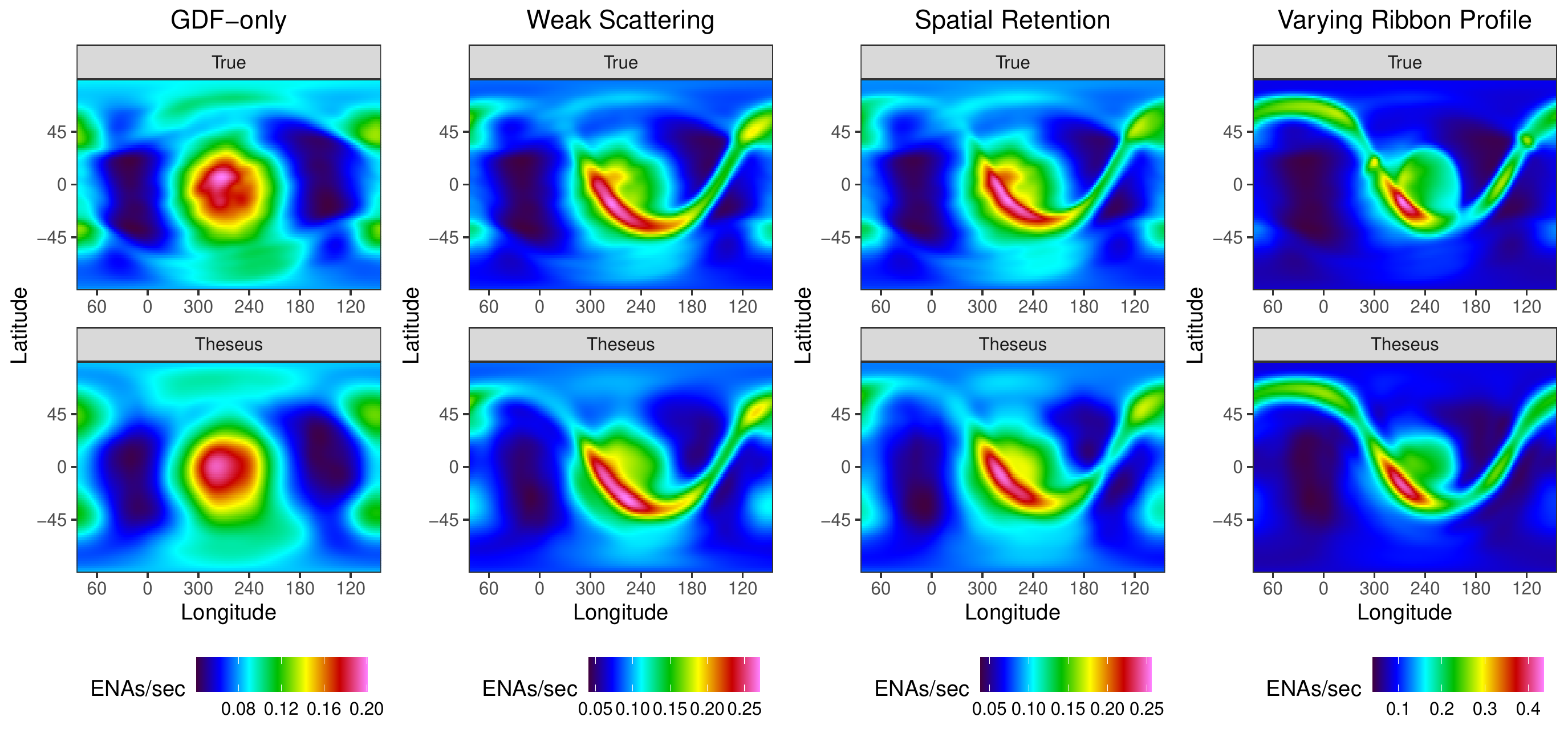}
\caption{\small The true and Theseus estimated blurred sky maps (rows) for the different ribbon models (columns) for ESA 6. Visually we see good agreement between Theseus and the true sky maps. ISOC does not distinguish between blurred and unblurred sky maps, treating them both as unblurred sky maps, hence why they are not shown.}
\label{fig:blurredestimatedmaps}
\end{figure}

Figure \ref{fig:blurredestimatedmaps} shows the Theseus estimated blurred sky maps and the true blurred sky maps for the different ribbon models corresponding to the simulated binned direct event data for ESA 6. 
Visually, there is good agreement between the Theseus estimated blurred sky maps and the true blurred sky maps.
It also appears that Theseus does a better job estimating the blurred sky maps than the unblurred sky maps.
This is not surprising, as the binned direct event data directly inform the blurred sky maps but only indirectly inform the unblurred sky maps.
Theseus was unable to capture the knots on the Varying Ribbon Profile's blurred sky map, an unsurprising finding given its inability to do so on the unblurred sky map.
ISOC's blurred sky map estimates are not shown because ISOC does not distinguish between blurred and unblurred sky map estimates. 
ISOC only estimates an unblurred sky map.

\subsubsection{Agreement with Simulated Binned Direct Event Data}
\label{subsubsec:datacomp}

To assess how well Theseus-estimated blurred sky maps and the ISOC-estimated unblurred sky maps agree with the simulated binned direct event data, we examine the probability integral transform (PIT) histograms \citep{gneiting2007probabilistic,diebold1998}. More concretely, let PIT($y(\bs{s}_i)$, $\hat{\lambda}(\bs{s}_i)$) be a random draw from Uniform($l(\bs{s}_i)$, $u(\bs{s}_i)$), where $l(\bs{s}_i)$ is the cumulative distribution function (cdf) of a Poisson evaluated at $y(\bs{s}_i) - 1$ with mean parameter $\hat{\lambda}(\bs{s}_i)$ (computed following Equation \ref{eq:lambdaestimate}) and $u(\bs{s}_i)$ is the cdf of a Poisson evaluated at $y(\bs{s}_i)$. Note that, if $y(\bs{s}_i) = 0$, then $l(\bs{s}_i) = 0$. If the binned direct event observations were conditionally independent draws from their generating data model with the blurred sky map estimate $\hat{\bs{\theta}}$ plugged in, we would expect the collection of PIT($y(\bs{s}_i)$, $\hat{\lambda}(\bs{s}_i)$) to follow a Uniform(0,1) distribution. 

In addition to PIT histograms, we assess the spatial distribution of the PIT values. If PIT($y(\bs{s}_i)$,$\hat{\lambda}(\bs{s}_i)$) $\overset{\text{iid}}{\sim}$ Uniform(0,1) and were randomly assigned to binned direct event locations $\bs{s}_i$, we would expect the mean and variance of the difference between nearest neighbor PIT values to be 0 and 1/6 (the mean and variance of the difference of two independent, Uniform(0,1) random variables). 

Table \ref{tab:obspval} shows summaries of PIT values and differences of nearest neighbor PIT values. The results are consistent with expectations under theory, indicating no serious lack-of-fit for either ISOC or Theseus. 
Overall, the Theseus PIT histograms do, however, have better agreement with nominal values than do the ISOC PIT histograms.
\begin{table}[ht]
\centering
\caption{\small Means and variances of PIT histograms (PIT) and differences of nearest-neighbor PIT values (PIT Differences). The nominal mean and variance for PIT histograms are 0.5 and 0.083, respectively. The nominal mean and variance for differences of nearest neighbor PIT values are 0 and 0.167, respectively. Bold values for each ribbon model indicate the method closest to nominal values. Results show no serious lack-of-fit for ISOC or Theseus, though Theseus has PIT values closer to the nominal means/variances than does ISOC.}
\begin{tabular}{ll|rr|rr}
  \hline
   & & \multicolumn{2}{c}{\textbf{PIT}} & \multicolumn{2}{|c}{\textbf{PIT Differences}} \\
\textbf{Ribbon Model} & \textbf{Method} & \textbf{Mean} & \textbf{Variance} & \textbf{Mean} & \textbf{Variance} \\ 
  \hline
\multirow{2}{*}{GDF-only} & ISOC & 0.495 & 0.080 & \textbf{-0.000} & 0.171 \\ 
   & Theseus & \textbf{0.500} & \textbf{0.083} & -0.001 & \textbf{0.168} \\ \hline
  \multirow{2}{*}{Weak Scattering} & ISOC & 0.494 & 0.081 & -0.001 & 0.172 \\ 
   & Theseus & \textbf{0.499} & \textbf{0.084} & \textbf{-0.000} & \textbf{0.168} \\ \hline
  \multirow{2}{*}{Spatial Retention} & ISOC & 0.494 & 0.081 & \textbf{-0.000} & 0.172 \\ 
   & Theseus & \textbf{0.499} & \textbf{0.083} & \textbf{0.000} & \textbf{0.168} \\ \hline
  \multirow{2}{*}{Varying Ribbon Profile} & ISOC & 0.493 & 0.081 & \textbf{-0.000} & 0.170 \\ 
   & Theseus & \textbf{0.499} & \textbf{0.084} & \textbf{0.000} & \textbf{0.166} \\ 
   \hline
\end{tabular}
\label{tab:obspval}
\end{table}

\begin{figure}[!ht]
\centering
\includegraphics[width=1\linewidth]{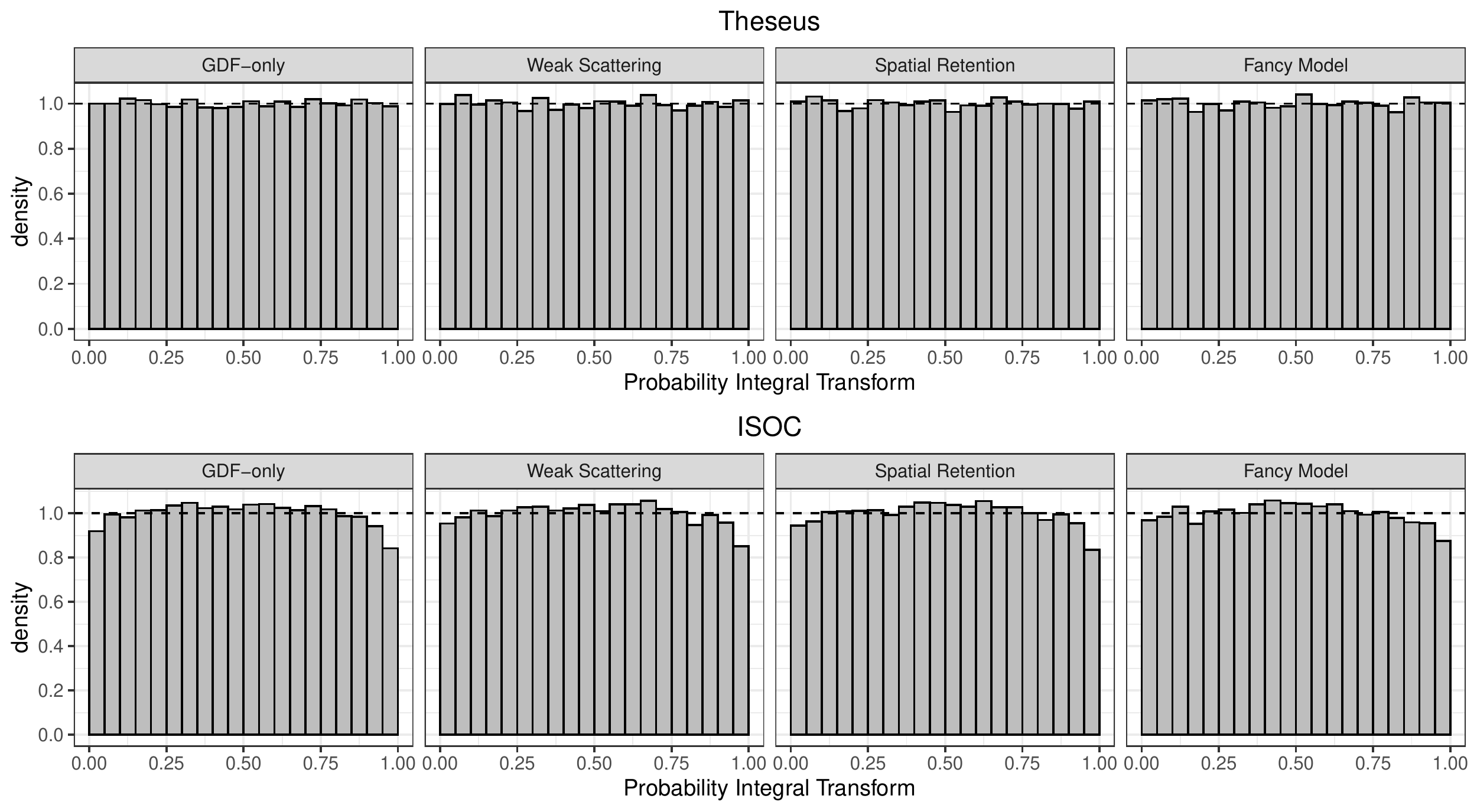}
\caption{\small PIT histograms for Theseus (top) and ISOC (bottom) for ESAs 2 through 6 displayed for each ribbon model. Histograms show no visual departure from uniformity for Theseus while ISOC shows a slight mounded shape for all ribbon models, indicating slight over-fitting.}
\label{fig:simpvals}
\end{figure}

Figure \ref{fig:simpvals} plots the PIT histograms for Theseus and ISOC for ESAs 2 through 6 and each ribbon model. PIT histograms show no visual departure from uniformity for Theseus while ISOC shows a slight mounded shape for all ribbon models, indicating a slight over-fitting with PIT values near 0 and 1 occurring less often than would be expected under nominal conditions.

\subsubsection{Unblurred Estimated Sky Map Comparisons to Simulated Sky Maps}
\label{subsubsec:pixelcomp}
To assess the quality of Theseus and ISOC unblurred sky map estimates, we compute the mean absolute percent error (MAPE) to assess the \emph{accuracy} of the point estimates, the continuous rank probability score (CRPS) to assess the \emph{skill} \citep{matheson1976scoring,gneiting2007strictly}, the 95\% confidence interval widths to asses the \emph{sharpness}, and the 95\% empirical coverage to assess the \emph{reliability}. 

More specifically and suppressing indexing for ESA, ribbon model, and estimation method, let 
\begin{itemize}
    \item $\Theta(\bs{s}_j)$ be the true simulated ENA rate of the pixel centered at $\bs{s}_j$ (referred to as pixel $j$)
    \item $\hat{\Theta}(\bs{s}_j)$ be the point estimate for the ENA rate of pixel $j$
    \item $\hat{q}_{0.025,j}$ and $\hat{q}_{0.975,j}$ are the 0.025 and 0.975 quantiles for the estimated distribution for pixel $j$, respectively.
\end{itemize}
For Theseus, $\hat{q}_{0.025,j}$ and $\hat{q}_{0.975,j}$ correspond to the 2.5 and 97.5 percentile bootstrap intervals. For ISOC, which estimates a mean and variance for each pixel, $\hat{q}_{0.025,j}$ and $\hat{q}_{0.975,j}$ are the corresponding quantiles of a Gaussian distribution with ISOC's estimated mean and variance. 

MAPE, 95\% confidence interval (CI) widths, and 95\% empirical coverage are computed as follows:
\begin{linenomath}\begin{align}
    \text{MAPE} &= \frac{100}{N_{p}}\sum_{j=1}^{N_p} |\Theta(\bs{s}_j) - \hat{\Theta}(\bs{s}_j)|/\Theta(\bs{s}_j)\\
    \text{95\% CI Width} &= N_p^{-1} \sum_{j=1}^{N_p} (\hat{q}_{0.975,j} - \hat{q}_{0.025,j})\\
    \text{95\% Empirical Coverage} &= \frac{100}{N_{p}} \sum_{j=1}^{N_p} \text{I}(\hat{q}_{0.025,j} \leq \Theta(\bs{s}_j) \leq \hat{q}_{0.975,j})
\end{align}\end{linenomath}
\noindent where $\text{I}()$ equals 1 if the argument is true and 0 otherwise. 

The CRPS is a function of the pixel value $\Theta(\bs{s}_j)$ and the estimated cdf of $\hat{\Theta}(\bs{s}_j)$, call it $F_{\hat{\Theta}(\bs{s}_j)}$. The CRPS is computed as
\begin{linenomath}\begin{align}
    \text{CRPS}\big(\Theta(\bs{s}_j), F_{\hat{\Theta}(\bs{s}_j)}\big) &= \int_{-\infty}^{\infty} \bigg(F_{\hat{\Theta}(\bs{s}_j)}(x) - \text{I}\big(\Theta(\bs{s}_j) \leq x\big)\bigg)^2 dx.
\end{align}\end{linenomath}

For ISOC, $F_{\hat{\Theta}(\bs{s}_j)}$ is the cdf of a Gaussian with corresponding mean and variance estimates. For Theseus, $F_{\hat{\Theta}(\bs{s}_j)}$ is an empirical distribution composed of the $N_B$ bootstrap estimates $\hat{\Theta}(\bs{s}_j)_b$. The CRPS for ISOC and Theseus were computed using functions within the \texttt{scoringRules} package \citep{scoringrules2019} in \texttt{R}.

Table \ref{tab:pixelcomp} presents the results comparing the unblurred sky map estimates to the true, unblurred sky maps. For each ribbon model, assessments are made for the whole map, but also partitioned into ribbon only and GDF only, allowing us to isolate sky map estimates to different parts of the simulated heliosphere. The partitioning of pixels can be visually seen in Figure \ref{fig:rcmap}.

\begin{table}[ht]
\caption{\small The mean absolute percent error (MAPE), 95\% confidence interval width (CI Width), 95\% empirical coverage (Coverage), and continuous rank probability score (CRPS) for ISOC and Theseus. Results are presented averaged over all ESAs and each ribbon model (Ribbon Model). Results are computed for the whole map, and also isolated to the ribbon only and the GDF only (Map). Bolded values indicate better results.}
\begin{tabular}{ll|cc|cc|cc|rr}
\hline
   & \textbf{Ribbon} & \multicolumn{2}{|c}{\textbf{MAPE (\%)}}  & \multicolumn{2}{|c}{\textbf{CI Width ($\times 10^{-2}$)}} & \multicolumn{2}{|c}{\textbf{Coverage (\%)}} & \multicolumn{2}{|c}{\textbf{CRPS $(\times 10^{-3})$}} \\ 
\textbf{Map} & \textbf{Model} & \textbf{ISOC} & \textbf{Theseus} & \textbf{ISOC} & \textbf{Theseus} & \textbf{ISOC} & \textbf{Theseus} & \textbf{ISOC} & \textbf{Theseus} \\ 
  \hline
\multirow{4}{*}{\shortstack{Whole\\ Map}} & GDF-only & 11.2 & \textbf{5.5} & 4.4 & \textbf{2.2} & 94.2 & \textbf{94.7} & 6.4 & \textbf{3.2} \\ 
                                   & Weak Scattering & 11.3 & \textbf{6.1} & 4.5 & \textbf{2.7} & 92.2 & \textbf{95.0} & 7.2 & \textbf{4.0} \\ 
                                 & Spatial Retention & 11.3 & \textbf{6.2} & 4.5 & \textbf{2.8} & 91.7 & \textbf{94.4} & 7.2 & \textbf{4.2} \\ 
                            & Varying Ribbon Profile & 11.9 & \textbf{7.0} & 4.6 & \textbf{3.3} & 87.4 & \textbf{91.7} & 9.3 & \textbf{5.8} \\ \hline
  \multirow{4}{*}{\shortstack{Ribbon\\ Only}} & GDF-only & 8.7  & \textbf{4.9} & 4.2 & \textbf{2.2} & 97.2 & \textbf{95.6} & 6.2  & \textbf{3.2} \\ 
                                       & Weak Scattering & 9.0  & \textbf{6.4} & 4.6 & \textbf{3.5} & 92.0 & \textbf{95.2} & 9.1  & \textbf{5.3} \\ 
                                     & Spatial Retention & 9.2  & \textbf{7.3} & 4.5 & \textbf{3.5} & 90.3 & \textbf{91.0} & 9.4  & \textbf{6.8} \\ 
                                & Varying Ribbon Profile & 10.8 & \textbf{8.4} & 5.0 & \textbf{4.6} & 78.2 & \textbf{85.5} & 15.5 & \textbf{10.8} \\ \hline
  \multirow{4}{*}{\shortstack{GDF\\ Only}} & GDF-only & 12.0 & \textbf{6.5} & 4.8 & \textbf{2.4} & 97.1 & \textbf{93.8} & 7.1 & \textbf{3.6} \\ 
                                    & Weak Scattering & 11.9 & \textbf{7.0} & 4.8 & \textbf{2.7} & 97.1 & \textbf{95.2} & 7.0 & \textbf{3.9} \\ 
                                  & Spatial Retention & 11.6 & \textbf{6.6} & 4.8 & \textbf{2.9} & 96.9 & \textbf{96.3} & 7.0 & \textbf{3.8} \\ 
                             & Varying Ribbon Profile & 12.2 & \textbf{8.1} & 4.9 & \textbf{3.2} & 96.2 & \textbf{94.1} & 7.6 & \textbf{4.9} \\ 
   \hline
\end{tabular}
\label{tab:pixelcomp}
\end{table}

As shown in Table \ref{tab:pixelcomp}, Theseus is more accurate, has better sharpness, has empirical coverage closer to nominal coverage, and has better skill for every considered ribbon model and all parts of the sky map, providing strong evidence that Theseus-estimated sky maps are better than ISOC-estimated sky maps.

Specifically, Theseus has better point estimates than ISOC for all ribbon models and all parts of the sky map, as measured by MAPE. This is strong evidence that Theseus is more \emph{accurate} than ISOC. Averaging over all ribbon models for the whole map, ISOC MAPE is about 85\% larger than Theseus MAPE.   

Theseus has lower CRPS than ISOC for all ribbon models and all parts of the sky map, providing strong evidence that Thesues has better \emph{skill} than ISOC. The CRPS for ISOC is approximately 80\% larger than the CRPS for Theseus, averaged over all ribbon models for the whole map. For both ISOC and Theseus, CRPS is larger for the part of the sky map with the ribbon compared to the GDF, indicating that estimating the ribbon is more challenging than the GDF.

\begin{figure}[!ht]
\centering
\includegraphics[width=1\linewidth]{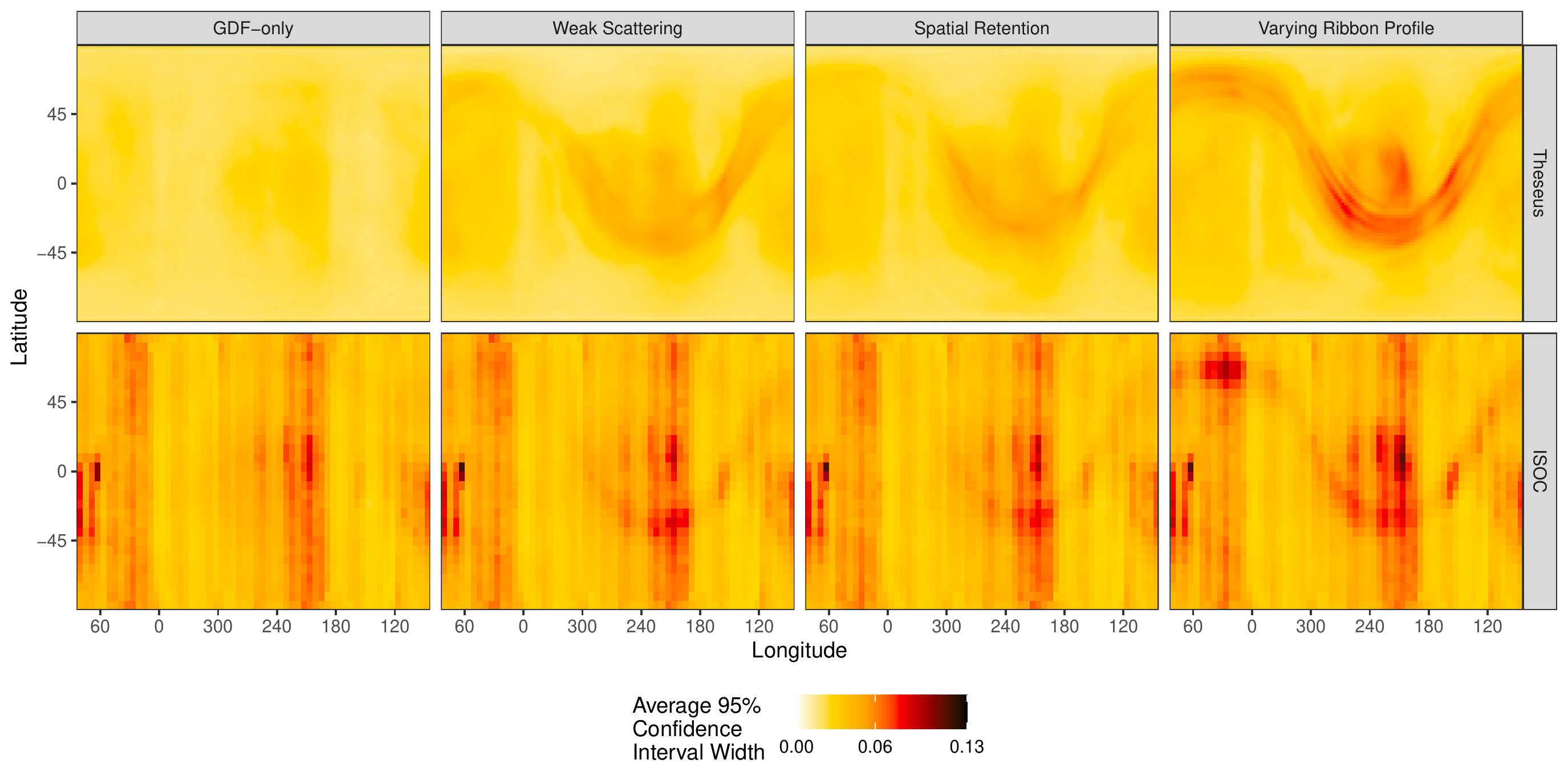}
\caption{\small Theseus (top row) and ISOC (bottom row) average 95\% confidence interval (CI) widths across all ESAs for all ribbon scenarios (columns). Theseus has narrower CI widths than ISOC, in general. The ribbon is present as larger CI widths for both Theseus and ISOC. Vertical streaks of larger CI widths are present in ISOC maps, corresponding to arcs with lower exposure times. These vertical stripes are not as present in Theseus sky maps because of the smoothing across larger spatial regions, while ISOC sky maps are made on a pixel-by-pixel basis. }
\label{fig:ciwidths}
\end{figure}

Theseus has narrower 95\% confidence interval widths than ISOC for all ribbon models and all parts of the sky map, indicating that Theseus has better \emph{sharpness} than ISOC (see Figure \ref{fig:ciwidths}). On average, ISOC's 95\% confidence interval widths are 65\% wider than Theseus's, averaged over all ribbon models for the whole map. As can be seen in Figure \ref{fig:ciwidths}, Theseus's confidence interval widths are largest over the ribbon while ISOC's confidence interval widths are largest over the ribbon and notably also along arcs with little to no exposure time, appearing as vertical streaks of wide CI widths.

Finally, Theseus had empirical coverage closer to nominal coverage than ISOC for all ribbon models and parts of the sky map, indicating that Theseus's uncertainty estimates are more \emph{reliable} than ISOC's.
Over the whole map, Theseus's empirical coverages are within 1\% of nominal for all ribbon models except the Varying Ribbon Profile model (with an empirical coverage of 91.7\%). 
ISOC's empirical coverages range from 87.4\% to 94.2\%, indicating under-coverage.
Over the GDF only portion of the sky map and for all ribbon models, empirical coverages were within 2.1\% of nominal coverage for both ISOC and Theseus, indicating reliable uncertainty estimates. 
Empirical coverages deviated some from nominal coverages over the ribbon only portion of the sky map, however, with empirical coverages dropping from the Weak Scattering model to the Spatial Retention model to the Varying Ribbon Profile model, coinciding with the ribbon shape becoming increasingly sharper and more asymmetric.
These results suggest that sharper and more asymmetric ribbon shapes are more challenging to estimate, likely related to the assumption made in Stage 2 of Theseus that the unblurred and blurred sky maps are ``close" to each other: an assumption that is most significantly violated in the Varying Profile Ribbon model with the sharpest and most asymmetric ribbon shape.

\subsubsection{Comparing the Ribbon Shapes}
\label{subsubsec:skew}
\begin{figure}[!ht]
\centering
\includegraphics[width=1\linewidth]{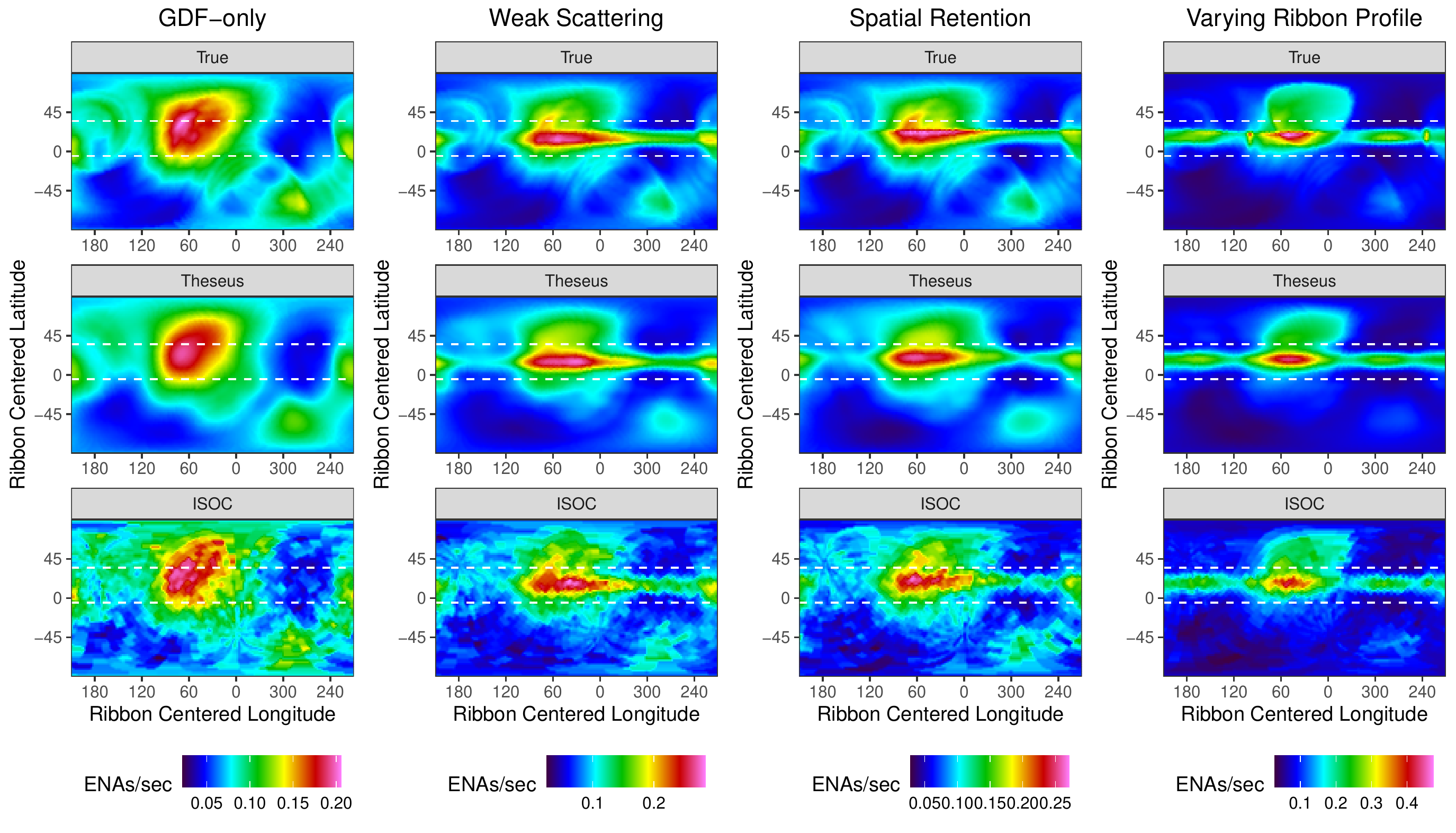}
\caption{\small The ribbon-centered projection of the true simulated sky maps (top row), along with the Theseus (middle row) and ISOC (bottom row) estimates for binned direct event data corresponding to ESA 6. In these ribbon-centered projections, the ribbon appears as a horizontal line. The nose of the heliosphere is also centered (near ribbon centered longitude 60). White, dashed horizontal lines outline the location of the ribbon. The ``Ribbon only" sky map segment in Table \ref{tab:pixelcomp} correspond to pixels between the two horizontal dashed lines. The ``GDF only" sky map segment are the pixels that fell above or below the horizontal dashed lines.}
\label{fig:rcmap}
\end{figure}

The shape of the ribbon is of scientific interest. 
In this section, we isolate and compare the ribbon profiles from the Theseus and ISOC sky maps to the truth.
Recall that the ribbon forms a nearly complete circular band across the sky.
In order to isolate and compare the ribbon profiles, we first project the sky maps into a ribbon-centered frame as shown in Figure \ref{fig:rcmap}.
In ribbon-centered frame, the ribbon appears as a horizontal streak.
We are able to do this ribbon-centering for simulated maps without error because we know the ribbon center location.\footnote{It is also of scientific interest, but outside the scope of this work, to estimate the ribbon center for non-simulated settings.}
The pixels in the ribbon-centered frame sky map between the dashed horizontal lines are the pixels summarized in the ``Ribbon-Only" section of Table \ref{tab:pixelcomp}.

\begin{figure}[!ht]
\centering
\includegraphics[width=1\linewidth]{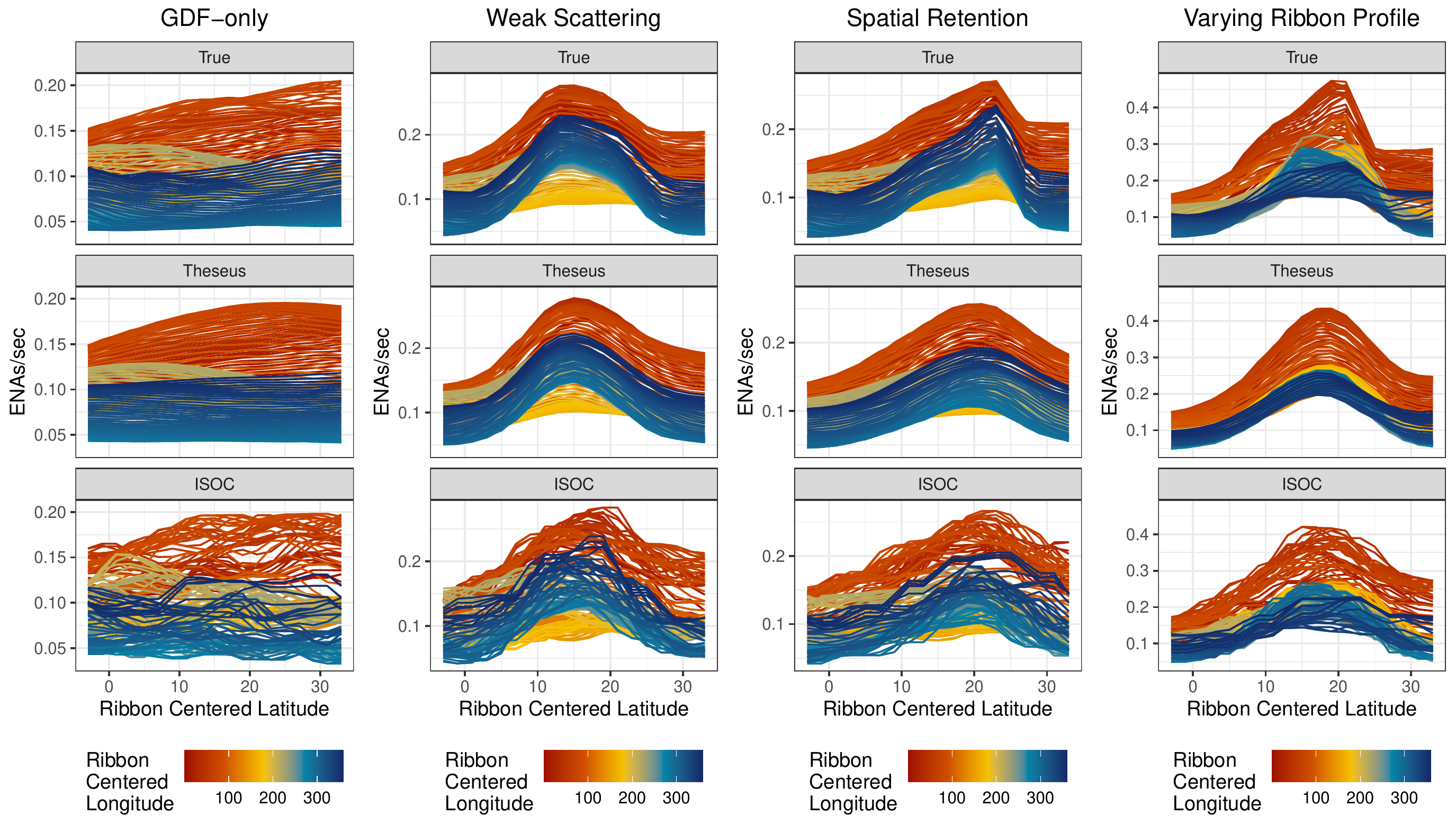}
\caption{\small Ribbon cross-sections for each ribbon model (columns) and sky map estimation method (rows). Each line within a panel corresponds to a ribbon-centered longitude (i.e., a vertical slice in Figure \ref{fig:rcmap}). We can see that Theseus better approximates the shape of the ribbon than does ISOC. Neither ISOC nor Theseus, however, completely captures the asymmetry of the ribbon for the Spatial Retention model or the Varying Ribbon Profile model.}
\label{fig:rcribbon}
\end{figure}

Figure \ref{fig:rcribbon} plots the ribbon profiles for the ESA 6 sky maps (i.e., the ENA rates for the vertical cross-sections between the horizontal dashed lines in Figure \ref{fig:rcmap}). It appears that Theseus better visually captures the shape of the ribbon than ISOC. 
Theseus captures the smoothly varying ribbon shapes from adjacent cross-sections, while ISOC ribbon shape estimates vary more erratically as a result of ISOC's pixel-by-pixel estimation procedure. 
Theseus also better captures the skew of the ribbon profiles than ISOC (see Table \ref{tab:skewmae}).
While Theseus visually does a better job capturing the ribbon shape than does ISOC, both Theseus and ISOC fail to capture some of the finer features of the ribbon, such as the sharp peak/pronounced asymmetry of the Spatial Retention model and the strong asymmetric features of the Varying Ribbon Profile. ISOC's failure to capture these sharp, asymmetric ribbon shapes is in part because it does not account for the IBEX-Hi PSF and is estimating a blurred sky map but treating it as an unblurred sky map. Theseus's failure to capture these sharp, asymmetric ribbon shapes is in part related to the assumption made in Theseus Stage 2 that the unblurred sky map is ``close" to the blurred sky map. While this assumption prevents unblurred sky map estimates from being wildly unreasonable (recall $\hat{\bs{\Theta}}_2$ and $\hat{\bs{\Theta}}_3$ in Figure \ref{fig:deconvolution}), it also prevents Theseus from capturing the parts of the unblurred sky map that deviate most dramatically from the blurred sky map: namely the ribbon. To better capture the ribbon in the Spatial Retention model and Varying Ribbon Profile, additional assumptions would likely need to be made about the shape of the ribbon; assumptions intentionally avoided for the discovery science setting we are operating within.

\begin{table}[!ht]
\centering
\caption{Mean absolute error (MAE) between the true ribbon skew and the Theseus and ISOC estimated ribbon skew, averaged over all cross-sections and all ESAs. Lower MAE is indicated in bold. Theseus has smaller MAE than ISOC for all ribbon models, indicating it better captures the skew of the ribbon than ISOC.}
\begin{tabular}{l|rr}
  \hline
 \textbf{Ribbon Model} & \textbf{ISOC} & \textbf{Theseus}  \\ 
 \hline
GDF-only & 0.053 & \textbf{0.040} \\ 
  Weak Scattering & 0.043 & \textbf{0.032} \\ 
  Spatial Retention & 0.057 & \textbf{0.048} \\ 
  Varying Ribbon Profile & 0.054 & \textbf{0.044} \\ 
   \hline
\end{tabular}
\label{tab:skewmae}
\end{table}

\section{Real Binned Direct Event Data Example}
\label{sec:realdataexample}

\begin{figure}[!ht]
\centering
\includegraphics[width=.45\linewidth]{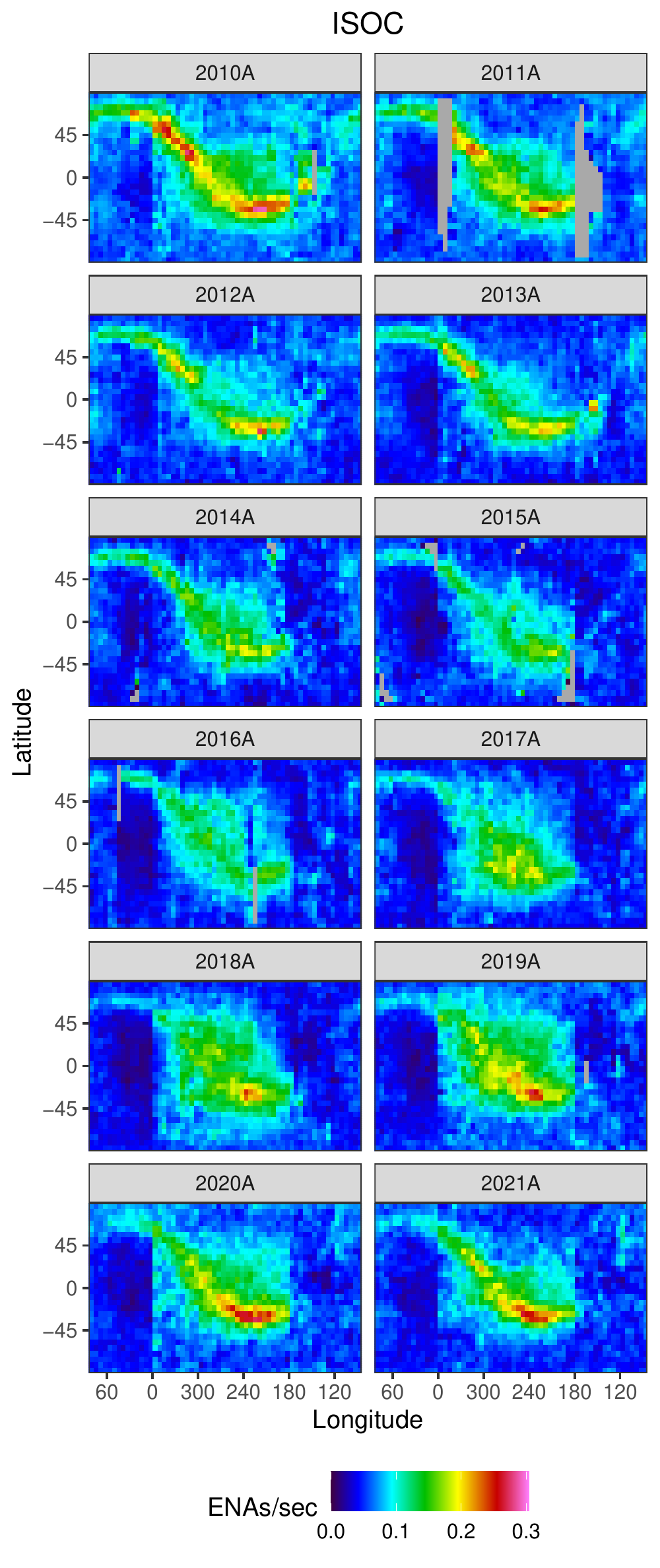}
\includegraphics[width=.45\linewidth]{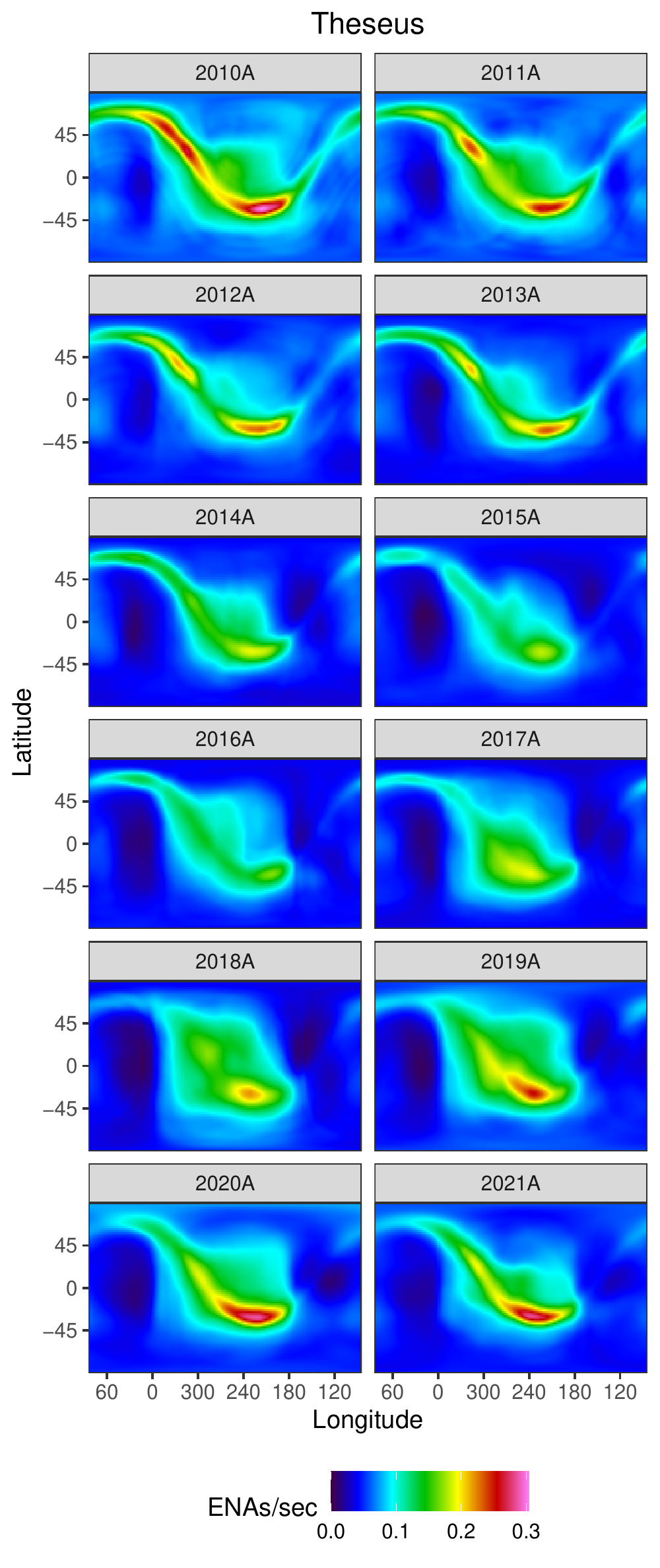}
\caption{ISOC (left) and Theseus (right) unblurred sky map estimates fit to real binned direct event data corresponding to ESA 4 collected over roughly the first half each calendar year (the ``A" maps). Grey pixels in the ISOC sky maps correspond to pixels ISOC does not estimate because no binned direct event data was collected sufficiently close to the pixel's center.}
\label{fig:realmaps}
\end{figure}

We fit Theseus and ISOC to the real binned direct event data collected by the IBEX-Hi instrument corresponding to the ESA 4 ``A" maps between the years 2010 to 2021 to illustrate how the current state-of-the-art ISOC sky maps compare to Theseus sky maps. Unblurred sky maps are shown in Figure \ref{fig:realmaps}. Theseus sky maps appears noticeably smoother than ISOC sky maps. Theseus sky maps are at a $2\degree$~pixel resolution and interpolate over regions of the sky while ISOC sky maps are made at a $6\degree$~resolution on a pixel-by-pixel basis and leave ``holes" in the sky maps where no binned direct event data are present.

\begin{figure}[!ht]
\centering
\includegraphics[width=.45\linewidth]{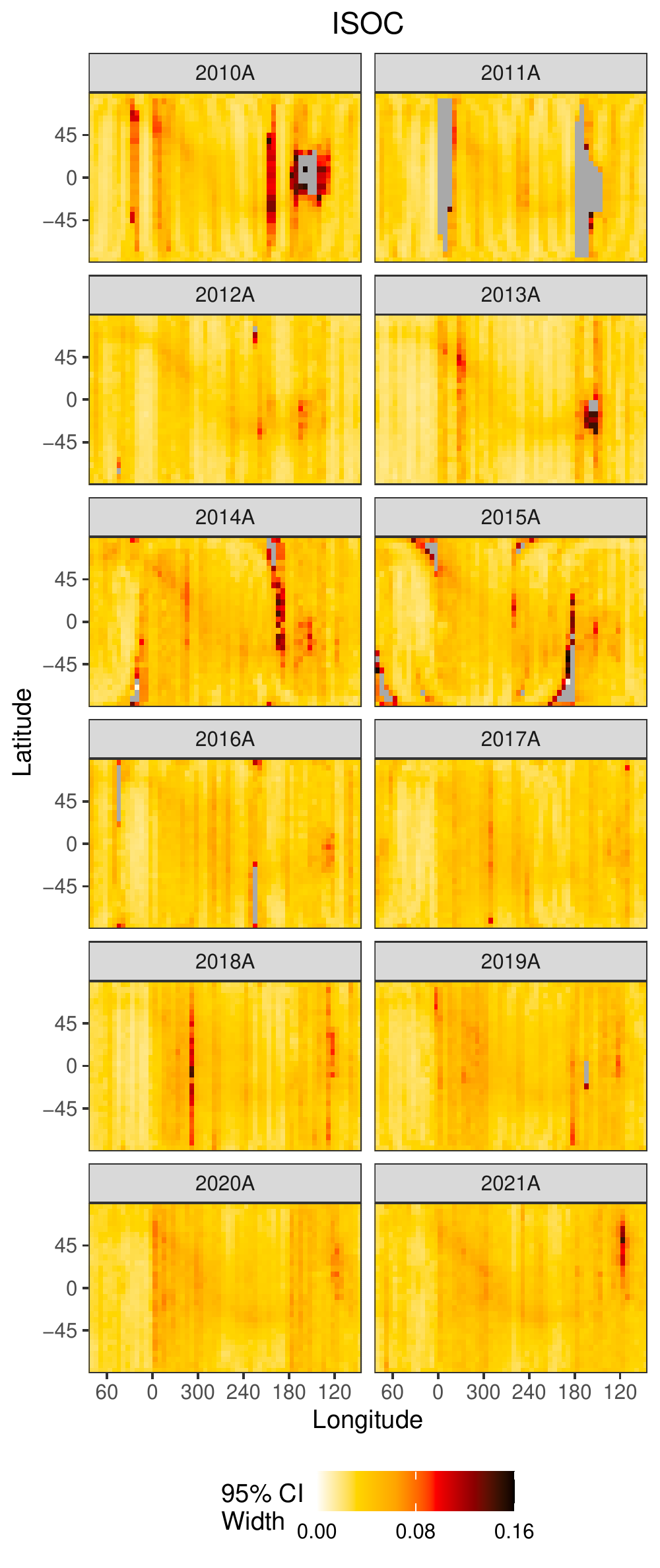}
\includegraphics[width=.45\linewidth]{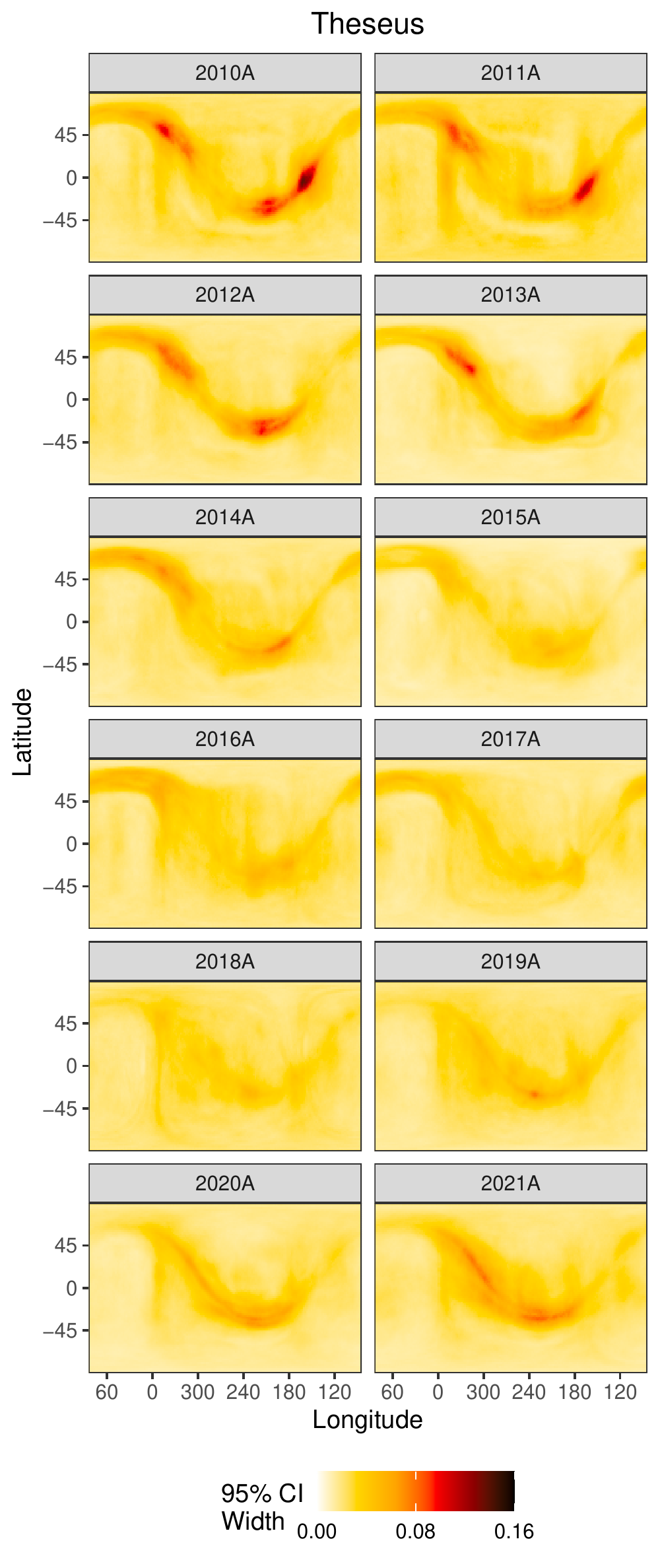}
\caption{\small ISOC (left) and Theseus (right) 95\% confidence interval widths (ENAs per second) corresponding to ESA 4. Theseus has narrower 95\% confidence intervals widths than ISOC overall. Grey pixels in the ISOC sky maps either represent pixels with no estimate or pixels with extremely large 95\% confidence interval widths that were truncated for ease of viewing.}
\label{fig:realmapsci}
\end{figure}

\begin{figure}[!ht]
\centering
\includegraphics[width=1\linewidth]{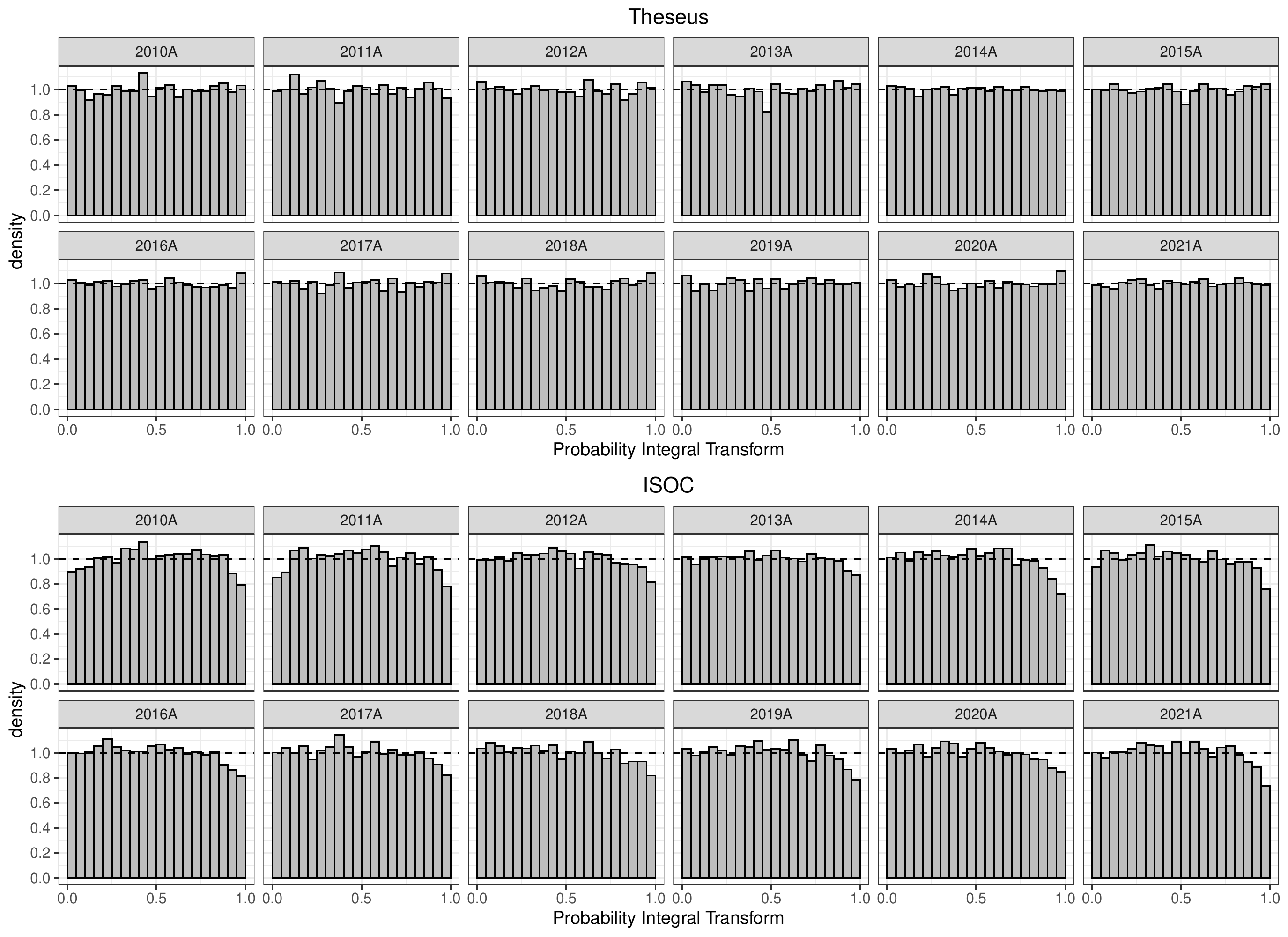}
\caption{\small Probability integral transform (PIT) histograms by years from 2010 to 2019 for ESA 4. The top set of histograms are for Theseus sky maps; the bottom for ISOC sky maps. No indication of non-uniformity is visible for Theseus PIT histograms. Slightly mounded shapes exists for 2010A and 2011A for ISOC maps. For later years, fewer PIT values near 1 are observed relative to what would be expected.}
\label{fig:realpithistograms}
\end{figure}

While we do not know if ISOC or Theseus sky maps better reflects the true, latent unblurred sky maps, we have reason to believe it is Theseus due to the similar relative performance between Theseus and ISOC we see on real data as we did on simulated data (recall Section \ref{sec:simstudy}). For simulated data where Theseus sky maps better captured simulated sky maps than did ISOC, we saw that Theseus had narrower 95\% confidence interval widths and PIT histograms better reflecting samples from a Uniform(0,1). We have the same findings with real data. For instance, ISOC's 95\% confidence interval widths are, on average, $\sim$50\% wider than Theseus's (Figure \ref{fig:realmapsci}). Furthermore, Theseus PIT histograms are consistent with a Uniform(0,1) for all sky maps, while ISOC PIT histograms show a lower frequency for PIT values near 1 for all sky maps than would be expected, and a similar mounded shape for 2010A and 2011A that we saw with the simulated sky maps (Figure \ref{fig:realpithistograms}).

Artifacts of the Theseus and ISOC sky map estimation approaches are present. In the ISOC sky maps, we see coarse pixelation which can obscure finer scale features. We can also see some vertical banding in locations where binned direct event data exposure times are lower (e.g., in 2014A near ecliptic longitude 190 and ecliptic latitude range 0 to 90). If no binned direct event data are sufficiently close to a pixel's center, ISOC will not provide an estimate for that pixel; those pixels appear as grey in Figure \ref{fig:realmaps}. 
For Theseus, some of the artificial banding caused by the PPR regression in step 1 of Theseus Stage 1 can be seen in the light blue bands running along the ribbon in the 2010A sky map and less so in 2011A sky map. While the artificial banding in Figure \ref{fig:realmaps} is relatively minimal, it can be more pronounced, especially in ESA 2 where less binned direct event data is available to estimate the sky maps (not shown).

There are numerous scientific studies that will directly benefit from the demonstrated improvements found in Theseus sky maps over the standard ISOC sky maps. In terms of ribbon science, as already discussed there is the improved ability to discern the shape of the ribbon profile, and knowing this has the potential to differentiate between competing ribbon formation theories.  Beyond this, the higher resolution of Theseus maps more accurately locates points along the ribbon in the sky.  This will allow a more accurate application than previously done of the astronomical method of parallax to determine the distance to the ribbon \citep{swaczyna2016parallax}. More accurately demarcating the circular path of the ribbon around the sky will allow a better determination of the location of the ribbon center than previously done \citep{funsten2013ribboncirc, dayeh2019ribbon}.  This is important because the ribbon center is associated with the direction of the interstellar magnetic field \citep[e.g.,][]{schwadron2009ribbon, zirnstein2016local}, which helps govern the shape of the heliosphere \citep{zirnstein2016ribbon} and is also a topic of great interest in galactic astronomy \citep[e.g.,][]{frisch2012ismf, frisch2022ismf}. And finally, the improved rendering of the GDF in Theseus sky maps will allow for more accurate determinations of the size and time variability of GDF features \citep[e.g.,][]{mccomas2013heliotail, dayeh2022heliotail}, which in turn helps constrain simulations of the heliosphere's dynamics and evolution, and allows for a more accurate determination of the three-dimensional shape of the heliosphere than done previously \citep{reisenfeld20213Dheliosphere}.

\clearpage
\section{Discussion}
\label{sec:discussion}

In this paper, we presented a novel sky map estimation procedure called Theseus. Theseus produces sky map estimates in two stages. In Theseus Stage 1, a blurred sky map is estimated from binned direct event data. In Theseus Stage 2, an unblurred sky map is estimated from the previously estimated blurred sky map. Bootstrapping is used to quantify pixel uncertainties. Over a variety of realistic simulation scenarios, we demonstrated that Theseus is better than the current state-of-the-art sky map estimation procedure used by ISOC with respect to half a dozen metrics. Specifically, Theseus had more uniform PIT histograms, smaller MAPE, narrower 95\% confidence interval widths, better empirical coverage, smaller CRPS, and more accurate ribbon skewness than ISOC sky maps. Finally, we fit Theseus and ISOC to real binned direct event data and showed that, like with the simulated data, Theseus had more uniform PIT histograms and narrower 95\% confidence interval widths, providing some assurance that Theseus sky maps are likely better estimates of the true, latent unblurred sky maps than are ISOC's.

While Theseus is a clear improvement over the current state-of-the-art used by the ISOC, it is by no means perfect. 
Though efforts to reduce banding artifacts in the sky map estimates were helpful, they did not completely remove them in all sky maps. 
The banding artifacts are caused by the PPR-generated candidates in step 1 of Theseus Stage 1 and can be amplified by the sharpening map of Theseus Stage 2. 
Even though they are relatively minor, the banding artifacts can be problematic for science as they might be confused as genuine heliospheric features rather than unfortunate methodological byproducts. 
In the future, there are numerous options to try to reduce the impacts of these artifacts, both through methodological modifications and data enhancements. No single option is perfect, but all are worth exploring. Methodologically, we could 
\begin{enumerate}
    \item replace the PPR-candidates in step 1 of Theseus Stage 1 with smoothed versions of themselves (e.g., we could replace $\dot{\bs{\theta}}_l$ with $g(\bs{K} \dot{\bs{\theta}}_l)$; a debiasing function $g()$ applied to the smoothed/blurred version of $\dot{\bs{\theta}}_l$). This should reduce the banding artifacts in the candidate blurred sky maps but possibly at the expense of capturing the sharpness of the ribbon peaks.
    \item consider alternatives (or additions) to PPR to include in step 1 of Theseus Stage 1 such as multivariate adaptive regression splines \citep{friedman1991multivariate}, neural networks \citep{rosenblatt1958perceptron,rumelhart1986learning}, local regression and likelihood methods \citep{loader2006local}, or fast Vecchia approximations of Gaussian processes \citep{katzfuss2021general,katzfuss2022scaled}. This may, however, just replace one artifact with another.
\end{enumerate}  
\noindent Alternatively, there is evidence to suggest that as the number of available ENA counts for binned direct event data increases, the banding artifacts fade (e.g., sky maps 2010A and 2011A have the lowest total number of counts of all sky maps in Figure \ref{fig:realmaps}). From a data enhancement perspective, we could
\begin{enumerate}
    \item incorporate a new data product. There is another data product related to the IBEX mission not used in this manuscript that would effectively triple the event rate for the binned direct event data.
    \item improve the pre-processing step. Recall that the standard IBEX data pre-processing step discards roughly two-thirds of the direct event data when constructing binned direct event data. Improving the pre-processing step in such a way that converts more direct event time segments into binned direct event data will increase total exposure time and may help reduce the banding artifacts.
\end{enumerate}

One of the most challenging and scientifically important features for Theseus (and ISOC) to capture is the ribbon shape. 
Recall that the ribbon was completely unexpected and only discovered after IBEX's launch. 
The IBEX-Hi's $6.5\degree$ FWHM PSF was anticipated to be, and is narrow enough to resolve GDF features. 
Resolving ribbon features, however, has proven more challenging. 
Assessing if Theseus sky maps are ``good enough" to discern between competing ribbon origin theories will depend on the theories and how similar their resulting ribbon profiles are. 
If, for instance, competing theories can be differentiated by their skew (as is the case with the Weak Scattering and Spatial Retention models: recall Figure \ref{fig:rcribbon}), then Theseus sky maps should be good enough to  contribute to the scientific process now.
If, however, two competing theories are quite similar and only deviate in, say, how peaked their ribbons are (imagine two slightly different versions of the Spatial Retention model), Theseus sky maps are likely not presently good enough for that level of discernment.

The challenge of improving ribbon shape estimation is further amplified in our discovery science setting, where few assumptions can be made \emph{a priori} about the shape of the ribbon. Were we willing and able to make additional assumptions about the shape of the ribbon, however, we would likely be able to unblur the blurred sky maps more accurately. For instance, \citet{kuusela2017shape} provides an illustration of how unblurring can be improved when shape constraints can be assumed. 

In the absence of additional ribbon shape assumptions, improved instrumentation with narrower PSFs may be needed to accurately resolve the ribbon. Lucky for us, the upcoming Interstellar Mapping and Acceleration Probe (IMAP) \citep{mccomas2018interstellar} mission will have instrumentation with a FWHM field-of-view roughly half that of IBEX-Hi. Theseus is well-positioned to analyze IMAP data upon arrival after its scheduled 2025 launch.

\section{Acknowledgements}
We thank C.C. Essix for her assistance and encouragement. 
Research presented in this manuscript was supported by the Laboratory Directed Research and Development (LDRD) program of Los Alamos National Laboratory (LANL) under project number 20220107DR and by the NASA IBEX Mission as part of the NASA Explorer Program (80NSSC20K0719).
Additionally, L.J.B. was supported by LANL's LDRD Richard Feynman Postdoctoral Fellowship (20210761PRD1), while E.J.Z. acknowledges support from NASA grant 80NSSC17K0597 in the development of the ribbon models.
Approved for public release: LA-UR-22-30879.

\clearpage
\appendix 
\input{Supplementary_Materials}

\newpage
\bibliographystyle{plainnat}
\bibliography{references.bib}

\end{document}

%% file: Supplementary_Materials.tex
\appendix

\maketitle 

\begin{center}
\title{\Large Supplementary Materials for ``Towards Improved Heliosphere Sky Map Estimation with Theseus"}\\
\author{Dave Osthus$^{1,*}$, Brian P. Weaver$^1$, Lauren J. Beesley$^{1,2}$, Kelly R. Moran$^1$, Madeline A. Ausdemore$^1$, Eric J. Zirnstein$^4$, Paul H. Janzen$^{5,6}$, Daniel B. Reisenfeld$^3$
\\
$^1$Statistical Sciences Group, Los Alamos National Laboratory, Los Alamos, New Mexico, USA\\
$^2$Information Systems and Modeling, Los Alamos National Laboratory, Los Alamos, New Mexico, USA\\
$^3$Space Science and Applications Group, Los Alamos National Laboratory, Los Alamos, New Mexico, USA\\
$^4$Department of Astrophysical Sciences, Princeton University, Princeton, New Jersey, USA\\
$^5$Department of Physics and Astronomy, University of Montana, Missoula, Montana, USA\\
$^6$The New Mexico Consortium, Los Alamos, New Mexico, USA\\
*Corresponding Author: Dave Osthus, dosthus@lanl.gov
}
\end{center}


\section{How to Construct $\bs{K}(\bs{s}_i)$}
\label{appendix:K}
In three steps, we describe below how to construct $\bs{K}(\bs{s}_i)$, an $N_p \times 1$ row vector of the point spread function (PSF) matrix $\bs{K}$ centered at \emph{geographic} coordinate\footnote{In the main paper, $\bs{s}_i$ is in an \emph{ecliptic} coordinate system. Ecliptic coordinates are mapped into geographic coordinates by subtracting $180\degree$ from the ecliptic longitude. Describing the construction of $\bs{K}(\bs{s}_i)$ in geographic rather than ecliptic coordinates removes a layer of translation.} location $\bs{s}_i$ and $N_p$ is the number of mutually exclusive and exhaustive sky map pixels. For a $2\degree$ sky map, $N_p = (180/2)*(360/2) = 16,200$. The same procedure is followed for every binned direct event data location $\bs{s}_i$ and pixel center location $\bs{s}_j$. 

In what follows, we will map between geographic coordinates $\bs{s}_i = (\text{lon}_i, \text{lat}_i)$, where $\text{lon}_i \in [-180,180)$ and $\text{lat}_i \in [-90,90]$, and \emph{spherical} coordinates $\bs{s}^s_i$, where $\bs{s}^s_i = (x_i, y_i, z_i)$ and $x_i^2 + y^2_i + z^2_i = 1$.
Mapping from geographic $\bs{s}_i$ to spherical $\bs{s}^s_i$ is done as follows:
\begin{linenomath}\begin{align}
    x_i &= \text{cos}\big(\frac{\pi}{180}~\text{lon}_{i}\big) \text{cos}\big(\frac{\pi}{180}~\text{lat}_{i}\big) \nonumber\\
    y_i &= \text{sin}\big(\frac{\pi}{180}~\text{lon}_{i}\big) \text{cos}\big(\frac{\pi}{180}~\text{lat}_{i}\big)\nonumber\\
    z_i &= \text{sin}\big(\frac{\pi}{180}~\text{lat}_{i}\big).\nonumber
\end{align}\end{linenomath}
\noindent Mapping from spherical $\bs{s}^s_i$ to geographic $\bs{s}_i$ is done as follows:
\begin{linenomath}\begin{align}
    \text{lon}_{i} &= \frac{180}{\pi}~\text{arctan}(y_{i}, x_i)\nonumber\\
    \text{lat}_{i} &= \frac{180}{\pi}~\text{arcsin}(z_{i})\nonumber.
\end{align}\end{linenomath}

\subsection{Step 1: Draw $M$ realizations from a 3-dimensional von Mises-Fisher distribution}
\label{subsubsec:makeXstep1}
The IBEX-Hi field of view is approximately $6.5\degree$~full width at half max (FWHM) \citep{funsten2009interstellar}. 
The relationship between a Gaussian distribution with standard deviation $\sigma$ and its FWHM is
\begin{linenomath}\begin{align}
\label{eq:fwhm}
    \text{FWHM} &= \sigma \big(2\sqrt{2~\text{ln}(2)}\big).
\end{align}\end{linenomath}
Equation \ref{eq:fwhm} implies that $\sigma = 2.76\degree$~or, equivalently, $\sigma_{\text{rad}} = \sigma \frac{\pi}{180} = 0.0482$ radians. 

In directional statistics (i.e., statistics for unit vectors), the $p$-dimensional von Mises-Fisher distribution is a popular distribution analogous to the $p$-dimensional multivariate normal distribution for data in $\mathbb{R}^p$. The real IBEX-Hi PSF has hexagonal rather than circular contours near the center of the distribution.
The contours, however, become more circular as a function of distance from distribution center \citep[][see Figure 4]{funsten2009interstellar}.
We approximate the real PSF with a von Mises-Fisher distribution with the same center and FWHM variability as the real PSF.

The 3-dimensional von Mises-Fisher distribution has the following density:
\begin{linenomath}\begin{align}
    \label{eq:vmf}
    f(\bs{x}|\bs{s}^s_i, \kappa) &= \frac{\kappa}{4 \pi \text{sinh}(\kappa)} \text{exp}\big(\kappa (\bs{s}^s_i)' \bs{x}\big) 
\end{align}\end{linenomath}
where $\kappa \geq 0$ is a scalar concentration parameter controlling the level of dispersion, $\bs{s}^s_i$ is a three dimensional unit vector corresponding to spherical location $\bs{s}^s_i$, and sinh() is the hyperbolic sine function.

We sample $M=100,000$ draws from the 3-dimensional von Mises-Fisher distribution with 
$\kappa = 1/\sigma_{\text{rad}}^2 = 430.43.$  We refer to the $m^{th}$ draw as $\bs{s}^s_{i,m} = (x_{i,m}, y_{i,m}, z_{i,m})'$.
Figures \ref{fig:makeXstep1} and \ref{fig:makeXstep1_v2} show samples $\bs{s}^s_{i,m}$ from the 3-dimensional von Mises-Fisher distribution.

\subsection{Step 2: Map the $M$ draws from spherical coordinates back to geographic coordinates}
\label{subsubsec:makeXstep2}
For each of the $M$ draws, we map $\bs{s}^s_{i,m}$ from spherical coordinates back to geographic coordinates $\bs{s}_{i,m} = (\text{lon}_{i,m}, \text{lat}_{i,m})'$. Figures \ref{fig:makeXstep2} and \ref{fig:makeXstep2_v2} shows the draws $\bs{s}_{i,m}$ in geographic coordinates.

\subsection{Step 3: Compute pixel probabilities}
\label{subsubsec:makeXstep3}
The last step to compute $\bs{K}(\bs{s}_i)$ is to calculate the proportion of the $M$ draws that fell within each of the $N_p$ pixels. 

Let $p(\bs{s}_j)$ be the sky map's j$^{\text{th}}$ rectangular pixel of width $\delta$ (in degrees) centered at location $\bs{s}_j = (\text{lon}_j, \text{lat}_j)'$.
We say a point $\bs{s}^* = (\text{lon}^*, \text{lat}^*)'$ is in pixel $p(\bs{s}_j)$ if $\text{lat}^* \in [\text{lat}_j - \delta/2, \text{lat}_j + \delta/2)$ and $\text{lon}^* \in [\text{lon}_j - \delta/2, \text{lon}_j + \delta/2)$, accounting for spherical wrapping..

Given the above notation, the value of the j$^{\text{th}}$ column of the $\bs{K}(\bs{s}_i)$ row vector, $\bs{K}(\bs{s}_i)_j$, is computed as follows:
\begin{linenomath}\begin{align}
\label{eq:makeXstep3}
    \bs{K}(\bs{s}_i)_j = M^{-1}\sum_{m=1}^M \text{I}\big(\bs{s}_{i,m} \in p(\bs{s}_j)\big)
\end{align}\end{linenomath}
where $\text{I}()$ is an indicator function equal to 1 if the argument is true and 0 otherwise.
By construction, each element $\bs{K}(\bs{s}_i)_j \geq 0$ and $\sum_{j=1}^{N_p} \bs{K}(\bs{s}_i)_j = 1$.
$\bs{K}(\bs{s}_i)_j$ is an estimate of the probability that a draw from a von Mises-Fisher will fall in pixel $j$.

\begin{figure}
     \centering
     \begin{subfigure}[b]{.32\textwidth}
         \centering
          \includegraphics[width=\textwidth]{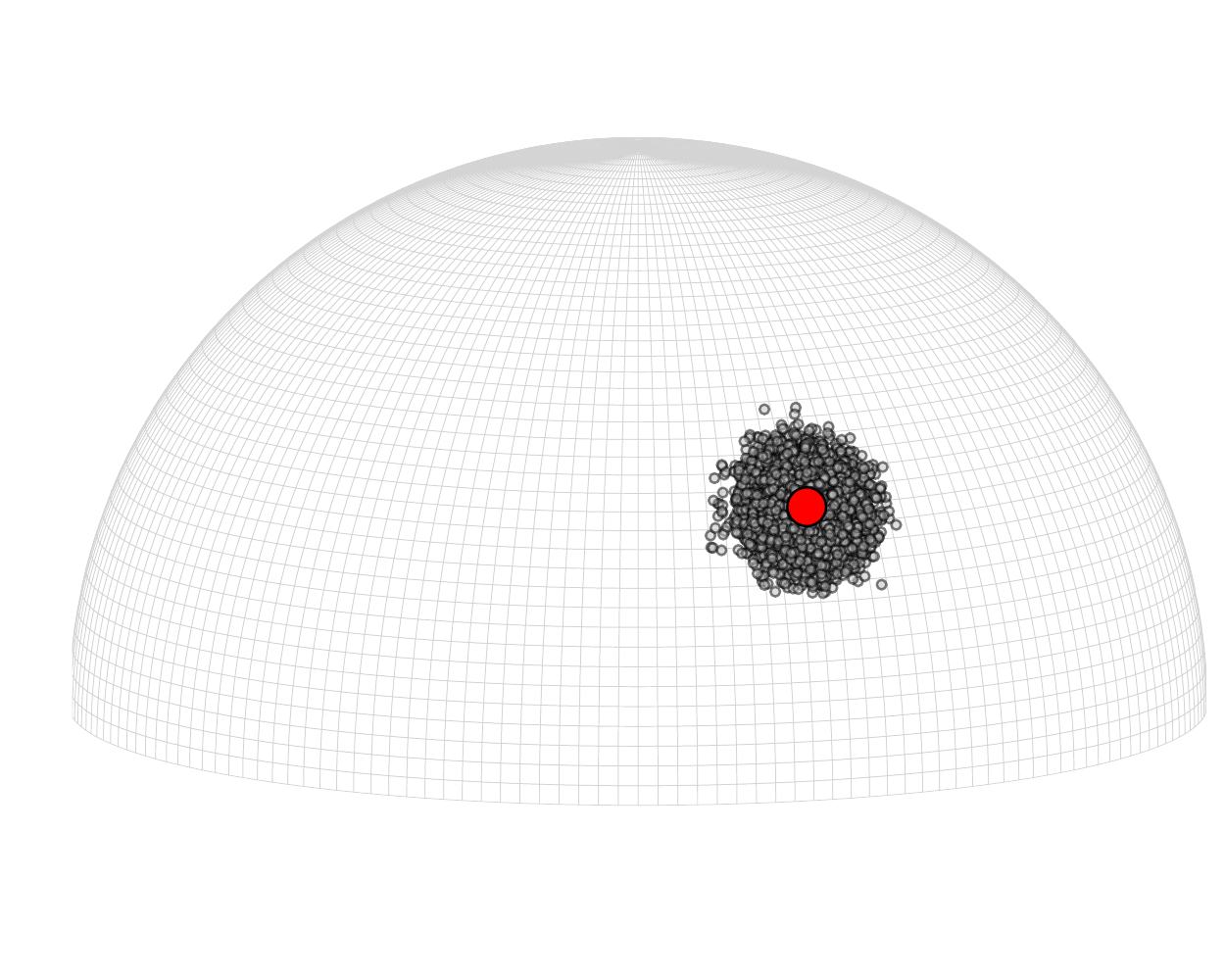}
         \caption{Step 1 corresponding to $\bs{s}_i = (30, 30)$.}
         \label{fig:makeXstep1}
     \end{subfigure}
     \hfill
     \begin{subfigure}[b]{0.32\textwidth}
         \centering
         \includegraphics[width=\textwidth]{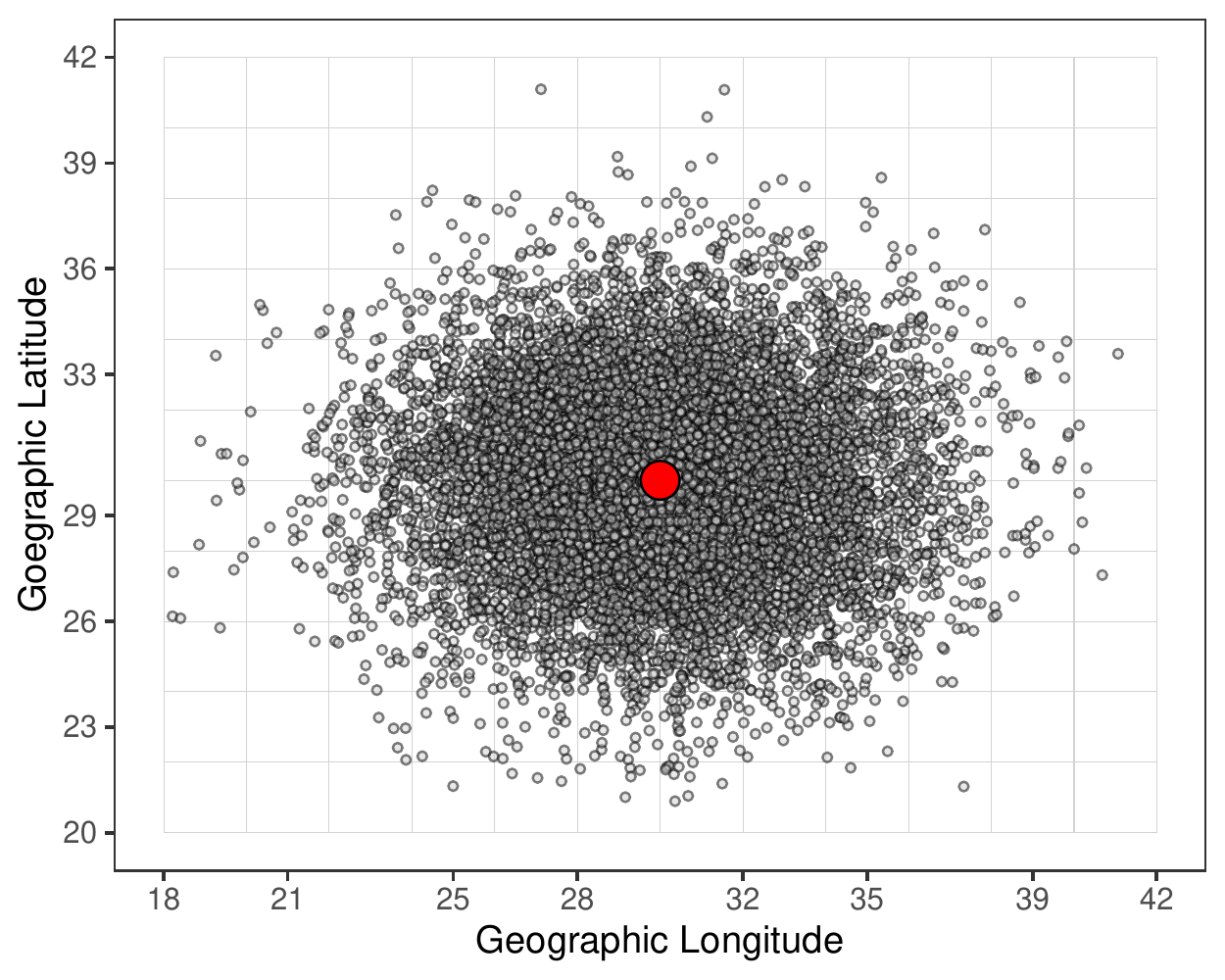}
         \caption{Step 2 corresponding to $\bs{s}_i = (30, 30)$.}
         \label{fig:makeXstep2}
     \end{subfigure}
     \hfill
     \begin{subfigure}[b]{0.32\textwidth}
         \centering
         \includegraphics[width=\textwidth]{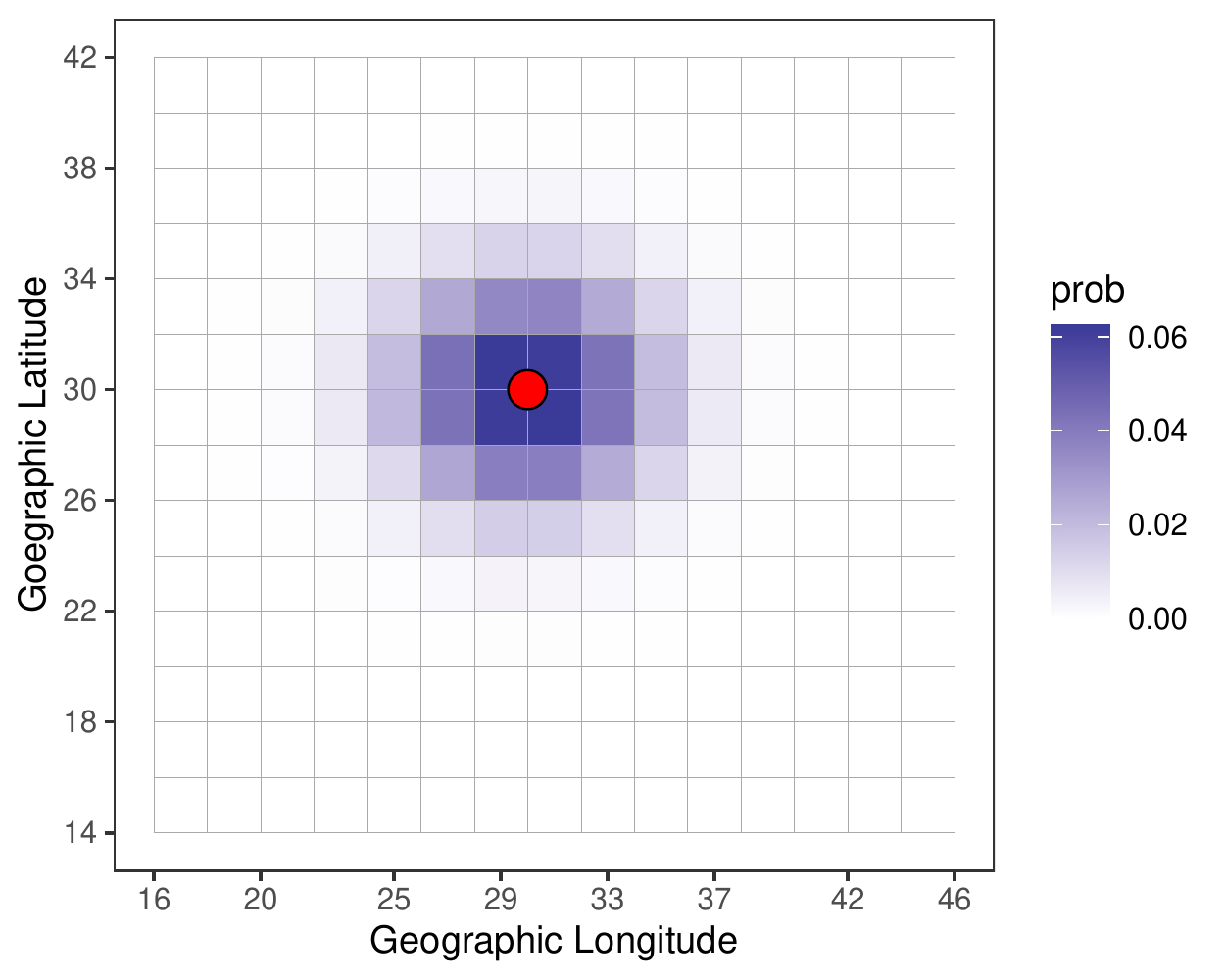}
         \caption{Step 3 corresponding to $\bs{s}_i = (30, 30)$.}
         \label{fig:makeXstep3}
     \end{subfigure}
     \hfill     
     \begin{subfigure}[b]{.32\textwidth}
         \centering
          \includegraphics[width=\textwidth]{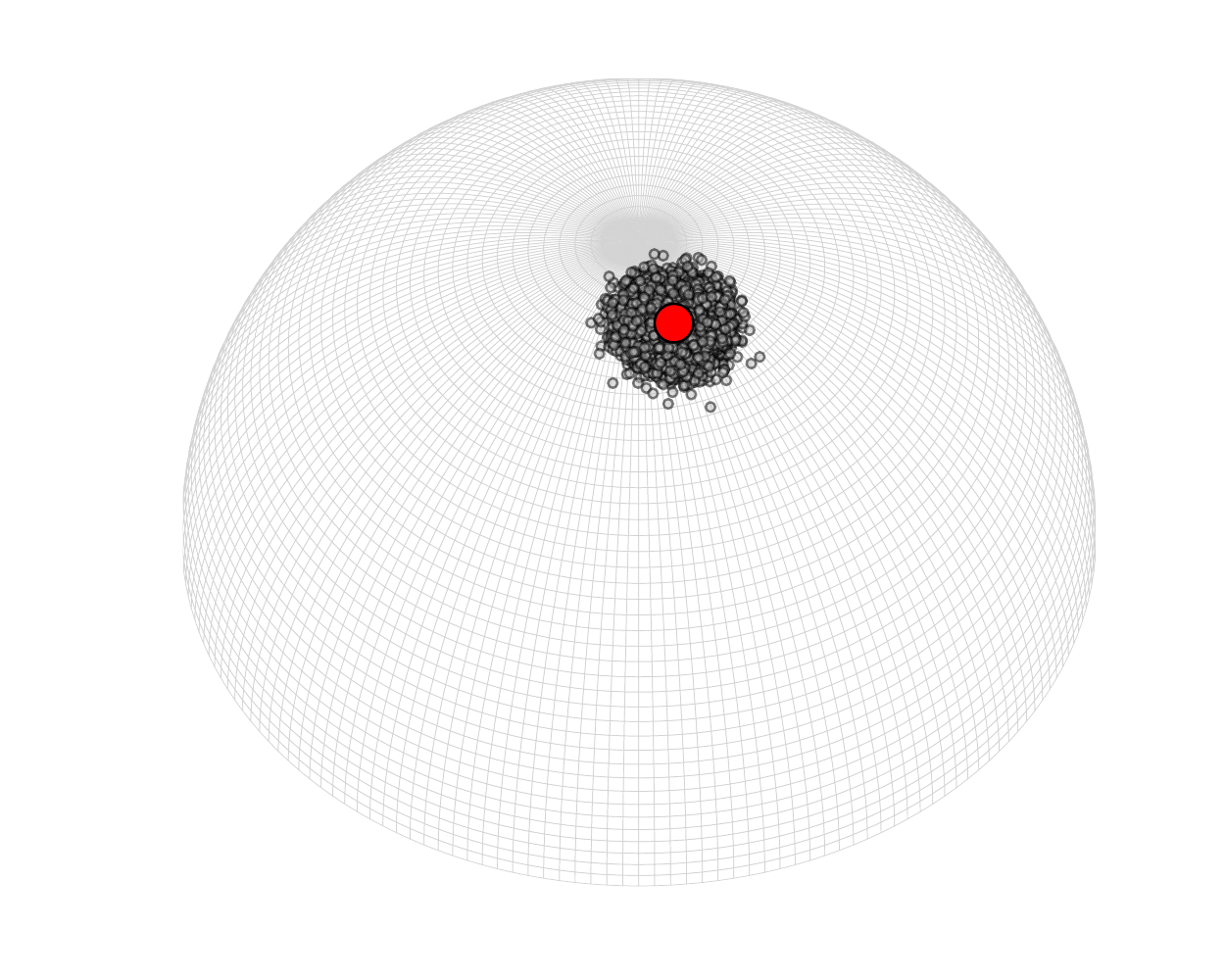}
         \caption{Step 1 corresponding to $\bs{s}_i = (30, 77)$.}
         \label{fig:makeXstep1_v2}
     \end{subfigure}
     \hfill
     \begin{subfigure}[b]{0.32\textwidth}
         \centering
         \includegraphics[width=\textwidth]{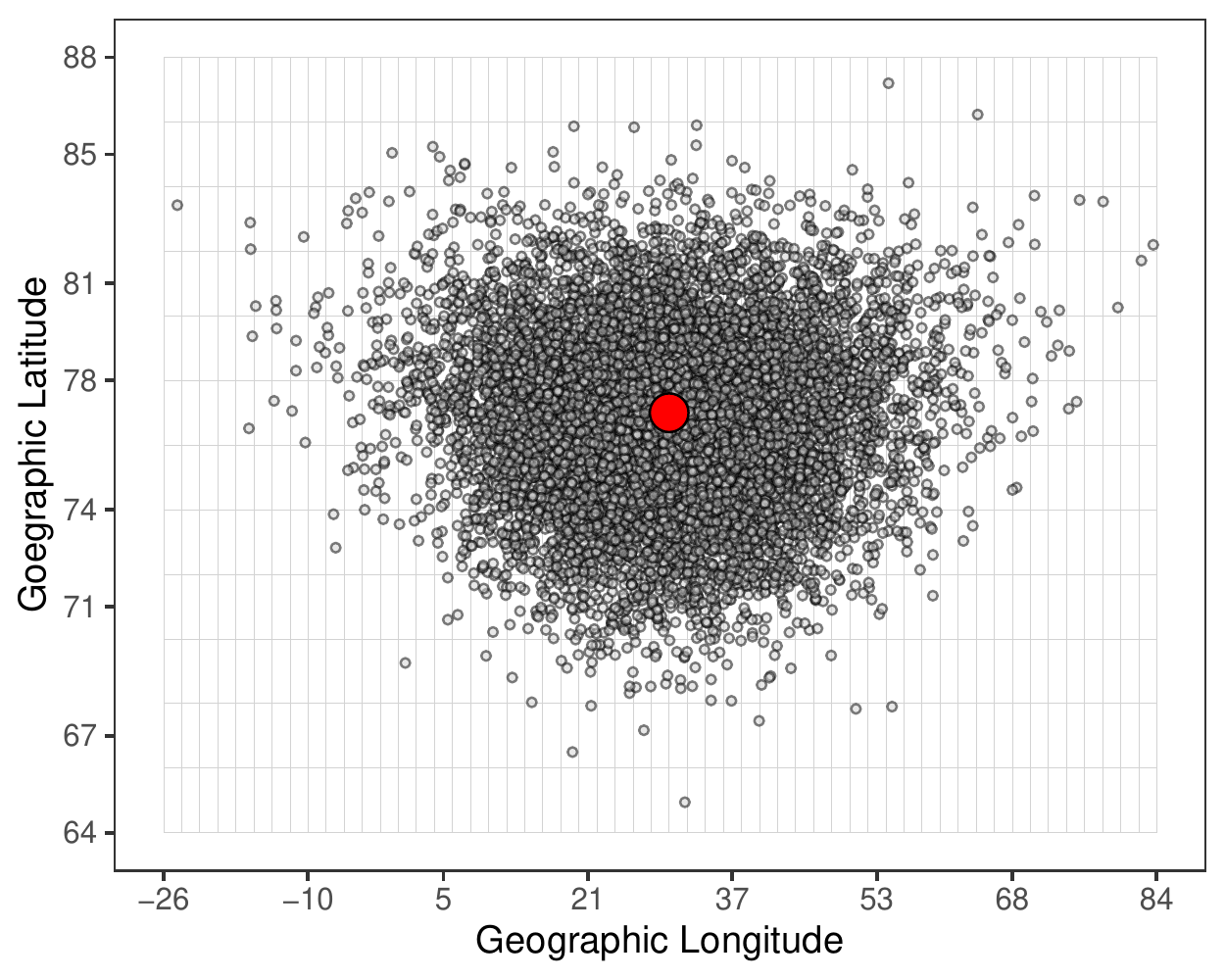}
         \caption{Step 2 corresponding to $\bs{s}_i = (30, 77)$.}
         \label{fig:makeXstep2_v2}
     \end{subfigure}
     \hfill
     \begin{subfigure}[b]{0.32\textwidth}
         \centering
         \includegraphics[width=\textwidth]{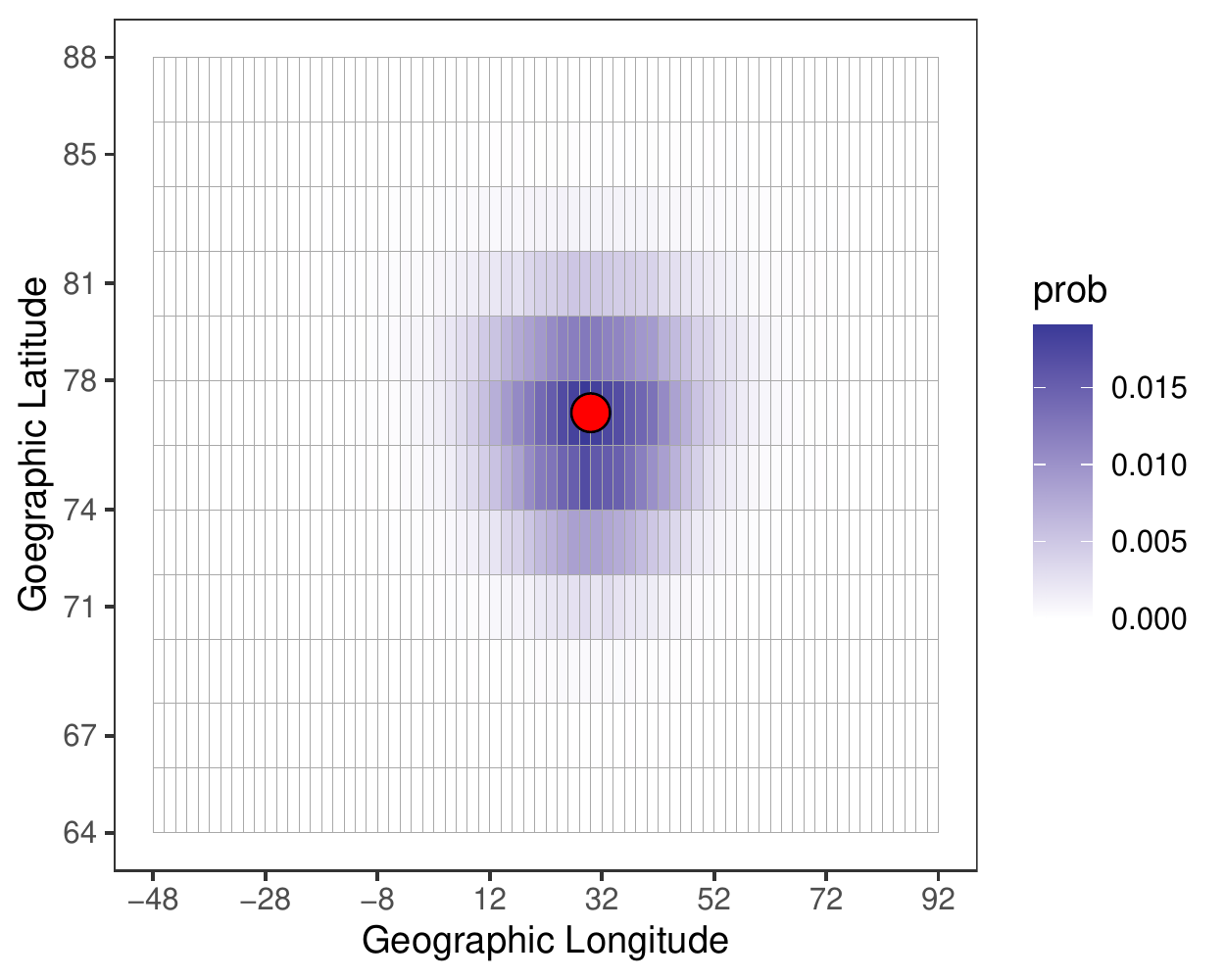}
         \caption{Step 3 corresponding to $\bs{s}_i = (30, 77)$.}
         \label{fig:makeXstep3_v2}
     \end{subfigure}
        \caption{Steps 1 through 3 constructing $\bs{K}(\bs{s}_i)$. The red point corresponds to the geographic longitude and latitude $\bs{s}_i = (30, 30)$ (top) and $\bs{s}_i = (30, 77)$ (bottom) in either spherical (Step 1) or geographic coordinates (Steps 2 and 3). The grey points are the draws $\bs{s}^s_{i,m}$ (Step 1) or $\bs{s}_{i,m}$ (Step 2) from the von Mises-Fisher distribution (10,000 draws shown). The coloring of the pixels (Step 3) correspond to the values of $\bs{K}(\bs{s}_i)$ based on $M=100,000$ draws.}
        \label{fig:makeX}
\end{figure}

Figures \ref{fig:makeXstep3} and \ref{fig:makeXstep3_v2} show $\bs{K}(\bs{s}_i)$ corresponding to $\bs{s}_i = (30,30)$ and $\bs{s}_i = (30,77)$, respectively. The number of pixels assigned non-zero proportions is higher near the poles (geographic latitude = -90 and 90) and lower near the equator (geographic latitude = 0) because the equal-area pixels in geographic coordinates do not have equal area in spherical coordinates. 

\clearpage
\section{GAM and PPR Details}
\label{appendix:stage1details}

Consider the basic model
\begin{linenomath}\begin{align}
    \label{eq:basiczmodel}
    z(\bs{s}_i) &= \theta(\bs{s}_i) + \epsilon(\bs{s}_i)
\end{align}\end{linenomath}
\noindent where $\bs{s}_i$ is the location of binned direct event datum $i$ in spherical coordinates, $z(\bs{s}_i) = y(\bs{s}_i)/t(\bs{s}_i) - b(\bs{s}_i)$, and $\epsilon(\bs{s}_i)$ is a mean 0 error term. 

The goal of step 1 in Theseus Stage 1 is to create $l=1,2,\ldots,N_L$ estimates of $\bs{\theta}$, $\dot{\bs{\theta}}_l$, referred to as candidates. $\bs{\theta}$ is a spatially continuous surface wrapped on a sphere. In this paper, we consider two different general approaches to estimate $\bs{\theta}$: generalized additive models (GAMs) and projection pursuit regression (PPR). For each of these approaches, we restrict the covariates to only include $\bs{s}_i$ (i.e., the estimated blurred sky map is a function of only spatial location).

\subsection{Generalized Additive Models}
\label{subsec:gams}
GAMs assume that $\theta(\bs{s}_i)$ in Equation \ref{eq:basiczmodel} can be represented as
\begin{linenomath}\begin{align}
    \label{eq:gam}
    \theta(\bs{s}_i) &= \sum_{k=1}^{N_K} b_k(\bs{s}_i) \alpha_k,
\end{align}\end{linenomath}
\noindent a linear combination of known basis functions of the argument $\bs{s}_i$.
The form of the basis function (e.g., cubic splines \citep{wahba1990spline} or P-splines \citep{eilers1996flexible}) and the number of knots $N_K - 2$ must be chosen. 
Once, however, they are chosen, Equation \ref{eq:gam} is a standard linear model. 
To protect from over-fitting, the parameter vector $\bs{\alpha} = (\alpha_1, \alpha_2, \ldots, \alpha_{N_K})'$ is estimated with a penalty on the second derivative of $\theta(\bs{s}_i)$, which penalizes ``wiggliness". 
This amounts to a minimization of a quadratic form, with respect to the parameter vector $\bs{\alpha}$ \citep[Section 3.2]{wood2006generalized}. 
Observation weights can also be provided, resulting in a minimization of a weighted quadratic form. 

For step 1 of Theseus Stage 1, we consider four combinations of two levels of weighting (no weighting and exposure time weighting) and two levels of basis functions (cubic splines and P-splines). The exposure time weighting gives more weight in the GAM fitting to locations that IBEX ``viewed" longer. The cubic splines and P-splines were somewhat arbitrarily chosen. Typically, one would select basis functions that would resemble features consistent with the same assumed features of the underlying function $\theta(\bs{s}_i)$. In our discovery science setting, however, there are few assumptions we are willing and able to make about the properties of the unblurred sky map, other than spatial continuity. 

We fit GAMs with six covariate components: univariate smooth functions for each of the three spherical coordinate directions and tensor smooths for each of the three two-way interactions of spherical coordinate directions. $N_K$ was set to 6 for each univariate basis and 36 for each two-way interaction basis; a relatively low number of basis functions in an attempt to speed up computations and prevent over-fitting. The GAM fitting was carried out using the \texttt{bam()} function in the \texttt{mgcv} package \citep{wood2016smoothing} in \texttt{R}: the analogue to \texttt{gam()} for large data sets.
The resulting estimated candidate sky maps $\dot{\bs{\theta}}_l$ are not constrained to be non-negative, though physically, ENA rates must be non-negative. 
All negative GAM-fitted ENA rates were truncated to 0.

\subsection{Projection Pursuit Regression}
\label{subsec:ppr}
PPR \citep{friedman1981projection} is similar to a GAM in that it is predicated on a linear combination of basis functions. It, however, deviates from GAMs in important ways. PPR follows a successive refinement approach, where a hierarchy of increasingly complex models are fit to the data until a stopping rule is reached. PPR constructs a smooth function of linear combinations of predictors (rather than smooth combinations of each predictor separately, as is the case with GAMs), called ridge functions. Next, PPR projects the inputs that maximize the unexplained variability by the current model fit and fits another ridge function to a linear combination of inputs. This procedure of projecting inputs into optimal unexplained directions is repeatedly done until a stopping rule is reached, where the unexplained variability is unable to be appreciably reduced. As with GAMs, the functional family of the smooth functions and the decision to weight or not weight the observations must be made. 

For step 1 of Theseus Stage 1, we considered four combinations of two levels of weighting (no weighting and exposure time weighting) and two levels of smooth functions (Friedman's super smoother \citep{friedman1984smart} and cubic smoothing splines with degrees of freedom chosen via generalized cross-validation). The PPR fitting was carried out using the \texttt{ppr()} function in the \texttt{stats} package in \texttt{R}. We fit PPR to the three spherical coordinate directions and set the maximum number of terms to 100; an intentionally large number allowing the internal stopping rule to terminate fitting, rather than prematurely doing so with the maximum number of terms is reached.
As with the GAMs, PPR provides no non-negativity constraint on the fitted ENA rates.
All negative PPR-fitted ENA rates were truncated to 0.

\clearpage
\section{Making ISOC Sky Maps}
\label{appendix:isoc}

Let
\begin{linenomath}\begin{align}
    y(\bs{s}_i)~|~t(\bs{s}_i),\theta(\bs{s}_i), b(\bs{s}_i) &\sim \text{Poisson}\big(t(\bs{s}_i)[\theta(\bs{s}_i) + b(\bs{s}_i)]\big)
\end{align}\end{linenomath}
\noindent where 
\begin{itemize}
    \item $\bs{s}_i$ is the ecliptic longitude/latitude location for observation $i=1,2,\ldots,N_o$
    \item $y(\bs{s}_i) \in 0,1,2, \ldots $ is the number of direct events (ENAs $+$ background particles) counted by IBEX-Hi
    \item $t(\bs{s}_i) > 0$ is the known exposure time (seconds)
    \item $b(\bs{s}_i) \geq 0$ is the unknown background rate (background particles/second) but with known mean $\mu_b(\bs{s}_i)$ and variance $\sigma^2_b(\bs{s}_i)$
    \item $\theta(\bs{s}_i) \geq 0$ is the unknown ENA rate (ENAs/second). (Recall that ISOC does not distinguish between a blurred and an unblurred sky map).
\end{itemize}
The ISOC sky map rendering methodology assumes $\theta(\bs{s}_i)$ is a random variable where
\begin{linenomath}\begin{align}
\label{eq:meanS}
\text{E}\big(\theta(\bs{s}_i)\big) &= \mu_{\theta}(\bs{s}_i)\\
\label{eq:varS}
\text{Var}\big(\theta(\bs{s}_i)\big) &= \sigma^2_{\theta}(\bs{s}_i)\\
\label{eq:covS}
\text{Cov}\big(\theta(\bs{s}_i), \theta(\bs{s}_{i'})\big) &= 0
\end{align}\end{linenomath}
\noindent for $\bs{s}_i \neq \bs{s}_{i'}$.

ISOC estimates $\mu_{\theta}(\bs{s}_i)$ with the method of moments
\begin{linenomath}\begin{align}
\label{eq:muhat}
\hat{\mu}_{\theta}(\bs{s}_i) &= \frac{y(\bs{s}_i)}{t(\bs{s}_i)} - \mu_b(\bs{s}_i),
\end{align}\end{linenomath}
\noindent and estimates $\sigma^2_{\theta}(\bs{s}_i)$ as
\begin{linenomath}\begin{align}
\label{eq:varhat}
\hat{\sigma}^2_{\theta}(\bs{s}_i) &= \frac{y(\bs{s}_i)}{t^2(\bs{s}_i)} + \sigma^2_{b}(\bs{s}_i).
\end{align}\end{linenomath}

Let $P(\bs{s}_j)$ be the random variable of the ENA rate corresponding to the pixel centered at location $\bs{s}_j$. The ISOC defines $P(\bs{s}_j)$ as a linear combination of random variables $\theta(\bs{s}_i)$:
\begin{linenomath}\begin{align}
\label{eq:pixdef}
P(\bs{s}_j) &= \sum_{i=1}^{N_j} w(\bs{s}_i, \bs{s}_j) \theta(\bs{s}_i)\\
w(\bs{s}_i, \bs{s}_j) &= \frac{t(\bs{s}_i) d(\bs{s}_i,\bs{s}_j)}{\sum_{i=1}^{N_j}t(\bs{s}_i) d(\bs{s}_i,\bs{s}_j)}\\
d(\bs{s}_i, \bs{s}_j) &= \bigg(1 - \frac{|\text{lat}(\bs{s}_i) - \text{lat}(\bs{s}_j)|}{7}\bigg) \bigg(1 - \frac{|\text{lon}(\bs{s}_i) - \text{lon}(\bs{s}_j)|}{7}\bigg)
\end{align}\end{linenomath}
\noindent where 
\begin{itemize}
    \item $N_j$ is the total number of binned direct event data observations in the weighted average, defined as all binned direct event data $i$ where $\bs{s}_i$ is within $6\degree$~ecliptic latitude and $7\degree$~ecliptic longitude of the pixel center $\bs{s}_j$
    \item $\text{lat}(\bs{s}_i)$ and $\text{lon}(\bs{s}_i)$ are the ecliptic latitude and longitude for location $\bs{s}_i$, respectively. The difference in longitudes, $|\text{lon}(\bs{s}_i) - \text{lon}(\bs{s}_j)|$, accounts for spherical wrapping.
    \item $d(\bs{s}_i, \bs{s}_j) \in [0,1]$ is an inverse distance between binned direct event data location $\bs{s}_i$ and pixel center $\bs{s}_j$.
    \item $w(\bs{s}_i, \bs{s}_j) \geq 0$ is the weight assigned to the pair ($\bs{s}_i, \bs{s}_j$) where $\sum_{i=1}^{N_j}w(\bs{s}_i, \bs{s}_j) = 1$. $w(\bs{s}_i, \bs{s}_j)$ is a function of exposure time of binned direct event datum $i$ and the distance between locations $\bs{s}_i$ and $\bs{s}_j$. The closer observation location $\bs{s}_i$ is to pixel center $\bs{s}_j$ and the larger the exposure time $t(\bs{s}_i)$ is, the more weight observation $i$ is assigned.
\end{itemize} 

ISOC computes the mean and variance for pixel $P(\bs{s}_j)$ as the mean and variance of linear combinations of assumed independent random variables:
\begin{linenomath}\begin{align}
\label{eq:meanpix}
\text{E}\big(P(\bs{s}_j)\big) &= \text{E}\bigg(\sum_{i=1}^{N_j} w(\bs{s}_i, \bs{s}_j) \theta(\bs{s}_i)\bigg) = \sum_{i=1}^{N_j} w(\bs{s}_i, \bs{s}_j) \text{E}\big(\theta(\bs{s}_i)\big) = \sum_{i=1}^{N_j} w(\bs{s}_i, \bs{s}_j) \mu_{\theta}(\bs{s}_i)\\
\label{eq:varpix}
\text{Var}\big(P(\bs{s}_j)\big) &= \text{Var}\bigg(\sum_{i=1}^{N_j} w(\bs{s}_i, \bs{s}_j) \theta(\bs{s}_i)\bigg) = \sum_{i=1}^{N_j} w^2(\bs{s}_i, \bs{s}_j) \text{Var}\big(\theta(\bs{s}_i)\big) = \sum_{i=1}^{N_j} w^2(\bs{s}_i, \bs{s}_j) \sigma^2_{\theta}(\bs{s}_i).
\end{align}\end{linenomath}

Finally, ISOC estimates $\text{E}(P(\bs{s}_j))$ and $\text{Var}(P(\bs{s}_j))$ by replacing $\mu_{\theta}(\bs{s}_i)$ in Equation \ref{eq:meanpix} with $\hat{\mu}_{\theta}(\bs{s}_i)$ in Equation \ref{eq:muhat} and replacing $\sigma^2_{\theta}(\bs{s}_i)$ in Equation \ref{eq:varpix} with $\hat{\sigma}^2_{\theta}(\bs{s}_i)$ in Equation \ref{eq:varhat}.

The above procedure is carried out for all pixels. If no binned direct event data locations $\bs{s}_i$ fall within $6\degree$~latitude and $7\degree$~longitude of pixel center $\bs{s}_j$, no estimate is returned. If only one observation is close enough to $\bs{s}_j$, a point estimate is returned but no variance estimate.